\documentclass[pre,twocolumn,nofootinbib,twoside,showpacs,superscriptaddress,tightenlines,floatfix]{revtex4-2}
\usepackage{dcolumn}
\usepackage{lipsum}
\usepackage{graphicx,amssymb,amsmath,epsf,bm}
\usepackage[utf8x]{inputenc}
\usepackage[T1]{fontenc}
\usepackage[normalem]{ulem}
\usepackage{color}
\usepackage{mathtools}
\usepackage{dsfont}
\usepackage{nicefrac}
\usepackage{hyperref}
\hypersetup{
    allcolors=blue,
    colorlinks=true,
    bookmarks=true,
    pdfpagemode=FullScreen,
}
\definecolor{nred}{RGB}{224,0,0}
\definecolor{nblue}  {RGB}{28,130,185}
\definecolor{dgreen} {RGB}{38,238,21}
\definecolor{norange}{RGB}{230,120,20}

\newcommand{\Tr}{{\rm Tr}}

\begin{document}


\title{Conformal invariance and composite operators: A strategy for improving the derivative expansion of the nonperturbative renormalization group}

\author{Bertrand Delamotte}
\affiliation{Sorbonne Universit\'e, CNRS, Laboratoire de Physique Th\'eorique de la Mati\`ere Condens\'ee, LPTMC, 75005 Paris, France}%

\author{Gonzalo De Polsi}
\affiliation{Instituto de F\'isica, Facultad de Ciencias, Universidad de la
	Rep\'ublica, Igu\'a 4225, 11400, Montevideo, Uruguay
}%

\author{Matthieu Tissier}
\affiliation{Sorbonne Universit\'e, CNRS, Laboratoire de Physique Th\'eorique de la Mati\`ere Condens\'ee, LPTMC, 75005 Paris, France}%

\author{Nicol\'as Wschebor}
\affiliation{Instituto de F\'isica, Facultad de Ingenier\'ia, Universidad de la
Rep\'ublica, J.H.y Reissig 565, 11000 Montevideo, Uruguay
}%

\begin{abstract}
  It is expected that conformal symmetry is an emergent property of many systems at their critical point. This imposes strong constraints on the critical behavior  of a given system. Taking them into account in theoretical approaches can lead to a better understanding of the critical physics or improve approximation schemes. However, within the framework of the non-perturbative or functional renormalization group and, in particular, of one of its most used approximation schemes, the Derivative Expansion (DE), non-trivial constraints only apply from third order (usually denoted $\mathcal{O}(\partial^4)$), at least in the usual formulation of the DE that 
  includes correlation functions involving only the order parameter.
  In this work, we implement conformal constraints on a generalized DE including composite operators and show that new
  constraints already appear at second order of the DE (or $\mathcal{O}(\partial^2)$). We show how these constraints can be used to fix nonphysical regulator parameters. 
\end{abstract}

\maketitle

\section{Introduction}

The Renormalization Group {\it \`a la} Wilson \cite{Wilson:1971dc,Wilson:1973jj} has been the theoretical framework for quantitatively describing critical phenomena as well as the presence of scale invariance and universality near critical points. Almost simultaneously with Wilson's original work,  Polyakov and Migdal \cite{Polyakov:1970xd,Migdal:1971xh} conjectured that conformal invariance holds at criticality and not just scale invariance. This assumption allowed to fully classify critical phenomena in two dimensions \cite{DiFrancesco:1997nk,Belavin:1984vu}. In recent years, the theoretical study of a significant number of critical systems has made considerable progress. In addition to an important advance in traditional methods such as the $\epsilon$-expansion (in which the order $\epsilon^7$ has recently been reached \cite{Schnetz:2016fhy,Kompaniets:2017yct,Shalaby:2020faz,Abhignan06,Shalaby:2020xvv}) or Monte-Carlo simulations \cite{Hasenbusch:2019jkj,Hasenbusch2005,Hasenbusch:2021tei}, two methodologies with less history made notable progress. One is known as the Conformal Bootstrap which, using conformal invariance, operator product expansion and unitarity, gives very stringent bounds on several exponents (and other critical parameters) in various models \cite{ElShowk:2012ht,El-Showk:2014dwa,Kos:2014bka,chester2019carving}. The other one is the Functional or Non-Perturbative Renormalization Group (FRG) \cite{Wetterich:1992yh,Ellwanger:1993kk,Morris:1993qb,
Delamotte:2007pf,Dupuis:2020fhh}. It is interesting to observe that these two methodologies are nothing but, on one side, the modern version of Wilson renormalization group and on the other side, of Polyakov and Migdal conformal symmetry analysis.

FRG-related methods, although at the heart of Wilson's initial ideas, were quickly overtaken as mainstream methods by a perturbative version of the Renormalization Group (RG) \cite{Guida:1998bx,Pelissetto02,ZinnJustin:2002ru}. The reason for this was the success of the $\epsilon$-expansion for the understanding of the Ising universality class and, more generally, of $O(N)$ models \cite{Wilson:1971dc,Pelissetto02,ZinnJustin:2002ru}. Only very recently, functional methods managed to catch up and surpass perturbative methods in accuracy in such models. Moreover, the FRG has shown a versatility for the formulation of non-perturbative approximations allowing one to study various problems that are clearly beyond the scope of perturbation theory. Just to mention some examples, it has been applied with great success to the study of the  Random Field Ising model (spontaneous supersymmetry breaking and the associated breaking of dimensional 
reduction in a nontrivial dimension) \cite{Tarjus:2004wyx,Tissier:2011zz}, 
the Kardar-Parisi-Zhang equation in dimensions larger than one (identification 
of the strong coupling fixed point) \cite{Canet10,Canet11,Canet11a},  the glassy phase of crystalline membranes \cite{Coquand2017},  systems showing different critical exponents in their high and low temperature phases \cite{Leonard15}, the phase diagram of reaction-diffusion systems \cite{Canet:2004je,Canet:2003yu}, the calculation of non-universal critical temperatures of spin systems \cite{Machado:2010wi}, the calculation of bound states masses in 3D Ising model near criticality \cite{Rose:2016wqz}, etc.
For a recent review on the FRG including other examples of application, see \cite{Dupuis:2020fhh}; for a pedagogical presentation of the method, see \cite{Delamotte:2007pf}.

In a similar way, methods based on conformal invariance, long before the relatively recent advances based on the Conformal Bootstrap \cite{ElShowk:2012ht,Kos:2014bka,chester2019carving,meneses2019structural}, allowed the development of an entire branch of mathematical physics oriented towards the study of two-dimensional conformal models \cite{DiFrancesco:1997nk}, starting from the seminal work \cite{Belavin:1984vu}.

The previous discussion highlights the fact that the various methods based on the RG (in their perturbative or non-perturbative variants), as well as the methods based on conformal invariance, have a long history. Paradoxically, however, there has been little dialogue between the two methodologies. From the viewpoint of methods based on the RG, the reason is clear: when implementing an approximation scheme in such a context, the existence of a critical point is expressed by the existence of a fixed point of the RG equations. Since only analytic and globally well-defined solutions of the fixed point equation are physically acceptable, this generically leads to a discrete set of isolated fixed points. These fixed points rigidly determine the associated critical (or multicritical) properties without ever using conformal invariance. From this perspective, conformal invariance appears as superfluous. At the level of exact equations, once dilatation invariance is imposed, the conformal symmetry should be fulfilled automatically \cite{delamotte2016scale,meneses2019structural}. However, because of unavoidable approximations, the conformal symmetry is, in practice, broken. It is expected that good approximation schemes would yield a small breaking of conformal symmetry. The reasons why conformal methods have failed to take advantage of the RG are symmetric. Both in the case of two-dimensional exact solutions and in the context of the Conformal Bootstrap, once the method in question has been employed, the solution is what it is and there is no room left to exploit the RG. Obviously, exceptions to this situation must be noted, notably for studying the Renormalization flow between two conformal fixed points (see, for example, \cite{cardy_1996}) but such exceptions only confirm the rule that only one family of methods is used to study the fixed points themselves.

For some years now, we have been studying the link between conformal invariance and FRG. In this context, it is natural (as for scale invariance) to analyze an infrared regularized version of conformal invariance which includes as a particular case the usual conformal invariance (in the limit where the infrared regulator vanishes). Initial studies have focused on trying to prove conformal invariance by exploiting the FRG equations plus specific properties of particular models \cite{delamotte2016scale,DePolsi2019}. More recently, some of the authors performed a first analysis showing one way to exploit conformal invariance in the Ising universality class in the context of the FRG \cite{Balog2020}. This study relied on the widely used Derivative Expansion (DE) which consists of projecting the Gibbs free energy on a functional form including terms with a given number of derivatives or less (see Sec.~\ref{secRes} for more details). Equivalently, the DE can be seen as a systematic expansion of the proper vertices regularized at small wave numbers. DE has been studied in great detail and an apparent convergence has been shown with the order of magnitude of successive terms being controlled by a parameter of the order of $\sim 1/4$ \cite{Balog:2019rrg,DePolsi2020,DePolsi2021}. In references \cite{Balog:2019rrg,DePolsi2020,DePolsi2021} only correlation functions of the local order parameter were analyzed. For these correlation functions it was first observed that conformal invariance is automatically satisfied at the fixed point at first and second order of that approximation scheme. The first non-trivial consequence of conformal invariance appears in the third order of the DE where all terms including four derivatives in the free energy are included. 

The first purpose of this paper is to generalize our previous work \cite{Balog2020} by studying correlation functions that involve composite operators. These correlation functions play an important physical role. They contain, in particular, the information concerning the various perturbations around the fixed point such as, for example, the critical exponents $\nu$ or $\omega$. In fact, considering composite operators in the context of FRG is quite natural since  they play a very important role in the conformal bootstrap program. This being said, even if studying such correlation functions in the context of FRG is not new at all \cite{Ellwanger:1993kk,Berges:2000ew, Dupuis:2020fhh}, applications are rather scarce in the context of statistical mechanics (see however \cite{Rose:2015bma,Rose:2021zdk}).

Our second purpose is to make use of this formalism to improve the accuracy of approximate FRG calculations. Approximations, be they implemented in a perturbative or nonperturbative framework, induce the breaking of many exact properties and as a consequence make physical quantities depend on the implementation of the approximate calculations. For instance, the critical exponents at one loop differ whether they are computed in the $\epsilon$-expansion (setting $\epsilon=1$ at the end of the calculation) or at fixed dimension in $d=3$  (in the massive zero momentum scheme for instance). Moreover, even at large orders, the perturbative calculation of critical exponents in $d=3$ needs to be supplemented by resummation techniques (Padé-Borel for instance) that unavoidably involve arbitrary parameters that must be fixed by one way or another and the dependence of the exponents on these parameters is in general not small. Thus physical predictions have spurious dependencies on nonphysical parameters. In the context of the FRG, a similar phenomenon occurs: while physical quantities (such as critical temperatures or exponents) should not depend on the choice of the coarse-graining procedure (regulating function), approximations induce a spurious dependence on this choice that must be fixed. This dependence has been studied rather extensively by the FRG community, see for instance \cite{Dupuis:2020fhh}, section 2.3.2. We show in this article that conformal invariance can help us choosing the regulating function and thus eliminating the arbitrariness induced by implementing the DE.

 In fact, by considering correlation functions that only include the local order parameter, conformal invariance only gives rise to non-trivial consequences at third order ($O(\partial^4)$) of the DE \cite{Balog2020}. Such order has been employed only rarely in the literature and only for O($N$) models, which significantly reduces the impact of the use of conformal invariance. In the present work, we show that in correlation functions including composite operators non-trivial information coming from conformal invariance is obtained from the second order ($O(\partial^2)$) of such an approximation scheme.  At odds with the third order, the second order  has been applied in an enormous diversity of systems \cite{Dupuis:2020fhh}, which qualitatively broadens the domain of application of conformal invariance in this context. 

The article is organized as follows. We first quickly recall the basics of FRG and discuss how composite operators are treated in this formalism. We then derive the Ward identities associated with the invariance under dilatations, which can be expressed as a fixed point of the RG transformation. Following the same path, we derive the Ward identity for conformal invariance in this framework. We then implement the dilatation and conformal Ward identities in a specific truncation, namely the DE at order $\mathcal O(\partial^2)$. We show that conformal invariance is not automatically realized at the fixed point of the FRG equation because of the approximation scheme considered and study the dependence of this spurious breaking of conformal Ward identity with the regulating function. This naturally leads to a criterion for choosing the regulator, that was coined the Principle of Maximal Conformality (PMC) in \cite{Balog2020}. We compare this criterion to the more widely used Principle of Minimal Sensitivity (PMS) and show that the two are compatible for the critical exponents $\nu$ and $\omega$.

\section{The functional renormalization group in the presence of a source for composite operators}\label{secFRG}

In a nutshell, the FRG is a framework which allows to incorporate progressively rapid fluctuations from microscopic to macroscopic scales in the computation of a Gibbs free energy. This is done by adding to the action $S$ a scale-dependent term $\Delta S_k$ that regulates infrared fluctuations (with respect to the momentum scale $k$) keeping fast fluctuations unaltered. We present here the method in the case of a scalar field with ${\mathbb Z}_2$ symmetry, but generalizing to more complex theories is straightforward.
It is convenient to take this regulating term in the form of a mass-like term, quadratic in the fields \cite{Polchinski:1983gv}: 
\begin{equation}\label{deltaSDirect}
    \Delta S_k[\varphi]=\frac 1 2 \int_{x,y}\varphi(x) R_k(x,y) \varphi(y),
\end{equation}
where $\int_x=\int d^dx$. The regulator $R_k$ is chosen to be invariant under translations and rotations, which means that its Fourier transform is just a function of $q^2$. For dimensional reasons one usually writes the regulator in direct space and in Fourier space as: \footnote{Notice that we employ the same symbol for the regulator $R_k$ and its Fourier transform. We employ the same criterion for the dimensionless counterpart $r$.}
\begin{equation}\label{eq:RegProf}
\begin{split}
    R_k(x,y)&=\alpha Z_k k^{2+d} r(k^2|x-y|^2),\\
    R_k(q^2)&=\alpha Z_k k^2 r(q^2/k^2),
\end{split}
\end{equation}
factorizing the overall scale of the regulator $\alpha$ and a field renormalization factor $Z_k$. For the regulating term to behave properly, we impose that the regulating function $r$ decreases faster than any power law for large arguments and tends to 1 for $q\ll k$.

The canonical approach in the FRG, since its development three decades ago \cite{Wetterich:1992yh,Ellwanger:1993kk,Morris:1993qb}, relies on adding this regulating term to the action in the functional integral in the presence of a source for the field. One can consider the same approach but also in the presence of a source $K(x)$ for a local composite operator \cite{Ellwanger:1993kk,Berges:2000ew,Pawlowski:2005xe,Dupuis:2020fhh,Rose:2015bma,Rose:2021zdk}, that we denote $\mathcal{O}(x)$. 
\begin{equation}\label{functIntegral}
	e^{W_k[J,K]}=\int \mathcal{D}\varphi e^{-S[\varphi]-\Delta S_k[\varphi]+\int_x J(x)\varphi(x)+\int_x K(x)\mathcal{O}(x)}
\end{equation}
Unless otherwise stated, average values will be taken with respect to this measure, in the presence of sources $J$ and $K$. It is implicitly assumed all along this article that the theory has been regularized somehow in the ultraviolet and we call $\Lambda$ the ultraviolet cut-off.

A generic local scalar composite operator can be expressed as a linear combination of scaling operators, which are local eigenperturbations of the RG flow around the fixed point. Since the fixed point of the RG flow corresponds to a scale-invariant theory, these operators are also eigenvectors of the dilatation transformation. They can therefore be classified according to their scaling dimension as well as their parity under $\mathbb{Z}_2$ transformations. The product of two such operators can also be expressed in that base, a procedure known as the Operator Product Expansion (OPE) \cite{Wilson1969,Zimmerman1973} first introduced by Wilson \cite{wilson1964products}. For example, under this assumption, a generic $\mathbb{Z}_2$-symmetric composite operator $\mathcal{O}(x)$ can be written as 
\begin{equation}\label{ope}
	\mathcal{O}(x)\sim\mathds{1}Z_0^{\mathcal{O}}(k)+\varphi_R^2(x)Z_1^{\mathcal{O}}(k)+\dots,
\end{equation}

\subsection{An example: the $\varphi^2(x)$ operator}
\label{phi2ex}
Consider, for the sake of clarity, the operator $\mathcal O(x)=\varphi^2(x)$. The  operator $\varphi^2_R(x)$ in Eq. \eqref{ope} is chosen such that $\left\langle\varphi^2_R\right\rangle=0$ and similarly for all other dilatation eigenvectors for any $k$ (except the identity). In this case,  $Z_0^{\varphi^2}(k)$ can be fixed as:
\begin{equation}
	Z_0^{\varphi^2}(k)=\left\langle\varphi^2(x)\right\rangle_{J=K=0},
\end{equation}
which is $x-$independent due to translation invariance. With the definition given in Eq. \eqref{ope} applied to this operator, we have that:

\begin{equation}\label{equivBphiRenandphi}
\begin{split}
    &\left\langle\varphi^2(x)\varphi^2(y)\right\rangle_{c,J=K=0}\sim
	(Z_1^{\varphi^2})^2\left\langle \varphi^2_R\,(x) \varphi^2_R\,(y)\right\rangle_{J=K=0} \\
  &\hspace{1cm}\sim\frac{(Z_1^{\varphi^2})^2}{|x-y|^{2D_{\varphi^2}}},\,\,\mathrm{for}\,\Lambda^{-1}\ll |x-y|\ll k^{-1},\xi
\end{split}
\end{equation}
where the $c$ subscript means that we are referring to the connected correlation function and $\sim$ stands for the leading behavior in the critical regime, corresponding to $|x-y|$ much larger than the microscopic length scale (e.g. lattice spacing $\sim \Lambda^{-1}$) and much smaller than both the correlation length and $k^{-1}$, the infrared scale provided by the regulator introduced in Eq.~\eqref{eq:RegProf}.  The last identity fixes the normalization of the operator $\varphi^2_R$. Eq.~\eqref{equivBphiRenandphi} means that, generically, and up to a normalization, all composite operators with the same symmetry properties as the $\varphi^2(x)$ operator have the same connected correlation functions at long distances.  In particular, for connected correlation functions, the operators $\varphi^2$ and $\varphi^2_R$ are proportional at long distances. The only exception to this rule is the case where $Z_1^{\varphi^2}$ vanishes. In this case, the first irrelevant operator cannot be neglected and must be considered, as discussed below.

Now, the singular part of the specific heat  is related to $W[J,K]=W_{k=0}[J,K]$ by $C_{\rm sing.}\propto\partial_T^2W$ and behaves according to the power law:
\begin{equation}
	C_{\rm sing.}\propto|T-T_c|^{-\alpha},
\end{equation}
which defines the critical exponent $\alpha$, not to be confused with the parameter $\alpha$ introduced in Eq. \eqref{eq:RegProf}. This power-law can be extracted from correlation functions for the composite operator in the following way. Reintroducing the temperature-dependence in the definition of the free energy, Eq.~\eqref{functIntegral}, we conclude that a derivative with respect to $T$ yields the integral over space of the Hamiltonian (or Euclidean action) density which behaves according to Eq.~\eqref{ope} and is therefore dominated by $\phi_R^2(x)$, up to an uninteresting additive constant. Since
\begin{equation}
	\left\langle \varphi^2\,(x)\varphi^2\,(y)\right\rangle_c=\frac{\delta^2 W}{\delta K(x)\delta K(y)}.
\end{equation}
and using the first equivalence in Eq.~\eqref{equivBphiRenandphi}, we conclude that:
\begin{equation}
	\int_{x,y}\left\langle\varphi^2\,(x) \varphi^2\,(y)\right\rangle_{c,J=K=0}\sim|T-T_c|^{-\alpha}.
\end{equation}
Now, due to translation invariance, one of the space integrals contributes to a volume factor while the other contributes as $\xi^{d-2D_{\varphi^2}}$. Using the scaling behavior $\xi\sim |T-T_c|^{-\nu}$ and the scaling law $\alpha=2-\nu d$ we conclude that 
\begin{equation}\label{Dphi2andCEnu}
	D_{\varphi^2}=d-\frac{1}{\nu}.
\end{equation}
This equation implies that $\varphi^2_R(x)$ is nothing but the most relevant even and non-trivial eigenperturbation of the dilatation operator around the fixed point as was stated initially.

\subsection{The FRG flow equation and fixed point perturbations}
Returning to the functional integral \eqref{functIntegral} in the general case of a composite operator $\mathcal O$ and taking the modified Legendre transform to the free energy $W_k[J,K]$ with respect of $J$ defines the scale-dependent effective action $\Gamma_k[\phi,K]$ as:
\begin{equation}
\label{eq_legendre}
	\Gamma_k[\phi,K]=-W_k[J,K]-\Delta S_k[\phi]+\int_x J(x)\phi(x),
\end{equation}
This unusual definition of the Legendre transform, with the subtraction of the regulating term, ensures that $\Gamma_k[\phi,K]$ interpolates between the bare action at scale $k=\Lambda$, $\Gamma_{k=\Lambda}[\phi,K]=S[\phi]+\int_x K(x)\mathcal{O}(x)$ and the effective action at $k=0$, i.e. $\Gamma_{k=0}=\Gamma[\phi,K]$. Also, $\phi(x)$ is defined, starting from Eq.~\eqref{functIntegral}, as 
\begin{equation}\label{orderParam}
\frac{\delta W_k}{\delta J(x)}=\langle
\varphi(x)\rangle_{J,K}\equiv\phi(x),
\end{equation}
where the differentiation with respect to $J(x)$ is taken at constant $K$. 
Deriving \eqref{eq_legendre} with respect to $K$ we obtain 
\begin{equation}\label{W1minusGamma1}
\frac{\delta W_k}{\delta K(y)}=-\frac{\delta \Gamma_k}{\delta K(y)}.
\end{equation}
In the previous equation, the differentiations are performed at fixed $J$ in the left hand side and at fixed $\phi$ in the right hand side.  In the following, we make no explicit reference  to the variable that are kept constant under differentiation when this is the expected one. 
Another thing to keep in mind is that when working with the effective action $\Gamma_k[\phi,K]$, the two-point function of the composite operator $\mathcal O$ is related to derivatives of $\Gamma_k$ in a nontrivial way \cite{Rose:2015bma,Rose:2021zdk}. To be specific, it can be easily verified that:
\begin{equation}
\begin{split}
\big\langle\mathcal O(x)\mathcal O(y)\big\rangle_c&=\frac{\delta^2 W_k}{\delta K(x)\delta K(y)}\\
&=-\frac{\delta^2 \Gamma_k}{\delta K(x)\delta K(y)}+\\
\int_{z,w}&\frac{\delta^2 \Gamma_k}{\delta K(x)\delta\phi(w)}G_k(w,z)\frac{\delta^2 \Gamma_k}{\delta\phi(z)\delta K(y)}, 
\end{split}
\end{equation}
where the propagator $G_k(x,y)$ is as usual:
\begin{equation}
    \left\langle\varphi(x)\varphi(y)\right\rangle_c=\frac{\delta^2 W_k}{\delta J(x)\delta J(y)}=\Big(\frac{\delta^2\Gamma_k}{\delta\phi\delta\phi}+R_k\Big)_{x,y}^{-1}.
\end{equation}
For completeness and, in the same manner, we can express the correlation function $\left\langle\varphi(x)\mathcal O(y)\right\rangle_c$ in terms of derivatives of the effective average action:
\begin{align}
\big\langle\varphi(x)\mathcal O(y)\big\rangle_c&=\frac{\delta^2 W_k}{\delta J(x)\delta K(y)}
=-\int_z\frac{\delta \phi(z)}{\delta J(x)}\frac{\delta^2 \Gamma_k}{\delta\phi(z)\delta K(y)}\nonumber\\
&=-\int_z G_k(x,z) \frac{\delta^2 \Gamma_k}{\delta\phi(z)\delta K(y)}.  
\end{align}

As previously stated, the idea behind the FRG is to progressively integrate fluctuations from scale $k=\Lambda$ to $k=0$. It can be shown (see for example \cite{Delamotte:2007pf}), that the equation  governing the evolution of $\Gamma_k$ reads:
\begin{equation}\label{WettEq}
	\partial_t \Gamma_k[\phi,K]=\frac{1}{2}\int_{x,y}\partial_t R_k(x,y)G_k(x,y),
\end{equation}
where $\partial_t=k\partial_k$.

When analyzing the critical properties of the system we must first consider dimensionless variables in order to find a fixed point of Eq. \eqref{WettEq}. To this end, we introduce:
\begin{equation}
\label{dimensionless}
   \begin{split}
       \tilde{x}&=k x,\\
       \tilde{\phi}(\tilde{x})&=Z_k^{\nicefrac{1}{2}}k^ {\frac{2-d}{2}}\phi(x),\\
       \tilde{K}(\tilde{x})&=k^{D_\mathcal{O}-d}K(x),
   \end{split} 
\end{equation}
where the tilde means that the variables are dimensionless and renormalized  and where the running anomalous dimension is
given by $\partial_t Z_k=-\eta_k Z_k$. At the fixed point, $\eta_k^*=\eta$, the anomalous dimension of the field,
yielding the usual scaling dimension of the field, $D_\varphi=(d-2+\eta)/2$.

The dimensionless version of Eq.~\eqref{WettEq} is:
\begin{equation}
\begin{split}
    \label{WettEqDless}
\partial_t \Gamma_k[\tilde{\phi},\tilde{K}]=&\int_{\tilde{x}}\frac{\delta \Gamma_k}{\delta \tilde{\phi}(\tilde{x})}\big(\tilde{x}^\nu\tilde{\partial}_\nu+D_\varphi\big)\tilde\phi(\tilde{x})\\ &-\int_{\tilde{x}}\tilde{K}(\tilde{x})\big(\tilde{x}^\nu\tilde{\partial}_\nu+D_{\mathcal{O}}\big)\frac{\delta \Gamma_k}{\delta \tilde{K}(\tilde{x})}\\+\frac{1}{2}\alpha\int_{\tilde{x},\tilde{y}}&\bigg[(d+2-\eta_k)r(|\tilde{x}-\tilde{y}|)\\&+|\tilde{x}-\tilde{y}|r'(|\tilde{x}-\tilde{y}|)\bigg]\tilde{G}_k(\tilde{x},\tilde{y}).
\end{split}\end{equation}
At a fixed point $\Gamma^\star$, the RG flow vanishes: $\partial _t\Gamma_k[\tilde\phi,\tilde K]=0$. As discussed below, the fixed point constraint can be interpreted as an invariance under dilatation regularized in the infrared.

\subsection{Eigenperturbations of flow equations around the fixed point}
\label{sec_eigenperturbations}

Let us now recall how the scaling operators are usually obtained, without making reference to the source $K$ (see for example \cite{delamotte2016scale}). Consider the fixed point of the flow equation, expressed in terms of {\it dimensionless} variables $\Gamma^{*}[\tilde\phi]=\Gamma^{*}[\tilde\phi,\tilde K=0]$. Let us consider a small perturbation around this fixed point
\begin{equation}
    \Gamma_k\big[\tilde{\phi}\big]=\Gamma^*\big[\tilde{\phi}\big]+\gamma_k[\tilde{\phi}],
\end{equation}
where $\Gamma^*[\tilde\phi]$ is a $t-$independent solution of Eq. \eqref{WettEqDless} \textit{with} $\tilde K=0$ and $\gamma_k[\tilde\phi]$ is small.  From now on, we only consider dimensionless quantities, and, consequently, for notation simplicity we omit the tilde notation. At leading order,  the flow equation for $\gamma_k[\phi]$ is linear and reads:
\begin{equation}\label{perturbaroundfixed}
 \begin{split}
\partial_t \gamma_k[\phi]+\int_x&\gamma_k^{(1)}[\phi;x] \big(D_\varphi+x_\mu \partial_\mu\big) \phi(x)
\\&=\frac{1}{2}\Tr\big[ \dot{R}_k G\gamma_k^{(2)}[\phi] G\big]    
 \end{split}
\end{equation}
where a matrix notation has been adopted and the trace denotes an integral over the volume.

A technical point must be explained here. In the previous equation and in the rest of the article, we choose a particular renormalization scheme where $\eta_k$ is fixed to its fixed point value, $\eta_k=\eta, \forall k$. This implies that the renormalization factor $Z_k$ in Eq. \eqref{dimensionless} is $Z_k^{\rm(present)}=(k/\Lambda)^{-\eta}$. In a more standard scheme, $\eta_k$ evolves all along the flow and becomes $\eta$ only asymptotically when $k\to 0$. We then call the renormalization factor $Z_k=Z_k^{\rm (standard)}$. We now show that both schemes are acceptable and can be explicitly related at least in the vicinity of a fixed point by changing the normalization of the $\phi$ field which is immaterial. 

The renormalization group flow for {\it dimensionful} quantities does not depend on the renormalization condition of the {\it dimensionless} field. Going from
one {\it dimensionless} flow to  another then corresponds to making the reparametrization
\begin{equation}
\label{reparam}
\phi^{\rm(present)}(x)= \Bigg(\frac{Z_k^{\rm(present)}}{Z_k^{\rm(standard)}}\Bigg)^{1/2} \phi^{\rm(standard)}(x).
\end{equation}
Since a reparametrization of the {\it dimensionless} field cannot change observables, the present choice is admissible, which proves that setting $\eta_k$ to its fixed point value is also admissible.

It should be pointed out, however, that the reparametrization in Eq. \eqref{reparam} can become singular far from the fixed point, making this scheme unsuitable for studying global flow properties. It is only suitable for studying
the neighborhood of the fixed point.
It has been successfully employed in the DE context in the past (see, for example, \cite{Morris:1994ie,Morris:1997xj}). In fact, as will be discussed below, from a numerical point of view, this scheme (implemented together with the analytical expression for the equation for eigenperturbation, instead of determining it by numerically perturbing the flow equation), is much more stable and precise than the standard one.

Since Eq. \eqref{perturbaroundfixed}  is linear in $\gamma_k[\phi]$ and has no explicit $t$-dependency, the set of solutions forms a vector space one basis of which shows an exponential dependence on $t$:
\begin{equation}
(\gamma_k)_i[\phi]=\exp(\lambda_i t) \hat{\gamma_i}[\phi]
\end{equation}
where  $\hat{\gamma_i}[\phi]$ is $t-$independent and fulfills the equation:
\begin{equation}
\label{perturbaroundfixed_bis}
    \frac{1}{2}\Tr\big[ \dot{R}_k G\hat{\gamma}^{(2)}_i[\phi] G\big]-\int_x\hat{\gamma}^{(1)}_i[\phi;x] \big(D_\varphi+x_\mu \partial_\mu\big) \phi=\lambda_i \hat{\gamma}_i[\phi].
\end{equation}
This equation can be seen as an eigenvalue problem because it is a linear equation in the functional $\hat{\gamma_i}[\phi]$ and its derivatives. The left-hand side of the above equation is usually called the {\it stability matrix}.
In most cases, one requires the eigenvectors $\hat{\gamma}_i[\phi]$ to be smooth in $\phi$, and this condition implies that the eigenvalue spectrum is discrete.

It is worth mentioning that equations (\ref{perturbaroundfixed}) and (\ref{perturbaroundfixed_bis}) are not valid only
for scalar or translation invariant perturbations but for {\it any} small perturbation around the fixed point. In particular, these could be perturbations transforming as a vector or as a tensor under space rotations and they could even be space in-homogeneous. Moreover, they do not need to be invariant under the internal symmetries of the fixed point. Let us point out, however, that \eqref{perturbaroundfixed} and \eqref{perturbaroundfixed_bis} being linear, the elements of the basis of solutions can be classified in multiplets according to the symmetries of the fixed point. For example, if the fixed point is $\mathbb{Z}_2$ symmetric the solutions are even or odd under this symmetry. In the same way, one can require any element of the basis of solutions to have a definite tensor structure (assuming the fixed point being translation and rotational invariant).

\subsection{The derivative expansion}
\label{sec_DE}
An exact solution to Eq. \eqref{WettEq} is generally out of reach, and some approximation is needed to make the calculations feasible. The most used approximation scheme in the framework of the FRG is the DE  which, in a nutshell, consists in truncating the momentum-dependence of the vertices up to the power $s$. This is usually referred to as the order $\mathcal{O}(\partial^s)$ of the DE.  The rationale behind this approximation is that we are mainly interested in the long-distance properties of the theory and expanding the vertices at small momenta seems reasonable. In fact, there are now many evidences that this is a very accurate and precise approximation scheme for $\mathbb{Z}_2$ and $O(N)$-invariant models (see for example, \cite{Berges2002,Canet2003b,Canet2003,Balog2019,DePolsi2020,DePolsi2021}). Moreover, it has been proven to be an extremely robust approximation in a large variety of problems as mentioned in the Introduction (for more examples, see \cite{Dupuis:2020fhh}).

For the Ising universality class  analyzed in this article, we use the DE at $\mathcal{O}(\partial^2)$ that we now describe  in some detail.

The first level of approximation is called \textit{Local Potential Approximation} (LPA) and it corresponds to the order $\mathcal{O}(\partial^0)$. It consists in taking as {\it ansatz} for the effective action the potential plus an unrenormalized kinetic term:
\begin{equation}\label{eq:ch2:isingLPA}
	\Gamma_k[\phi]=\int_x \biggl\{\frac{1}{2}\big(\nabla\phi\big)^2+U_k(\phi)\biggr\},
\end{equation}
with $U_k(\phi)$ an even function.
At $\mathcal{O}(\partial^2)$, $\Gamma_k[\phi]$ is approximated by:
\begin{equation}\label{eq:ch2:isingOD2}
\Gamma_k[\phi]=\int_x \biggl\{\frac{Z_k(\phi)}{2}\big(\nabla\phi\big)^2+U_k(\phi)\biggr\},
\end{equation}
where now both $U_k(\phi)$ and $Z_k(\phi)$ are even functions.

The FRG equations have a dressed one-loop structure where all propagators are regularized in the infrared by the regulator $R_k(q)$, Eqs.~\eqref{deltaSDirect} and \eqref{functIntegral}, ensuring the smoothness of the vertices as functions of momenta and the existence of the DE. In addition, the RG flow equations of  $U_k$ and $Z_k$ involve  $\partial_t R_k(q)$ in Fourier space, Eq. \eqref{WettEq}, which implies that the integral over the internal momentum is dominated by the momentum range $q \lesssim k$ due to the rapid decay of the regulator profile with $q$. The reason for the success of the DE is strongly related to this decoupling of low and high momenta (with respect to the regulator scale $k$).

Out of criticality, the first pole in the $p^2$ complex plane of the function $\Gamma^{(2)}(p,m)=\Gamma_{k=0}^{(2)}(p,m)$ is located  either at $p^2=4 m^2$ or $p^2=9 m^2$ depending on whether the system is in the broken or symmetric phase, with $m$ the renormalized mass, that is, the inverse of the correlation length.
Therefore, an expansion of $\Gamma_{k=0}^{(2)}(p,m)$ in powers of $p^2/m^2$ must have a radius of convergence of order 4$-$9. Since for $k\neq 0$ the presence of the regulator acts as a mass term for small momenta, that are effectively decoupled from large momenta in flow equations even at criticality, it is expected that the DE converges with a radius of convergence in $p^2/k^2$ of the order of $4-9$. Moreover, loop integrals being dominated in flow equations by the range $q^2 \lesssim k^2$,  one can estimate that successive orders of the DE are suppressed by a factor $\lambda\simeq 1/9-1/4$ \cite{Balog2019}. This estimate is confirmed by an empirical analysis of the Ising and $O(N)$ universality classes \cite{Balog2019,DePolsi2020,DePolsi2021}. More recently, this was improved by taking into account the dependence of $\lambda$ on the regulator profile \cite{DePolsi2022}.

It is worth emphasizing that the DE is a low-momentum expansion performed  in the presence of the regulator, so that the critical physics is recovered in the momentum range $k \ll p \ll \Lambda$. However, universal and non-universal quantities defined at zero momenta such as critical exponents can also be obtained in the range of validity of the DE, that is, $p\ll k$ \cite{Berges2002}.

In previous studies \cite{Balog2019,DePolsi2020}, it has been shown that the convergence of DE can be improved by tuning the overall scale of the regulator, that is, the parameter $\alpha$ in Eq.~\eqref{eq:RegProf}, (and more generally by changing the profile of the regulating function) according to a criterion known as the Principle of Minimal Sensitivity (PMS). The PMS states that since any physical quantity $Q$ should be independent of the shape of the regulator, a property which is violated by any approximation scheme, the optimal choice for $\alpha$ corresponds to $\alpha_{\rm opt}$ such that $dQ/d\alpha=0 $ for $\alpha=\alpha_{\rm opt}$.

Later on, an alternative criterion for fixing regulator parameters was discussed in the context of conformal symmetry \cite{Balog2020} at order $\mathcal{O}(\partial^4)$ of the DE for the Ising universality class. It was proposed to fix $\alpha$ in such a way that conformal symmetry is best verified, which is the exact physical scenario (see for example \cite{delamotte2016scale,ElShowk:2012ht}). This criterion was denoted "Principle of Maximal Conformality" (PMC). It was shown there that both criteria turn out to be generally equivalent.

This correlation between PMS and PMC was later explained \cite{DePolsi2022}, or at the very least inferred, to be a consequence of the fact that, given a family of regulators, the expansion parameter $\lambda$ of the DE, which depends on the actual family of regulator chosen, becomes the smallest possible for some values of $\alpha$. This happens due to an interplay between two opposing effects where the regulator finds an optimal zone behaving like a typical massive theory, favored by small values of $\alpha$, and the momenta contributing to the flows being small compared to this mass, favored by large values of $\alpha$. This implies an enhancement in the quality of the DE and, consequently, a lessening of the dependence of quantities with the regulator scale $\alpha$.

\section{Scale and conformal symmetry within the functional renormalization group}\label{secSCSFRG}
\subsection{Ward identities for dilatation and conformal invariance}
\label{sec_WIDI}
In this section, we derive the Ward identities for dilatation and conformal transformations in the presence of a source for a composite operator. 
The variations of the field under these transformations read:
\begin{equation}
\begin{split}
\delta_{\rm dil}\,\varphi(x)&=\epsilon \big(x_\mu \partial_\mu+D_\varphi\big) \varphi(x),\\
\delta_{\rm conf}\, \varphi(x)&=\epsilon_\mu \big(x^2\partial_\mu-2x_\mu x_\nu\partial_\nu-2x_\mu D_\varphi\big) \varphi(x).\\
\end{split}
\end{equation}
In this study, we restrict ourselves to a particular class of composite operators $\mathcal{O}(x)$, usually denoted as {\it primary}, which transform in the same way as the field itself, except for the value of its dimension:
\begin{equation}
\begin{split}
\delta_{\rm dil}\, \mathcal{O}(x)&=\epsilon \big(x_\mu \partial_\mu+D_{\mathcal{O}}\big) \mathcal{O}(x).\\
\delta_{\rm conf}\, \mathcal{O}(x)&=\epsilon_\mu 
\big(x^2\partial_\mu-2x_\mu x_\nu\partial_\nu-2x_\mu D_{\mathcal{O}}\big)\mathcal{O}(x).
\end{split}
\end{equation}

As usual (see, {\it e.g.} \cite{Rosten:2014oja,Sonoda:2015pva,delamotte2016scale,Rosten:2016zap,Balog2020}), one can derive the following modified Ward identities for dilatation:
\begin{equation}\label{Warddilat}
\begin{split}
\int_x \Big\{&\Gamma_k^{(1,0)}(x) \big(D_\varphi+x_\mu \partial_\mu\big) \phi(x)\\&-K(x)\big(D_{\mathcal{O}}+x_\mu \partial_\mu\big) \Gamma_k^{(0,1)}(;x)\Big\}\\&=-\frac{1}{2}\int_{x,y} \dot{R}_k(x-y) \Big(\Gamma_k^{(2,0)}+R_k\Big)^{-1}_{x,y},
\end{split}
\end{equation}
 and for special conformal transformations
\begin{equation}\label{Wardconformal}
\begin{split}
&\int_x\Big\{\Gamma_k^{(1,0)}(x) \big(x^2\partial_\mu-2x_\mu x_\nu\partial_\nu-2x_\mu D_\varphi\big) \phi(x)\\&-K(x)\big(x^2\partial_\mu-2x_\mu x_\nu\partial_\nu-2x_\mu D_{\mathcal{O}}\big) \Gamma_k^{(0,1)}(;x)\Big\}\\&=\frac{1}{2}\int_{x,y} (x_\mu+y_\mu)\dot{R}_k(x-y) \Big(\Gamma_k^{(2,0)}+R_k\Big)^{-1}_{x,y},
\end{split}
\end{equation}
where
\begin{equation}
\begin{split}
   \Gamma_k^{(n,m)}&(x_1,\cdots,x_n;y_1,\cdots,y_m)=\\& \frac{\delta^{n+m} \Gamma_k}{\delta\phi(x_1)\cdots\delta\phi(x_n)\delta K(y_1)\cdots\delta K(y_m)}.
\end{split}
\end{equation}
and
\begin{equation}
(2d-2D_\varphi+2x^2\partial_{x^2})R_k (x^2)=\partial_t R_k (x).
\end{equation}
Notice that the Ward identity for dilatation modified by the presence of the regulator is nothing but the fixed point equation obtained by setting $\partial_t \Gamma_k[\phi,K]=0$ in Eq. \eqref{WettEqDless}.

The Ward identity for scale invariance shows two important structural properties that we call triangular structures for reasons that we make clear below. Let us discuss the first one. The Ward identity for $\Gamma^{(n,0)}$ evaluated at $K=0$ does not depend on $\Gamma^{(n',m)}$ with $m>0$.  More generally, the Ward identity for $\Gamma^{(n,m)}$ at $K=0$ does not depend on $\Gamma^{(n',m')}$ with $m'>m$.  The same applies to the Ward identity for conformal transformations. This first triangular structure proves very useful in actual calculations. In particular, for studying the vertices $\Gamma^{(n,1)}$  it is
enough to include $K(x)$ linearly in the effective action. This is not
an approximation because higher orders in $K$ do not have any impact on lower orders.
\subsection{Relation between composite operators and eigenvectors of the stability matrix}
 

\label{2triangprop}

The standard way of computing the critical exponent $\nu$ in perturbation theory relies on introducing the composite operator $\phi_R^2$. In contrast, $\nu$ is in general obtained by diagonalizing the stability matrix, \eqref{perturbaroundfixed_bis} that naively does not involve composite operators. The two approaches are actually equivalent, as we recall now.


Let us define
\begin{equation}\label{eq:GammaHat}
\hat{\Gamma}[\phi]=\int_{x}\frac{\delta\Gamma_k}{\delta K(x)}\Big|_{K=0}
\end{equation}
One can deduce the dilatation Ward identity for $\hat{\Gamma}[\phi]$ by differentiating Eq.\eqref{Warddilat} with respect to $K(y_1)$:
\begin{equation}\label{Warddilat01}
\begin{split}
\int_x\Gamma_k^{(1,1)}&(x;y_1) \big(D_\varphi+x_\mu \partial_\mu\big) \phi(x)-\\&K(x)\big(D_{\mathcal{O}}+x_\mu \partial_\mu\big) \Gamma_k^{(0,2)}(;x,y_1)\\&-\big(D_{\mathcal{O}}+{y_1}_\mu \partial_\mu^{y_1}\big)\Gamma_k^{(0,1)}(;y_1)\\&=\frac{1}{2}\Tr\big[\dot{R}_k G\Gamma_k^{(2,1)}(;y_1)G\big]
\end{split}
\end{equation}   
By evaluating this expression at $K=0$ and integrating it over $y_1$ we obtain the dilatation equation for $\hat{\Gamma}[\phi]$:
\begin{equation}\label{hatGammadilat}
\begin{split}
\int_x&\hat{\Gamma}^{(1)}(x) \big(D_\varphi+x_\mu \partial_\mu\big) \phi(x)
\\&-\big(D_{\mathcal{O}}-d\big)\hat{\Gamma}=\frac{1}{2}\Tr\big[ \dot{R}_k G\hat{\Gamma}^{(2)} G\big].
\end{split}
\end{equation}

By comparing Eq.~(\ref{hatGammadilat}) with Eq.~(\ref{perturbaroundfixed_bis}) we
conclude that $\hat{\Gamma}[\phi]$ is an eigenperturbation of $\Gamma_k$ around the fixed point with $\lambda_{\mathcal{O}}=D_{\mathcal{O}}-d$. Here $\lambda_{\mathcal{O}}$ is the eigenvalue of the stability matrix associated with $\hat{\Gamma}=\int_x \mathcal{O}(x)$. It is a particularly simple perturbation, because it is a scalar under isometries. Let us notice that  Eqs.~\eqref{hatGammadilat} and \eqref{perturbaroundfixed_bis} are identical only if the renormalization scheme where $\eta_k=\eta$ has been chosen, a choice which is always possible close to the fixed point as explained below Eq. \eqref{perturbaroundfixed}. 

Eq. (\ref{hatGammadilat}) implies that, when considering a small translational and rotational invariant perturbation of $\Gamma_k$ around the fixed point solution, one obtains the spectrum of standard scalar perturbations of the fixed point, which is {\it decoupled} from perturbations that are inhomogeneous. This corresponds to the fact that vertices $\Gamma^{(n,1)}$ (at zero $K$) with zero momentum associated to $K$ or, equivalently, $x-$independent $K$ perturbations, do not depend on those vertices at non-zero $K-$momentum.

Let us note that there are two trivial perturbations that can be considered. The first one consists simply in a perturbation proportional to the identity $\mathcal{O}(x)=\mathds{1}$. In that case,  $D_{\mathds{1}}=0$, or, equivalently, $\lambda_{\mathds{1}}=-d$ (see Eq.~\eqref{perturbaroundfixed_bis}). The second one corresponds to the most relevant {\it odd} local operator, which is nothing but $\mathcal{O}(x)=\varphi(x)$. In that case, consistently, $D_{\mathcal{O}}=D_{\varphi}$, or, equivalently, $\lambda_{\varphi}=-(d+2-\eta)/2$.

Even eigenperturbations are, in a sense, more standard perturbations because they have the same symmetry group as the fixed point. In particular, if we focus on the most relevant eigenperturbation, one has that $\lambda_{\varphi^2}=-\nu^{-1}=D_{\varphi^2}-d$ in agreement with Eq.~\eqref{Dphi2andCEnu}. In a similar way, if we consider the first irrelevant even operator that we will denote as $\mathcal{O}^{(2)}$, one extracts the correction to scaling exponent $\omega=\lambda_{\mathcal{O}^{(2)}}=D_{\mathcal{O}^{(2)}}-d$.

Equation \eqref{hatGammadilat} for $\hat{\Gamma}[\phi]$ is the first equation in a tower of equations with a triangular structure for the following generalization:
\begin{equation}
\hat{\Gamma}_{\mu_1,\dots\mu_n}[\phi]=\int_x x_{\mu_1}\dots x_{\mu_n}\frac{\delta\Gamma_k}{\delta K(x)}\Big|_{K=0}
\end{equation}
These functionals are tensors not necessarily translational invariant. In some particular
cases, however, they can be translational invariant and some contractions of them can be scalars, as we discuss below. The important point is that we can deduce their dilatation Ward
identities in the same way as for $\hat{\Gamma}[\phi]$, which leads to:
\begin{equation}
\label{eq_gammahatmu1mu2}
\begin{split}
\int_x\hat{\Gamma}_{\mu_1\dots\mu_r}^{(1)}(x) \big(D_\varphi+&x_\mu \partial_\mu\big) \phi(x)
-\big(D_{\mathcal{O}}-d-r\big)\hat{\Gamma}_{\mu_1\dots\mu_r}\\=\frac{1}{2}&\Tr\big[ \dot{R}_k G\hat{\Gamma}_{\mu_1\dots\mu_r}^{(2)} G\big]
\end{split}
\end{equation}
One observes that $\hat{\Gamma}_{\mu_1,\dots\mu_n}[\phi]$ is a tensor eigenperturbation of the fixed
point with eigenvalue $D_{\mathcal{O}}-d-r$. Eq. (\ref{eq_gammahatmu1mu2}) reveals a second triangular structure because it relates $\hat\Gamma$ vertices with the same number of $\mu$ indices. For $r=0$, this corresponds to the well-known fact that homogeneous (that is, $x$-independent) eigenperturbations are decoupled from inhomogeneous ones.

One can make the same analysis for the Ward identity for special conformal transformations. In that case, however, the different $\hat{\Gamma}_{\mu_1\dots\mu_n}$ are coupled. For example, taking one derivative with respect to $K(y)$, then evaluating at $K=0$ and integrating over $y$ leads to the equation:
\begin{equation}\label{hatGammaconf}
\begin{split}
\int_x\hat{\Gamma}^{(1)}&(x)\big(x^2 \partial_\mu-2x_\mu x_\nu \partial_\nu-2 D_\varphi x_\mu\big) \phi(x)\\
&+2\big(D_{\mathcal{O}}-d\big)\hat{\Gamma}_\mu=\frac{1}{2}\Tr\big[\dot{\bar{R}}_{k,\mu} G\hat{\Gamma}^{(2)} G\big],
\end{split}
\end{equation}
where we introduced the definition  $\dot{\bar{R}}_{k,\mu}$ as the matrix notation for the function $(x_\mu+y_\mu)\dot{R}_{k}(x-y)$. 
One observes that the equation for $\hat{\Gamma}$ involves also $\hat{\Gamma}_\mu$. In general, the equation for $\hat{\Gamma}_{\mu_1\dots\mu_n}$ includes also $\hat{\Gamma}_{\mu_1\dots\mu_{n+1}}$.

Despite the fact that the conformal identity mixes different $\hat{\Gamma}_{\mu_1\dots\mu_n}$ one can see that the structure presented implies, {\it a priori}, an infinite set of linear constraints (one for each eigenperturbation). This infinite set of constraints suggests that most of those identities are somehow redundant but, for the moment, it is not obvious how.

\section{The derivative expansion with composite operators}\label{secRes}

In the presence of a composite operator for a scalar operator with $\mathbb{Z}_2$ symmetry, one can extend our usual procedure for the DE. Following the general discussion presented in Sect.~\ref{secSCSFRG}, the most general {\it ansatz} at order ${\cal O}(\partial^2)$ including up to linear terms in $K(x)$ is
\begin{align}\label{ansatz}
& \Gamma_k[\phi,K]=\int_x\Big\{U_0(\phi)+\frac 1 2 \big(Z_0(\phi)\big)(\partial_\mu\phi)^2+K(x) U_1(\phi)\nonumber\\
 &+\frac{K(x)}{2} Z_1(\phi)(\partial_\mu\phi)^2- Y(\phi) \partial^2 K(x) \Big\}.
\end{align}
As a consequence of the first triangular property discussed in Sect.~\ref{sec_WIDI}, terms of higher degree in $K$ do not appear in the RG flows of the five functions involved in the {\it ansatz} (\ref{ansatz}). This is why they have not been included in \eqref{ansatz}.


Following the general strategy described in Sect.~\ref{sec_DE}, we can derive the flow equations for the five functions appearing in Eq.~\eqref{ansatz}.\footnote{In this work, we use the usual strategy (sometimes called {\it full Ansatz}) of keeping, in a given diagram, all terms generated by the truncation, at odds with the {\it strict} DE where one keeps only terms involving up to 2 powers of the momenta (in general up to order $n$ if one considers the $\mathcal O(\partial^n)$, may they be internal or external, see \cite{Balog2019}.}
The flow equation for the dimensionless potential $U_0(\phi)$ can be obtained from the Eq.~\eqref{WettEqDless} evaluated in a uniform field. One obtains, as usual,
\begin{equation}\label{V0eq}
\partial_t U_0(\phi)+d U_0(\phi)-\phi D_\varphi U_0'(\phi)=\frac 1 2\int_q \dot{R}_k(q) G(q).
\end{equation}
In practice, it is more convenient to work with the derivative of this equation with respect to $\phi$, obtaining a constraint for $U_0'(\phi)$.

The flow of $Z_0(\phi)$ can be obtained, as usual, from the flow equation for $\Gamma^{(2,0)}(p)$, taking a derivative with respect to $p^2$ and then setting $p=0$. One then obtains:
\begin{equation}\label{Z0eq}
\begin{split}
\partial_t Z_0&(\phi)-\big(2 D_\varphi+2-d\big)Z_0(\phi)-\phi D_\varphi Z_0'(\phi)=\\-&\frac 1 2\int_q \dot{R}_k(q) G^2(q)
\Big\{Z_0''(\phi)-\\
&2\big[U_0'''(\phi)+q^2 Z_0'(\phi)\big]^2\big[G'(q^2)+2 \frac{q^2}{d}G''(q^2)\big]
-\\&\frac{8}{d} q^2 Z_0'(\phi)\big[U_0'''(\phi)+q^2 Z_0'(\phi)\big]G'(q^2)
-\\
&2 G(q^2)Z_0'(\phi)\Big[ 2  U_0'''(\phi)+q^2 Z_0'(\phi)(2+\frac 1d)\big)\Big]\Big\}.
\end{split}
\end{equation}
These flow equations for $U_0(\phi)$ and $Z_0(\phi)$ are the standard ones (those without the composite operator source, see for example \cite{Dupuis:2020fhh}).
As mentioned before, these equations become the dilatation Ward identities at the fixed point in the presence of an infrared regulator that, consistently with the first triangular property described at the end of Sect. \ref{sec_WIDI}, do not involve the functions $U_1$, $Z_1$ and $Y$.

The flow equations for the functions $U_1$, $Z_1$ and $Y$ are given in App.~\ref{App:WardsEquations}. 
Because of the second triangular property discussed in Sect.~\ref{2triangprop}, the fixed-point equations for $U_1$, $Z_1$ do not depend on $Y$. This implies that, if one is only interested in the invariance under dilatations, one obtains a closed subset of functions by considering $U_0(\phi)$, $Z_0(\phi)$, $U_1(\phi)$ and $Z_1(\phi)$.

Let us now discuss the information that can be extracted from conformal invariance.
In App.~\ref{gamma2confdil}, it is proven that conformal Ward identity applied to $\Gamma_k^{(2,0)}(p;\phi)$ does not have more information than dilatation Ward identity applied to this same vertex.\footnote{This result can be easily generalized to $O(N)-$invariant models.} Notice that this result is trivially satisfied at $\phi=0$ and $k=0$, but it is not so for $k\neq0$ and $\phi\neq 0$.
The conformal Ward identity for the vertex $\Gamma_k^{(1,1)}(p;\phi)$ leads to a non-trivial constraint at order $\mathcal{O}(\partial^2)$ that reads:
\begin{equation}
\label{conf_eq_v2}
C_L(\phi)=C_R(\phi),
\end{equation}
where
\begin{equation}
C_L(\phi)=\big(4D_{\mathcal{O}}-4-2d\big)Y'(\phi)-2\phi D_\varphi Z_1(\phi),
\end{equation}
and
\begin{equation}
\begin{split}
&C_R(\phi)
=\int_q \dot{R}_k(q) G^2(q^2)
\Big\{Z_1'(\phi)-G(q^2)\times \\
& \Big[3Z_0'(\phi)U_1''(\phi)+Z_1(\phi)\Big(\frac{2+4d}{d}q^2Z_0'(\phi)+U_0'''(\phi)\Big)\Big]-\\&\frac{4}{d} q^2 G''(q^2) \big(U_0'''(\phi)+q^2 Z_0'(\phi)\big)\big(U_1''(\phi)+q^2 Z_1(\phi)\big)-\\
&2G'(q^2) \Big[U_1''(\phi)\Big(U_0'''(\phi)+\frac{1+d}{d}q^2 Z_0'(\phi)\Big)
+\\&q^2 Z_1(\phi)\Big( \frac{3+d}{d}U_0'''(\phi)+\frac{4+d}{d}q^2Z_0'(\phi)\Big)\Big]\Big\}
\end{split}
\end{equation}
This identity now couples the function $Y$ to the others.

Note that in the absence of a source for composite operators, it is necessary to go to order $\mathcal{O}(\partial^4)$ to get the first nontrivial constraint from conformal symmetry, see \cite{Balog2020}.
The fact that in the presence of composite operators the first conformal constraint appears already at order $\mathcal{O}(\partial^2)$ enlarges considerably the relevance of this symmetry as for the DE because  the $\mathcal{O}(\partial^4)$ is very cumbersome and has only been implemented in O($N$) models unlike $\mathcal{O}(\partial^2)$ \cite{Dupuis:2020fhh}.

\subsection{Results}

We now show how the regulator can be chosen by imposing that the breaking of conformal invariance is as small as possible, that is, by imposing the PMC. In the following we restrict ourselves to   the exponential regulator, 
\begin{equation}
\label{expreg}
     R_k(q)=\alpha Z_k \exp(-q^2/k^2)
\end{equation}
and we fix the overall scale $\alpha$ by the PMC. 

We start by looking  for fixed point solutions and eigenperturbations of Eqs.~(\ref{V0eq},\ref{Z0eq},\ref{V1eq},\ref{Z1eq},\ref{Yeq}). As discussed above, we only find solutions for a discrete set of eigenvalues $D_{\mathcal O}$ and we focus on the first three that are associated with even perturbations $\mathcal{O}^{(1)}=(\varphi^2)_R$, $\mathcal{O}^{(2)}$ and $\mathcal{O}^{(3)}$. They correspond to the critical exponents, $D_{\mathcal{O}^{(1)}}=D_{\varphi^2}=d-\nu^{-1}$, $D_{\mathcal{O}^{(2)}}=d+\omega$ and $D_{\mathcal{O}^{(3)}}=d+\omega_2$ were $\omega$ and $\omega_2$ control the non-analytic leading corrections to scaling.

\begin{figure}[h!]
    \centering
    \includegraphics[width=\columnwidth]{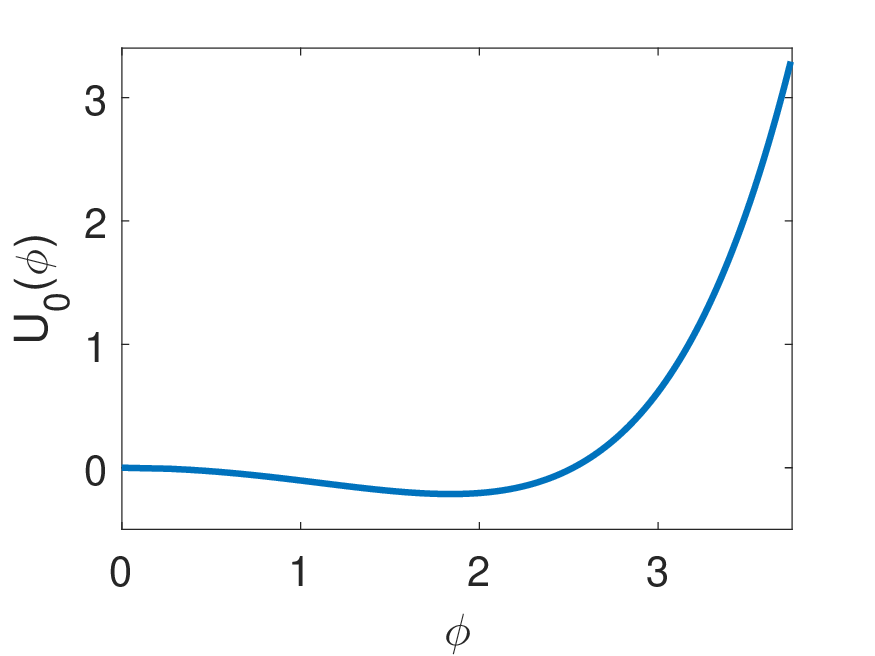}
    \includegraphics[width=\columnwidth]{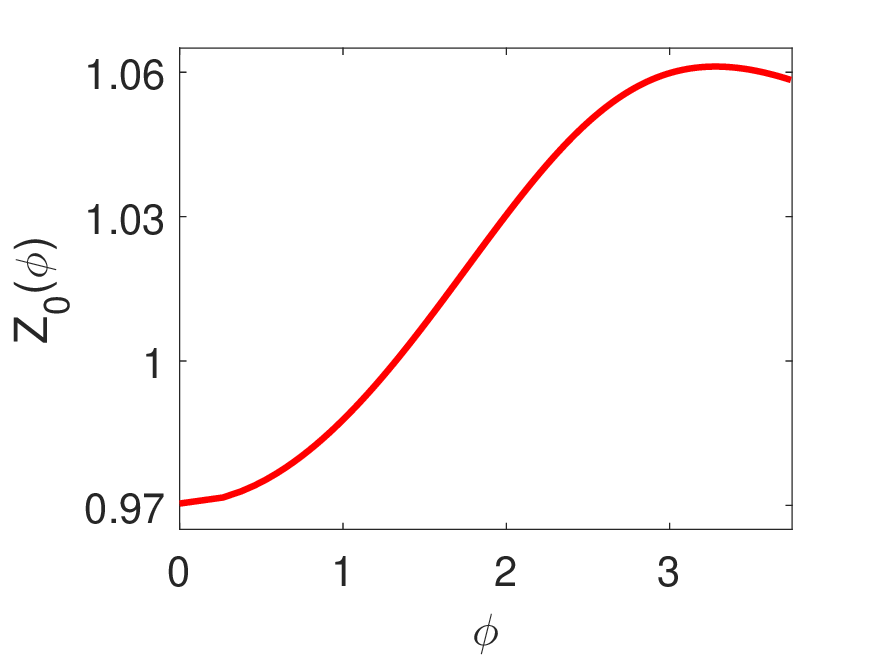}
    \caption{Typical solutions for the Wilson-Fisher fixed point using the regulator given by Eq.~\eqref{expreg} with $\alpha=2$. }
    \label{fig:fixedPoint}
\end{figure}

In Fig.~\ref{fig:fixedPoint} we show the typical Wilson-Fisher fixed point solution of equations \eqref{V0eq} and \eqref{Z0eq} in three dimensions.\footnote{Physical quantities that are obtained in the limit $k\to 0$, be they universal (critical exponents, etc) or not (critical temperatures, etc) should of course be independent of the renormalization scheme, which means in our case of the regulator. There are however other quantities that do depend on the renormalization scheme such as the beta-functions in perturbation theory or fixed point functions in FRG such as $U_0,\ Z_0,\ U_1,\ Z_1$ and $Y$. This dependence exists in the exact regularized theory, that is, is not induced by approximations. This is why fixed-point functions are shown for one given value of $\alpha$ for illustrative purpose.}

At large field, the right hand side of Eqs. \eqref{V0eq} and \eqref{Z0eq} are negligible which allows to find the fixed point behavior of these functions in this limit:
\begin{align}\label{largeRhoV0}
    U_0'(\phi)&\underset{\phi\gg 1}{\sim} A_{U_0}\phi^{(d+2-\eta)/(d-2+\eta)},\\
\label{largeRhoZ0}
    Z_0(\phi)&\underset{\phi\gg 1}{\sim} A_{Z_0}\phi^{-2\eta/(d-2+\eta)},
\end{align}
where we used that $D_\varphi=(d-2+\eta)/2$. The constants $A_{U_0}$ and $A_{Z_0}$ are not fixed by the large field behavior found above but can be determined by solving Eqs. \eqref{V0eq} and \eqref{Z0eq} for all values of the field.
Now, as previously shown, conformal Ward identity yields no new constraint for $\Gamma_k^{(2,0)}$.
 
The same argument applied to $U'_1(\phi)$, $Z_1(\phi)$ and $Y'(\phi)$ yields: 
\begin{align}\label{largeRhoV1}
    U_1(\phi)&\underset{\phi\gg 1}{\sim} A_{U_1}\phi^{2D_{\mathcal{O}}/(d-2+\eta)},\\
\label{largeRhoZ1}
    Z_1(\phi)&\underset{\phi\gg 1}{\sim} A_{Z_1}\phi^{2(D_{\mathcal{O}}-d-\eta)/(d-2+\eta)},\\
\label{largeRhoY}
    Y(\phi)&\underset{\phi\gg 1}{\sim} A_{Y}\phi^{2(D_{\mathcal{O}}-2)/(d-2+\eta)},
\end{align}
where once again, the constants $A_{U_1}$, $A_{Z_1}$ and $A_{Y}$ are fixed\footnote{Recall that the global normalization of eigenperturbations is arbitrary so only two ratios of these three constants are meaningful.} by the low to intermediate field behavior that depends on the loop term. 
The conformal Ward identity \eqref{conf_eq_v2} yields here a new constraint because its expansion at large field implies that $A_{Z_1}$ and $A_{Y}$ are not independent:
\begin{equation}
    A_{Z_1}D_{\mathcal O}^2=(D_{\mathcal O}-2)(2D_{\mathcal O}-d-2)A_{Y}.
\label{eq_exact}
\end{equation}
Observe that this constraint is exact and is not altered by the next orders of the DE. However, since $A_{Z_1}$ and $A_{Y}$ are already fixed from dilatation identities, this constraint is not exactly fulfilled and the \textit{l.h.s.} of the conformal constraint \eqref{conf_eq_v2} behaves at large field as \begin{equation}
C_L(\phi) \propto \phi^{(2D_{\mathcal{O}}-d-2-\eta)/(d-2+\eta)},
\end{equation}
which grows at large fields if $D_{\mathcal{O}}-d=\lambda > (2-d+\eta)/2$, as is the case for all irrelevant eigenperturbations of the model including, in particular, the first irrelevant eigenperturbation associated with $\omega$. In the present model, the above inequality is not fulfilled for the relevant even eigenperturbation which implies that the associated Ward identity tends to zero at large fields. This brings the relevant eigenperturbation constraint artificially to zero at large fields although it is just a matter of a negative power dominating this regime. 
It is convenient, both for the relevant and irrelevant eigenperturbations,  to get rid of this large field behavior by normalizing \eqref{conf_eq_v2} appropriately.
We therefore define  
\begin{equation}\label{normalizeConf}
   f(\phi,\alpha)= [C_L(\phi)-C_R(\phi)] (1+\phi^2/\phi_0^2)^{\frac{2+D_\phi-D_{\mathcal{O}}}{2D\phi}},
\end{equation}
so that, by construction, the function $f$ approaches a constant at large field. In the previous equation, $\phi_0$ is the minimum of the potential, $U'_0(\phi_0)=0$. We add this characteristic field parameter so that the normalization factor tends to 1 at small field. One can observe that when $\phi\to 0$ any dependence on $\phi_0$ disappears. However, by re-scaling with the minimum of the potential we obtain an approximate $\alpha-$independent behavior at large fields.

As was done in \cite{Balog2020}, instead of using the PMS one could use conformal symmetry in order to eliminate the spurious dependence of physical quantities, such as critical exponents, on non-physical parameters, such as $\alpha$, defining the regulator. The idea is to fix them to their value for which conformal symmetry is best verified.  
As in Ref.~\cite{Balog2020} we refer to this as the \textit{Principle of Maximal Conformality} (PMC). In the following, we therefore study how the function $f$ defined in \eqref{normalizeConf} depends on $\alpha$. Note that its normalization is different from the one used in Ref.~{\cite{Balog2020}.

\subsubsection{Leading eigenperturbation and critical exponent $\nu$}

We first consider the most relevant eigenperturbation, associated with the critical exponent $\nu$.
The typical fixed point solutions of the functions $U_1(\phi)$, $Z_1(\phi)$ and $Y(\phi)$ are shown in Fig.~\ref{fig:eigenSolutionlambda1} where the normalization is fixed by the condition $U''_1(0)=1$.
\begin{figure}[h!]
    \centering
    \includegraphics[width=\columnwidth]{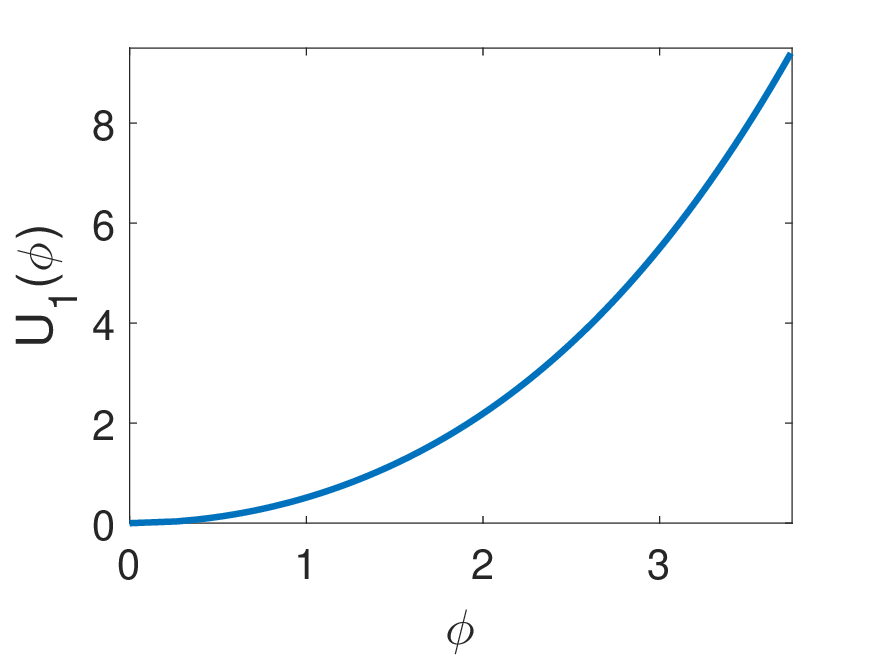}
    \includegraphics[width=\columnwidth]{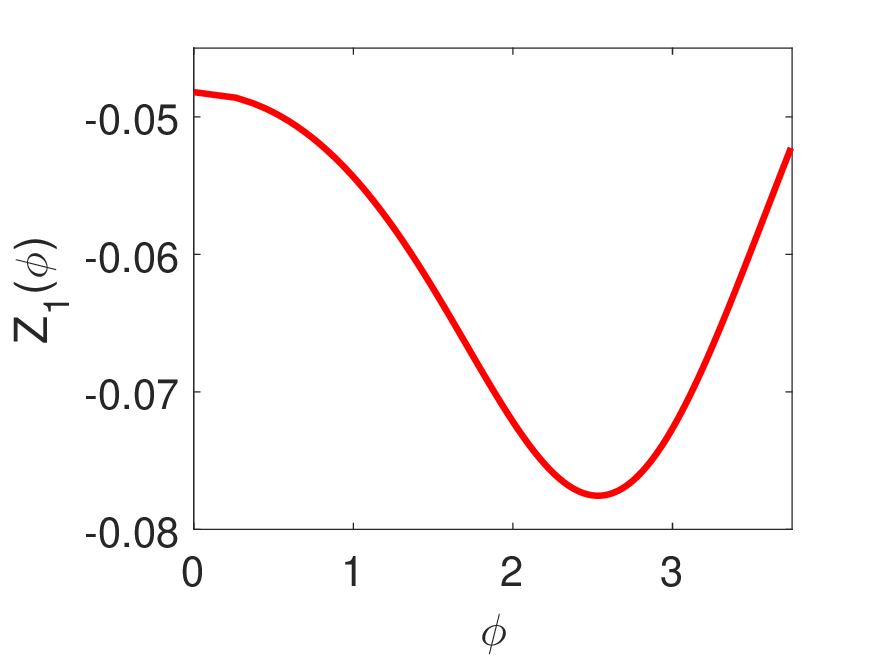}
    \includegraphics[width=\columnwidth]{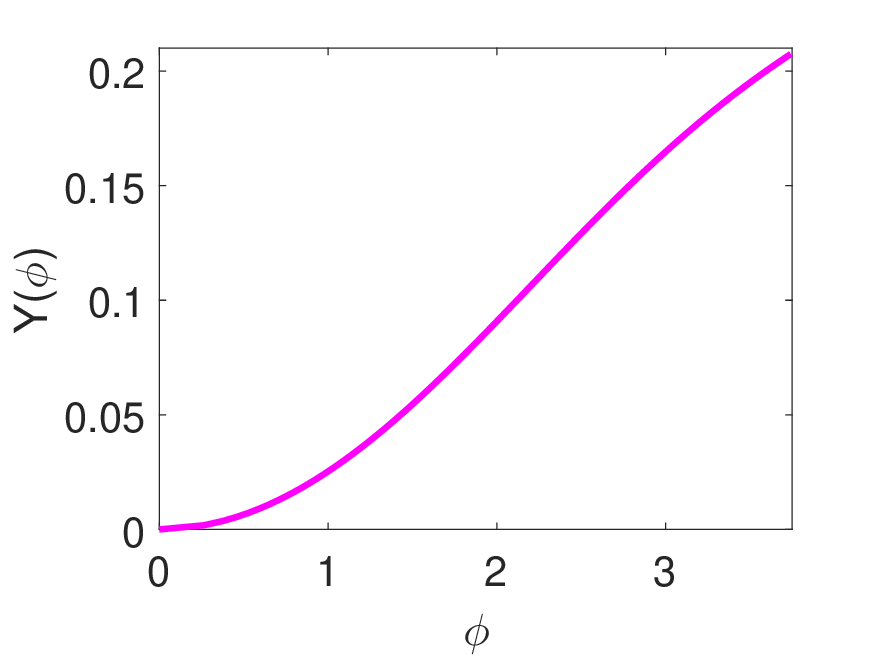}
    \caption{Typical solution for the leading eigenperturbation (associated with the critical exponent $\nu$) at the Wilson-Fisher fixed point using the regulator given by Eq.~\eqref{expreg} with $\alpha=2$.}
    \label{fig:eigenSolutionlambda1}
\end{figure}

 In Fig.~\ref{fig:nuvsalpha}, we show the dependence of the critical exponent $\nu$ with the overall scale of the regulator $\alpha$. We find that the PMS occurs for $\alpha=1.3$ for which $\nu=0.6280$. This is to be compared with the very precise result of the Conformal Bootstrap method $\nu^{\rm(CB)}=0.629971(4)$ \cite{Kos:2014bka}. 
\begin{figure}[h!]
    \centering
    \includegraphics[width=\columnwidth]{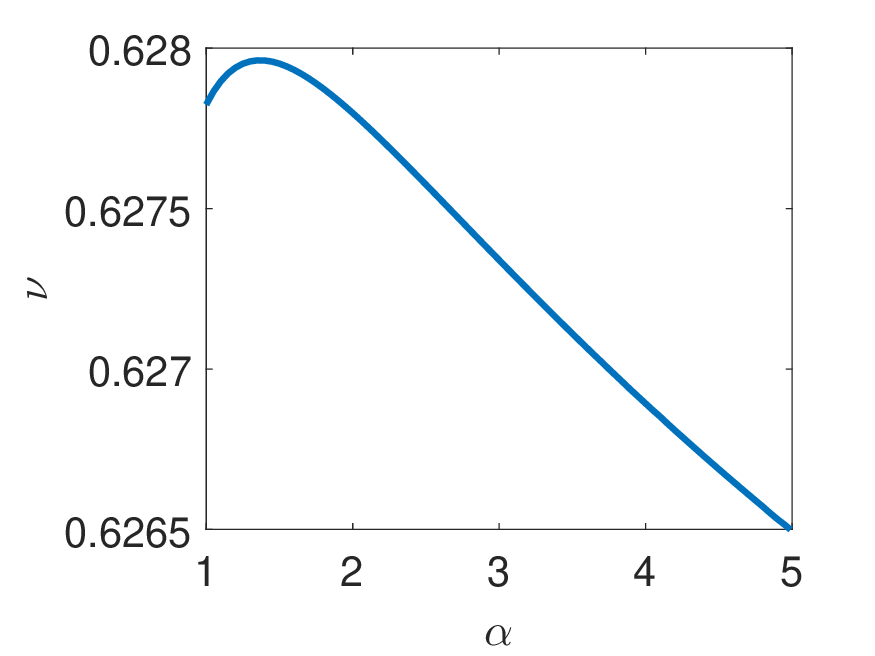}
    \includegraphics[width=\columnwidth]{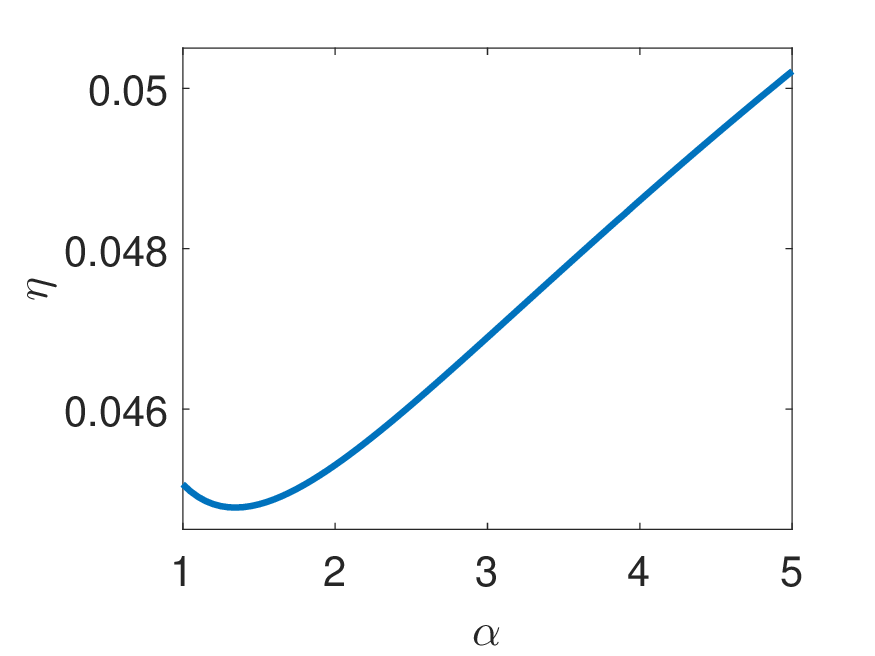}
    \caption{Critical exponents $\nu$ and $\eta$ as  functions of the regulator parameter $\alpha$ defined in Eq.~\eqref{expreg}. The PMS correspond to the local extrema of the curves, which  both occur for $\alpha= 1.35(5)$.}
    \label{fig:nuvsalpha}
\end{figure}

 In Fig.~\ref{fig:confConstnu}, we show the behavior of the function $f(\phi,\alpha)$ defined in Eq.~\eqref{normalizeConf} as a function of $\phi$ and $\alpha$. 
\begin{figure}[h!]
    \centering
    \includegraphics[width=\columnwidth]{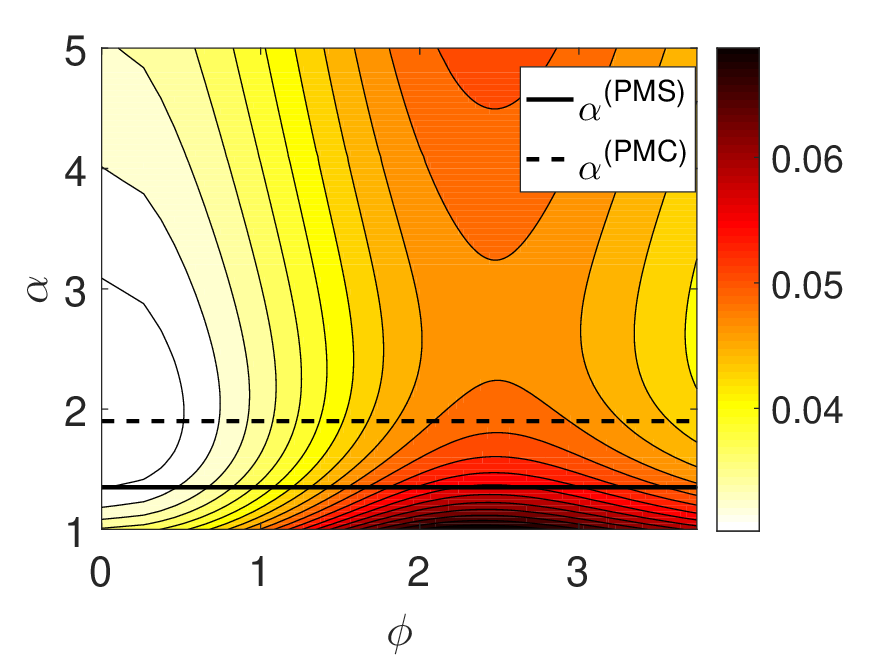}
    \caption{Contour plot of the function $|f|=-f$ defined in Eq.~\eqref{normalizeConf} as a function of the field $\phi$ (vertical axis) and $\alpha$ (horizontal axis) for the eigenperturbation associated with $\nu$.  The solid and dashed lines correspond to the values of $\alpha$ which fulfill the PMS and PMC criteria  respectively. }
    \label{fig:confConstnu}
\end{figure}
It is striking that the function tends to a quite small constant in the large field limit, which indicates that the exact constraint \eqref{eq_exact} is  reproduced with good accuracy.

We define $\alpha^{\rm(PMC)}(\phi)$ as the value of $\alpha$ that satisfies:
\begin{equation}
    |f(\phi,\alpha^{\rm(PMC)}(\phi))|=\min_{\alpha}|f(\phi,\alpha)|.
    \label{eq_criterion}
\end{equation}
Note that this definition makes $\alpha^{\rm(PMC)}$ a $\phi$-dependent quantity as seen in Fig.~\ref{fig:confConstnu}. We have checked that varying $\phi$ induces a variation of $\alpha^{\rm(PMC)}(\phi)$ in the range $[1.9,2.65]$. As seen in Fig. \ref{fig:nuvsalpha}, this leads to a variation of $\nu$ and $\eta$ respectively of  $0.1\%$ and  $3\%$, well below the error bars of the present approximation. We therefore implement the PMC at $\phi=0$ in what follows.


In Fig.~\ref{fig:confConstnu}, $\alpha^{\rm(PMC)}_\nu=1.9$ is marked with a dotted line which is not too far from $\alpha^{\rm(PMS)}_\nu\approx 1.35$ (for the regulator given in equation \eqref{expreg}). We obtain:
\begin{equation}
\begin{split}
   \nu^{(PMS)}\approx &\, 0.6280\\
   \nu^{(PMC)}\approx &\, 0.6278.
\end{split}
\end{equation}
These results imply that 
\begin{equation}\label{nuPMSvsnuPMC}
    \nu^{(PMS)}-\nu^{(PMC)}\approx 0.0002\ll \Delta\nu^{(2)},
\end{equation}
where $\Delta\nu^{(2)}=0.0027$ is the error associated with the critical exponent $\nu$ computed at order $\mathcal{O}(\partial^2)$ of the DE, see \cite{Balog2019,DePolsi2020}. This clearly shows that as was the case for order $\mathcal{O}(\partial^4)$ of the DE discussed in \cite{Balog2020}, the PMS and PMC are also compatible at order $\mathcal{O}(\partial^2)$. 

\subsubsection{Eigenperturbation for the critical exponent $\omega$}

The critical exponent $\omega$ can be obtained as the eigenvalue associated with the least irrelevant even eigenperturbations. 
For this eigenperturbation, a typical solution for the functions $U_1(\phi)$, $Z_1(\phi)$ and $Y(\phi)$ are shown in Fig.~\ref{fig:eigenSolutionlambda2} where we normalize these functions with the condition $U''_1(0)=1$. 
\begin{figure}[h!]
    \centering
    \includegraphics[width=\columnwidth]{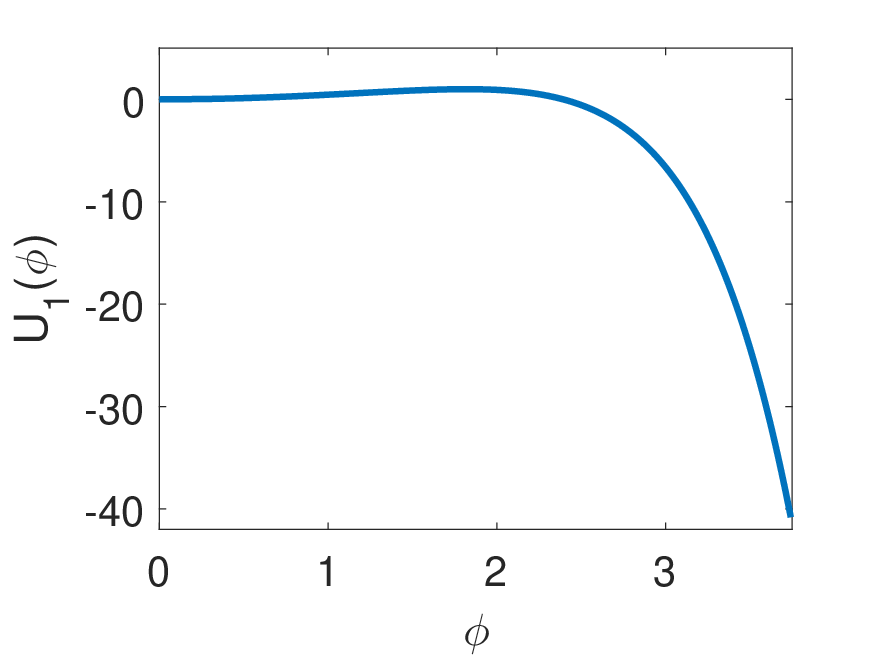}
    \includegraphics[width=\columnwidth]{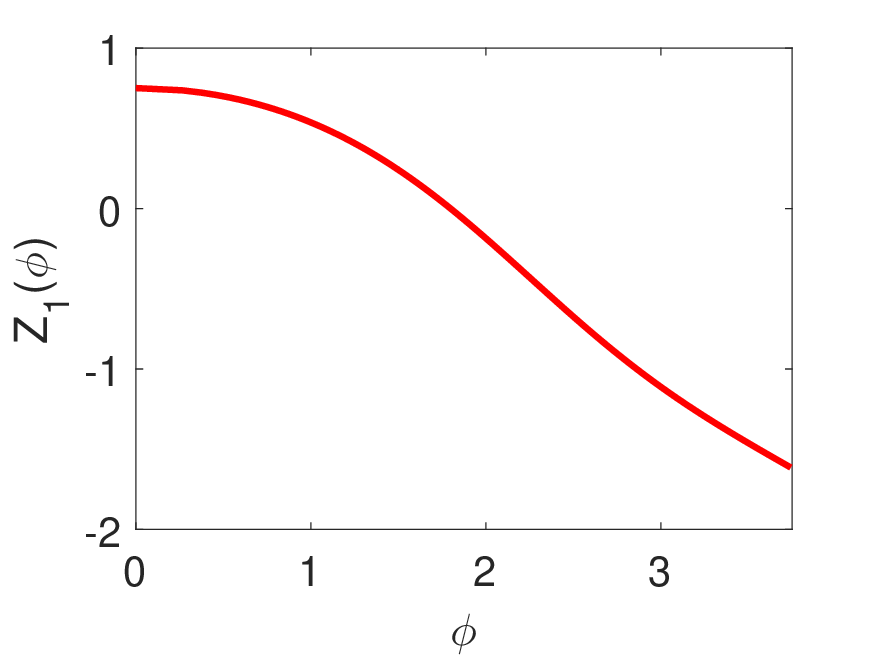}
    \includegraphics[width=\columnwidth]{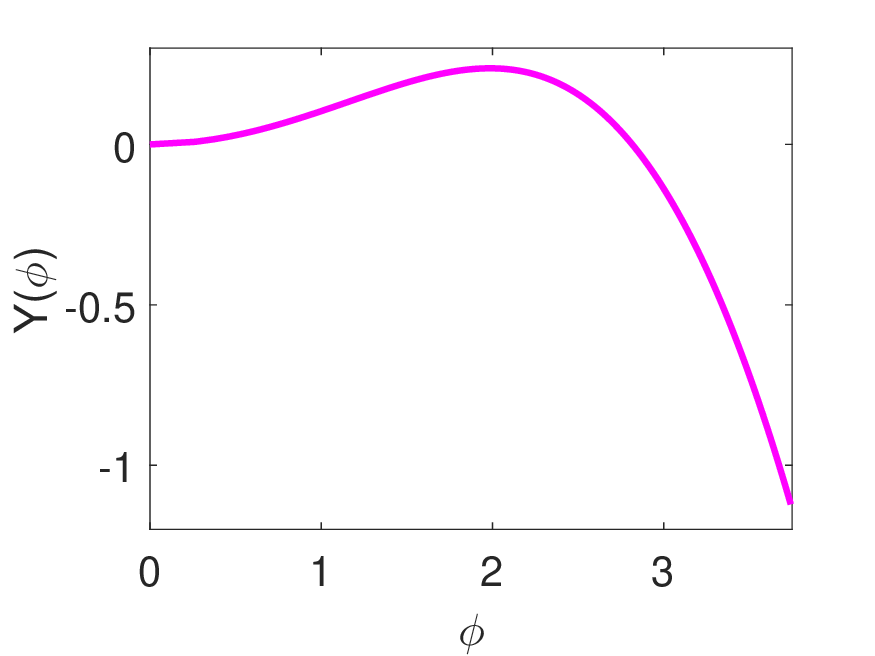}
    \caption{Typical solution for the least irrelevant eigenperturbation (associated with the critical exponent $\omega$) at the Wilson-Fisher fixed point using the regulator given by Eq.~\eqref{expreg} with $\alpha=2$.}
    \label{fig:eigenSolutionlambda2}
\end{figure}
At large fields, we observe a stronger increase of the function $|U_1|$ for $\omega$ than for $\nu$ (see Fig.~\ref{fig:eigenSolutionlambda1}) which is expected because of the asymptotic behavior given in Eq.~\eqref{largeRhoV1}.

In Fig.~\ref{fig:omegavsalpha}, we show the dependence of $\omega$ on $\alpha$ at order $\mathcal{O}(\partial^2)$ of the DE. The PMS leads to the value $\alpha=1.45$ for which $\omega=0.8483$. This is to be compared with the very precise result of the Conformal Bootstrap method $\omega^{\rm(CB)}=0.82968(23)$ \cite{Simmons-Duffin:2016wlq}.

\begin{figure}[h!]
    \centering
    \includegraphics[width=\columnwidth]{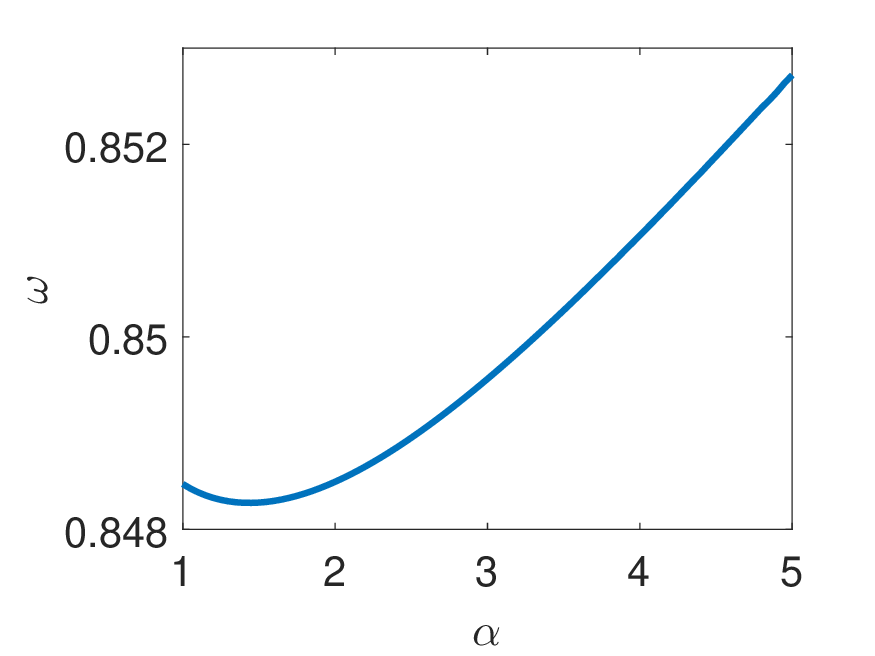}
    \caption{Critical exponent $\omega$ as a function of the regulator parameter $\alpha$ in Eq.~\eqref{expreg}. The PMS corresponds to the minimum of the curve that occurs at $\alpha= 1.45(5)$.}
    \label{fig:omegavsalpha}
\end{figure}
In Fig.~\ref{fig:confSolutionlambda2}, we show the behavior of $f(\phi,\alpha)$ for this first irrelevant eigenperturbation. The function $f$ saturates at large values of $\phi$  to values that are still small but bigger than for the perturbation associated with $\nu$. This indicates that, for the first irrelevant perturbation, conformal invariance is less accurately satisfied by the $\mathcal{O}(\partial^2)$ approximation than for $\nu$. Nonetheless, a regime in $\alpha$ occurs for which the conformal Ward identity is still approximately satisfied and we can observe a somewhat similar behavior as in the case of the  exponent $\nu$ (compare with Fig.~\ref{fig:confConstnu}).

As for $\nu$ we apply the PMC, Eq.~\eqref{eq_criterion},  at $\phi=0$ which leads to $\alpha^{\rm(PMC)}=2.45$ and $\omega=0.8489$. We have checked that the  variations of $\omega$ induced by the variations of $\alpha^{\rm(PMC)(\phi)}$ are again below error bars. 
Although now the values of $\alpha^{\rm(PMS)}_{\omega}$ and $\alpha^{\rm(PMC)}_{\omega}$ differ substantially, it is important to realize that the two criteria lead to similar values of $\omega$:
\begin{align}
    \omega^{\rm(PMS)}&= 0.8483 \nonumber\\
    \omega^{\rm(PMC)}&= 0.8489.  
\end{align}
As a consequence,
\begin{equation}
 \omega^{\rm(PMC)}-\omega^{\rm(PMS)}=0.0006\ll \Delta \omega^{(2)}
\end{equation}
where $\Delta \omega^{(2)}=0.055$ is the error estimate of the calculation of $\omega$ at order $\mathcal{O}(\partial^2)$ (see \cite{DePolsi2020}).

\begin{figure}[h!]
    \centering
    \includegraphics[width=\columnwidth]{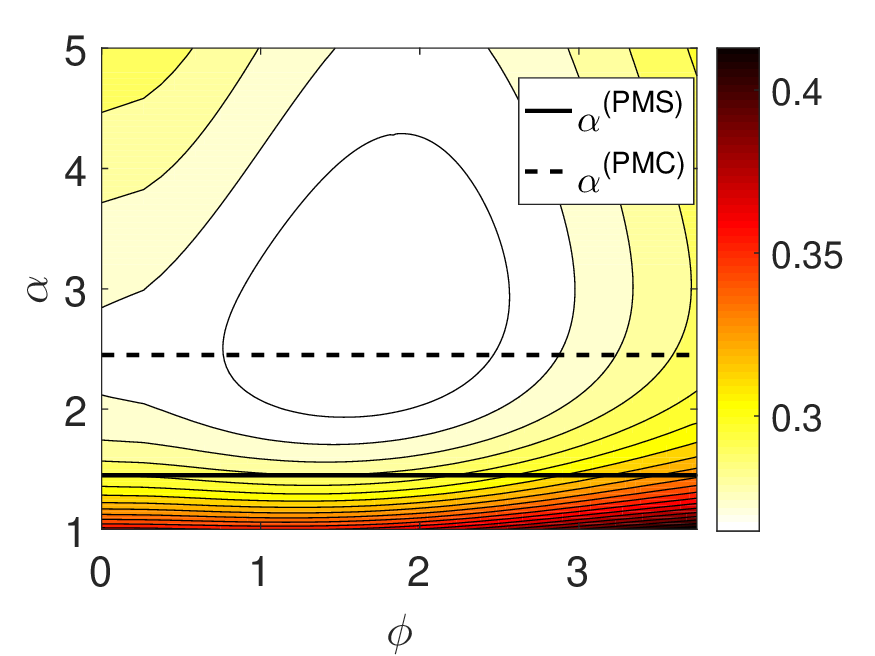}
    \caption{Contour plot of the function $f$ (see the text) as a function of the field $\phi$ (vertical axis) and $\alpha$ (horizontal axis) for the eigenperturbation associated with $\omega$. The solid and dashed lines correspond to the values of $\alpha$ which fulfill the PMS and PMC criteria respectively.}
    \label{fig:confSolutionlambda2}
\end{figure}

\begin{figure}[h!]
    \centering
    \includegraphics[width=\columnwidth]{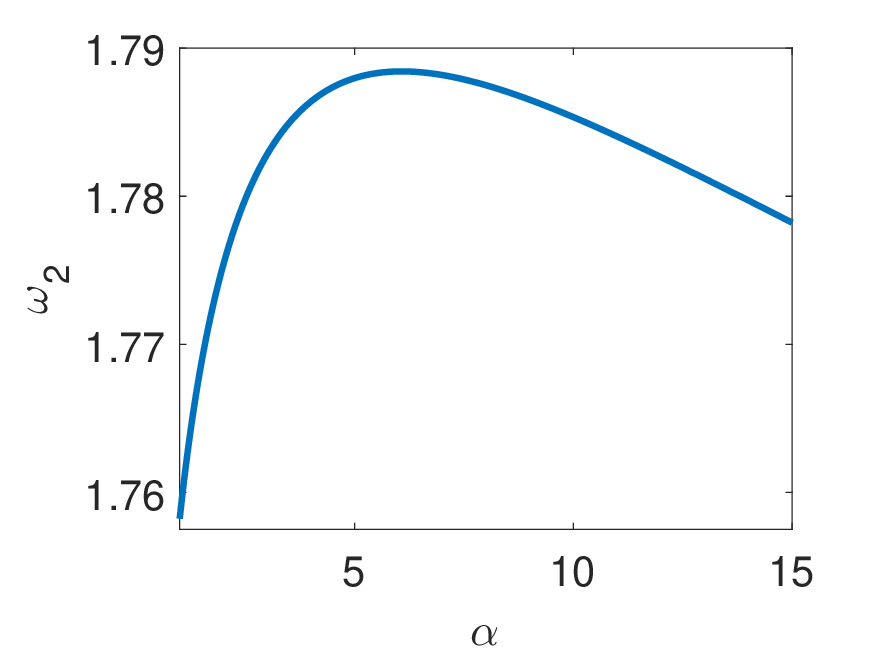}
    \caption{Critical exponent $\omega_2$ as a function of the regulator parameter $\alpha$ using the regulator given by Eq.~\eqref{expreg}.}
    \label{fig:omega2vsalpha}
\end{figure}

\subsubsection{Eigenperturbation associated with the critical exponent $\omega_2$}

Finally, we discuss the second irrelevant even eigenperturbation, associated with the critical exponent $\omega_2$. We emphasize on two methodological points. First, we stress that the most common method for deriving the critical exponents consists in computing numerically the stability matrix (see the discussion in \ref{sec_eigenperturbations}) and then computing  the eigenvalues of this matrix. This approach is known to lead to more and more noisy results when one considers more and more irrelevant operators. By comparing this method with the method outlined in the present article where the critical exponents are computed from the composite operators, we conclude that the latter is more stable than the former in agreement with the conclusions of \cite{wink23}. It also agrees with the results which are very stable numerically obtained long time ago by a similar method, for example, in Ref.~\cite{Morris:1994ie}. We found, in particular, that the second method is much less sensitive to a change in the grid size in $\phi$. 

For the $\omega_2$ eigenperturbation, the typical fixed point solutions for the functions $U_1(\phi)$, $Z_1(\phi)$ and $Y(\phi)$ has a similar form than for previously discussed eigenperturbations for $\nu$ and $\omega$. Due to this similarity they do not bring new information and we do not present their corresponding plots here. Let us comment, however, that once again the large $\phi$ behavior is steeper than for more relevant eigenperturbations. In Fig.~\ref{fig:omega2vsalpha} the behavior of $\omega_2$ as a function of $\alpha$ is shown. One obtains the value
$\omega_2^{(PMS)}=1.79$ to be compared with the conformal bootstrap result  \cite{ElShowk:2012ht} $\omega_2=1.67(11)$.
Unfortunately, we were not able to employ the method introduced in \cite{DePolsi:2020pjk} to estimate error bars for this exponent. The reason is that the method requires the comparison of estimates for an exponent at two consecutive orders of the DE but the quantity $\omega_2$ is spoiled at order LPA. One can understand it in the following way: in $d=4$ the exponents can be obtained by dimensional analysis. The leading relevant exponent is $-2$ (corresponding to $\nu=1/2$). The second even eigenperturbation is marginal, giving $\omega=0$. The two next even perturbations are degenerate in $d=4$ and correspond to the operators $\phi^6$ and $\phi^2 (\nabla \phi)^2$, having exponent $2$. Now, one of these can only be constructed with derivatives, so only one of the two eigenperturbation is properly captured at order LPA. We verified numerically at order $\mathcal{O}(\partial^2)$ that one of them has exponent below 2 for $d\sim 3$ and the other an exponent above 2. Only the most irrelevant is described satisfactorily at order LPA and it does not correspond to $\omega_2$.

Following the same line of arguments as for the other perturbations, we now consider the function $f(\phi,\alpha)$ which is represented in
Fig.~\ref{fig:eigenSolutionlambda3}. It tends to a bigger constant at large fields, which shows that conformal invariance is not that well retrieved for this perturbation.

\begin{figure}[h!]
	\centering
	\includegraphics[width=\columnwidth]{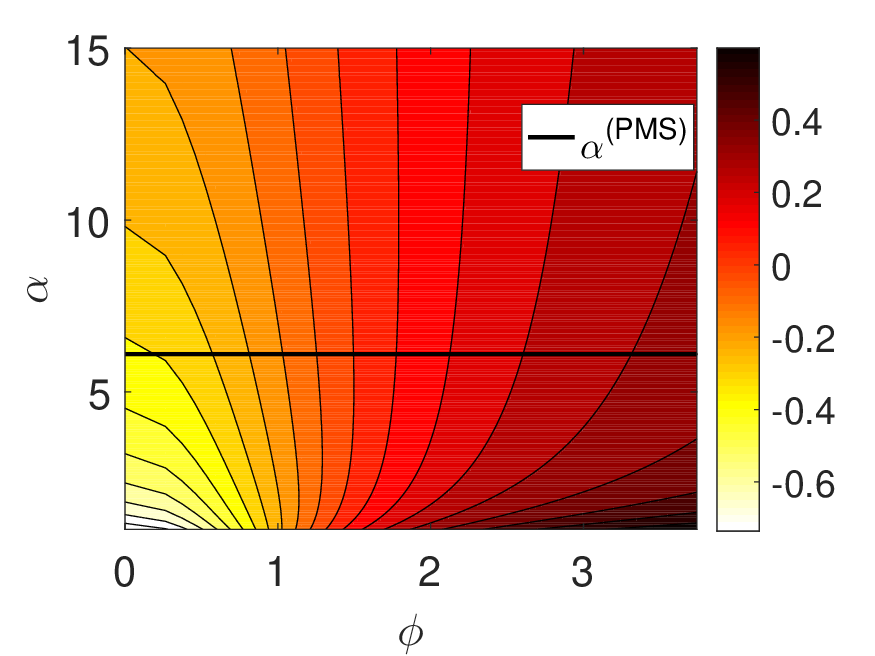}
	\caption{Contour plot of the function $f$ (see the text) as a function of the field $\phi$ (vertical axis) and $\alpha$ (horizontal axis) for the eigenperturbation associated with $\omega_2$. The solid line corresponds to the values of $\alpha$ which fulfill the PMS criterion for the exponent $\omega_2$.}
	\label{fig:eigenSolutionlambda3}
\end{figure}

Focusing on the behavior of $f(\phi=0,\alpha)$ (shown in Fig.~\ref{fig:eigenSolutionlambda3phi0}) to select the normalization constant, we observe that there is not a PMC value but that the curve tends to flatten at large values of $\alpha$. This suggests to take larger values of $\alpha$ as an optimum. At the same time, one must observe that the dependence on $\alpha$ seems to be relatively mild and taking very large values of $\alpha$ or values near to the PMS give results that are probably compatible (even if, as explained before, we do not have for the moment error bars for $\omega_2$). In this sense, the estimate of a PMC in this case is not possible, but this can be seen as the result of a relatively small dependence on $\alpha$. Additionally, it may be the case that the physics related to $\omega_2$ is not adequately described to order $\mathcal{O}(\partial^2)$, this being the first order capable of calculating this quantity. Therefore, we cannot rule out the possibility that the consideration of a higher order of the DE allows us to use the PMC criterion.

\begin{figure}[h!]
    \centering
    \includegraphics[width=\columnwidth]{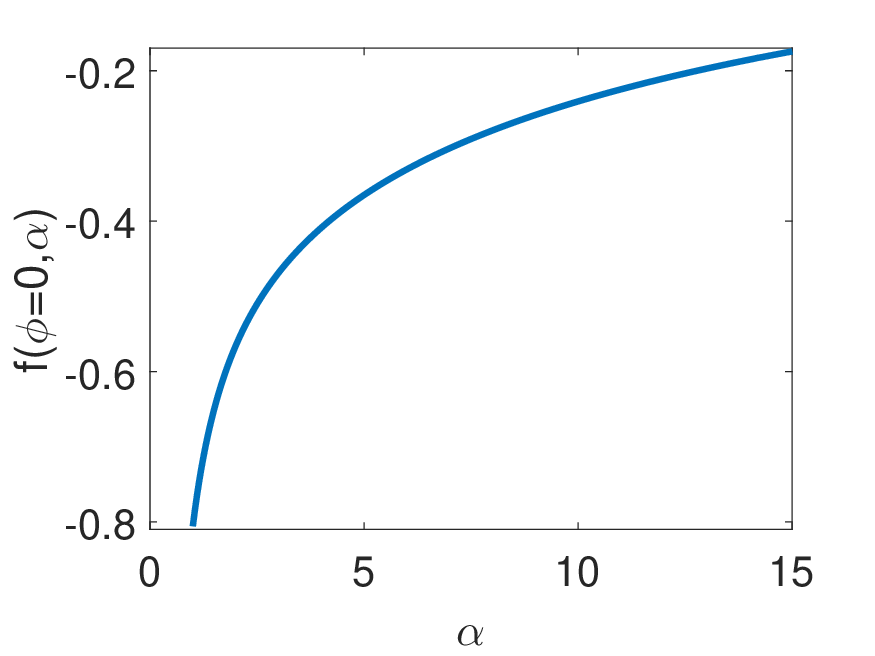}
    \caption{Plot of the function $f(\phi=0,\alpha)$ (see the text) as a function of the regulator parameter $\alpha$ for the eigenperturbation associated with $\omega_2$ and for the regulator given by Eq.~\eqref{expreg}.}
    \label{fig:eigenSolutionlambda3phi0}
\end{figure}

\section{Conclusions}\label{secConcl}

In this article, we have considered the Ward identities associated with conformal and dilatation symmetries in the formalism of the NP-FRG for the Ising universality class. On top of the field $\phi$, we have introduced sources for composite operators and  expanded the equations at order $\mathcal O(\partial^2$) of the DE, which is a widely used approximation scheme. We have shown that conformal invariance provides non-trivial information as early as this order, unlike when only the $\phi$ field is considered, where the order $\mathcal O(\partial^4$) is required to obtain new information from conformal invariance.

Our numerical resolution of the system of equations shows that, at the fixed point of the renormalization group ({\it i.e.} when the Ward identity for dilatations is fulfilled), the Ward identity for conformal invariance is violated. Since conformal transformations are known to be realized at the critical point of the Ising model, we can interpret the violation of the conformal Ward identity as a consequence of our approximations. This leads to a criterion for optimizing our results: we choose the regulator leading to the smallest breaking of the conformal Ward identity, a criterion called the principle of maximal conformality. 

We have tested this idea for the eigenperturbations associated with the critical exponents $\nu$, $\omega$ and $\omega_2$. In the first two cases, the regulator that is retained by the principle of maximal conformality is in good agreement with the one selected by the the widely used principle of minimal sensitivity. In the last eigenpertubation considered, corresponding to the critical exponent $\omega_2$, we find that the breaking of conformal invariance is rather independent of the parameter $\alpha$  (at least in the family of regulators that we have considered) which makes the criterion less selective and, in a way, not very useful. 
In comparing the three perturbations, we clearly see that the violations of conformal invariance are bigger and bigger when we consider more and more irrelevant perturbations.

This study represents a very significant progress over the previous study that considered conformal invariance in the NPRG formalism because it greatly extends the domain of applicability of this methodology. Indeed, although in the present study we have focused on the Ising universality class, which provides an excellent benchmark, the $\mathcal{O}(\partial^2)$ order of the DE, given its relative simplicity, has been applied to a huge variety of physical problems (see, e.g., \cite{Dupuis:2020fhh}). In contrast, higher orders of the DE have only been able to be implemented in the simplest models corresponding to the Ising universality class \cite{Canet2003b,Balog:2019rrg} or $O(N)$ models \cite{DePolsi:2020pjk,Peli:2020yiz,DePolsi2021}. That is, obtaining nontrivial information from conformal invariance at order $\mathcal{O}(\partial^2)$ of the DE then becomes a tool that can be widely employed.

The present work opens a large number of perspectives. The natural next step is to address the case of $O(N)$ models, which is currently under investigation by some of the authors. While such models have already been abundantly analyzed in the literature using FRG methods, they are a natural testing arena for new methodologies such as those presented here. A second natural extension of this work and one that we plan to analyze in the future is to repeat the present analysis to order $\mathcal{O}(\partial^4)$. This allows us to address two different issues. On the one hand, new constraints will surely appear whose behavior deserves to be analyzed. On the other hand, it is to be expected that the constraints coming from the special conformal transformations appearing at order $\mathcal{O}(\partial^2)$ will be satisfied with better precision than in the present work.
By employing the methods used in \cite{DePolsi:2020pjk} the analysis of the $\mathcal{O}(\partial^4)$ order would allow one to calculate error bars for the realization of such conformal identities and determine whether they are satisfied (within that margin of error). It should be noted that this would be a test of the emergence of conformal invariance at the critical point {\it without imposing such symmetry}. Another possible extension of the present work is to apply it to different approximations schemes such as the one considered in \cite{Blaizot2006,Benitez:2011xx,Benitez2012}. Last but not least, the present work opens a more ambitious line of work tending to take advantage of conformal invariance not only for the purpose of optimizing an approximation (such as DE) but also including higher order contributions thanks to the requirement of conformal invariance.

\acknowledgements

We are very grateful to Gilles Tarjus for valuable
comments on the manuscript. G. De Polsi and N. Wschebor thank for the
support of the Programa de Desarrollo de las Ciencias B\'asicas
(PEDECIBA). This work received the support of the French-Uruguayan
Institute of Physics project (IFU$\Phi$) and from the
grant of number FCE-1-2021-1-166479 of
the Agencia Nacional de Investigaci\'on e Innovaci\'on (Uruguay) and from the grant of number FVF/2021/160 from the Direcci\'on Nacional de Innovaci\'on, Ciencia y Tecnolog\'ia (Uruguay).

\appendix

\section{Ward identities for vertices for dilation and conformal invariance in presence of a composite operators}
\label{App:WardsEquations}

In this appendix we obtain the expressions for the Ward identities for dilatation and conformal invariance for different vertices in Fourier as well as their expressions in terms of the DE {\it ansatz} \eqref{ansatz}.

\subsection{Dilation and conformal Ward identities for the different vertices}

\subsubsection{Equation for $\Gamma_k^{(1,0)}$}

Let us differentiate Eq.~\eqref{Warddilat} with respect to $\phi(x_1)$:
\begin{equation}\label{Warddilat10}
\begin{split}
\int_x&\Gamma_k^{(2,0)}(x,x_1) \big(D_\varphi+x_\mu \partial_\mu\big) \phi(x) \\
 &-K(x)\big(D_{\mathcal{O}}+x_\mu \partial_\mu\big) \Gamma_k^{(1,1)}(x_1;x)\\
&+\big(D_\varphi-d-{x_1}_\mu \partial_\mu^{x_1}\big)\Gamma_k^{(1,0)}(x_1)\\
&=\frac{1}{2}\Tr\big[ \dot{R}_k G\Gamma_k^{(3,0)}(x_1) G\big],
\end{split}
\end{equation}
Here a matrix notation has been employed where a matrix contraction corresponds
to an integral over internal coordinates.

We evaluate this identity at $K=0$, in a uniform field $\phi(x)=\phi$ and for $x_1=0$. 
We then perform a Fourier transform of this expression and find:
\begin{equation}
\big(D_\varphi-d\big)\Gamma_k^{(1,0)}+\phi D_\varphi \partial_\phi\Gamma_k^{(1,0)}=\frac{1}{2} \Tr\bigg[\dot{R}_k G^2 \Gamma_k^{(3,0)}(q,-q)\bigg],
\end{equation}
where we used
\begin{align}\label{idGamma0momentum}
	&\Gamma_k^{(n+1)}(p_1,\cdots,p_{n-1},0)=\partial_\phi\Gamma_k^{(n)}(p_1,\cdots,p_{n-1}),\\
 &\Gamma_k^{(2,0)}(q=0) =\int_x\Gamma_k^{(2,0)}(0,x)
\end{align}

\subsubsection{Dilatation equation for $\Gamma_k^{(1,1)}$}

Now we differentiate Eq.~\eqref{Warddilat10} with respect to $K(y_1)$.
\begin{equation}\label{Warddilat11}
\begin{split}
&\int_x\Gamma_k^{(2,1)}(x,x_1;y_1) \big(D_\varphi+x_\mu \partial_\mu\big) \phi(x)\nonumber\\
&-K(x)\big(D_{\mathcal{O}}+x_\mu \partial_\mu\big) \Gamma_k^{(1,2)}(x_1;x,y_1) \nonumber\\
&+\big(D_\varphi-D_{\mathcal{O}}-d-{x_1}_\mu \partial_{x_1^{\mu}}-{y_1}_\mu \partial_{y_1^{\mu}}\big)\Gamma_k^{(1,1)}(x_1;y_1) \nonumber\\
&=\Tr\big[ \dot{R}_k G\bigg(\frac{1}{2}\Gamma_k^{(3,1)}(x_1;y_1)-\Gamma_k^{(3,0)}(x_1)G\Gamma_k^{(2,1)}(;y_1)\bigg) G\big].
\end{split}
\end{equation}

As before, we evaluate this identity for $K=0$, in a uniform field $\phi(x)=\phi$ and for $y_1=0$. We then  perform a Fourier transform and we obtain:
\begin{align}\label{dilG11}
&\big(D_\varphi-D_{\mathcal{O}}+p_\mu\partial_{p_\mu}+\phi D_\varphi \partial_\phi\big)\Gamma_k^{(1,1)}(p)\nonumber\\
&=\Tr\bigg[\dot{R}_k G^2 \bigg(\frac{1}{2}\Gamma_k^{(3,1)}(p;-p)-\Gamma_k^{(3,0)}(p) G \Gamma_k^{(2,1)}(;-p)\bigg)\bigg].
\end{align}

\subsubsection{Dilatation equation for $\Gamma_k^{(2,0)}$ and  $\Gamma_k^{(2,1)}$}

The same procedure repeated yields the following expressions for the vertices of interest $\Gamma_k^{(2,0)}$ and  $\Gamma_k^{(2,1)}$:
\begin{align}\label{eq:G20Dil}
&\big(2 D_\varphi-d+p_\mu\partial_{p_\mu}\big)\Gamma_k^{(2,0)}(p)+\phi D_\varphi \partial_\phi\Gamma_k^{(2,0)}(p)\nonumber\\
&=\Tr\bigg[\dot{R}_k G^2 \bigg(\frac{1}{2}\Gamma_k^{(4,0)}(p,-p)-\Gamma_k^{(3,0)}(p) G \Gamma_k^{(3,0)}(-p)\bigg)\bigg],
\end{align}

\begin{equation}
\begin{split}
\big(\phi D_\varphi\partial_\phi&+2D_\varphi-D_{\mathcal{O}}+p_\mu\partial_{p_\mu}+l_\mu\partial_{l_\mu}\big)\Gamma_k^{(2,1)}(p,l)\\
&=\Tr\bigg[\dot{R}_k G^2 \bigg(\frac{1}{2}\Gamma_k^{(4,1)}(p,l;-p-l) \\
&-\Gamma_k^{(3,1)}(p;-p-l) G \Gamma_k^{(3,0)}(l)\\
&-\Gamma_k^{(3,1)}(l;-p-l) G \Gamma_k^{(3,0)}(p,-q)\\
&-\Gamma_k^{(4,0)}(p,l) G \Gamma_k^{(2,1)}(;-p-l)\\
&+\Gamma_k^{(3,0)}(p) G \Gamma_k^{(3,0)}(l) G \Gamma_k^{(2,1)}(;-p-l)\\
&+\Gamma_k^{(3,0)}(p) G \Gamma_k^{(2,1)}(;-p-l) G \Gamma_k^{(3,0)}(l)\\
&+\Gamma_k^{(3,0)}(l) G \Gamma_k^{(3,0)}(p) G \Gamma_k^{(2,1)}(;-p-l)\bigg)\bigg]
\end{split}
\end{equation}

\subsubsection{Conformal equation for $\Gamma_k^{(1,1)}$ and $\Gamma_k^{(2,0)}$}
In the same manner, we could compute the Ward identities for the different vertices. However, since we are only interested in the one corresponding to $\Gamma_k^{(1,1)}$, we shall not go any further.
Starting from Eq.~\eqref{Wardconformal}, differentiating it with respect to $\phi(x_1)$ and $K(y_1)$ and then evaluating on $K=0$ and uniform field $\phi(x)=\phi$, we obtain:
\begin{equation}
\begin{split}
-\big(x_1^2\partial_{x_1^{\mu}}&-2{x_1}_\mu{x_1}_\nu\partial_{x_1^{\nu}}-2(d-D_\varphi){x_1}_\mu\\
&+y_1^2\partial_{y_1^{\mu}}-2{y_1}_\mu{y_1}_\nu\partial_{y_1^{\nu}}-2D_{\mathcal{O}}{y_1}_\mu\big) \Gamma_k^{(1,1)}(x_1,y_1) \\
&-2\phi D_\varphi\int_x x_\mu \Gamma_k^{(2,1)}(x,x_1;y_1)\\
&=\Tr\bigg[\dot{\bar{R}}_{\mu k} G^2 \bigg(-\frac{1}{2}\Gamma_k^{(3,1)}(x_1;y_1)\\
&+\Gamma_k^{(2,1)}(y_1) G \Gamma_k^{(3,0)}(x_1)\bigg)\bigg]
\end{split}
\end{equation}
where we introduced the notation ${\bar{R}}_{\mu k} (u,v)\equiv R_k (u,v)(u_\mu + v_\mu)$. Following the same steps as before, this is putting $y_1=0$ and performing a Fourier transform obtain:

\begin{equation}
\begin{split}
&\big(p_\mu\partial_{p_\nu}^2-2p_\nu\partial_{p_\nu}\partial_{p_\mu}-2D_{\varphi}\partial_{p_\mu}\big)\Gamma_k^{(1,1)}(p)\\
&-2\phi D_\varphi\partial_{q_\mu}\Gamma_k^{(2,1)}(q,p)\big|_{q=0}\\
&=\int_q\dot{R}_k(q) G^2(q)\big(\partial_{q_\mu}+\partial_{q'_\mu}\big)
\bigg(-\frac{1}{2}\Gamma_k^{(3,1)}(p,q,q')\\
&+\Gamma_k^{(3,0)}(p,q) G(p+q)\Gamma_k^{(2,1)}(q',p+q)\bigg)\bigg|_{q'=-q}.
\end{split}
\end{equation}

One can proceed in a similar manner to obtain other conformal Ward identities. For example, let us consider the identity for $\Gamma_k^{(2,0)}$. Taking functional derivatives with respect to $\phi(x_1)$ and $\phi(x_2)$, the same steps as before yield:
\begin{equation}\label{eq:G20Conf}
\begin{split}
&\big(p_\mu\partial_{p_\nu}^2-2p_\nu\partial_{p_\nu}\partial_{p_\mu}-2D_{\varphi}\partial_{p_\mu}\big)\Gamma^{(2,0)}(p)\\
&-2\phi D_\varphi\partial_{q_\mu}\Gamma_k^{(3,0)}(q,p)\big|_{q=0}\\
 &=\frac{1}{2}\int_{q} \dot{R}(q) G^2(q)\big(\partial_{q_\mu}+\partial_{q'_\mu}\big)\bigg(-\Gamma_k^{(4,0)}(p,q,q')\\
 &+\Gamma_k^{(3,0)}(p,q) G (p+q)\Gamma_k^{(3,0)}(p+q,q')\\
 &+\Gamma_k^{(3,0)}(q,p+q') G (p+q')\Gamma_k^{(3,0)}(p,q')\bigg)\bigg|_{q'=-q}
\end{split}
\end{equation}

\subsection{Ward identities with the derivative expansion at order $\mathcal{O}(\partial^2)$}

Let us now deduce the Ward identities for the various functions appearing
in the {\it ansatz} \eqref{ansatz}. The identities for $U_0(\phi)$ and $Z_0(\phi)$
are the standard ones (those without the composite operator source). We recall that for those functions, the conformal identity gives exactly the same equations that dilatation symmetry. We now deduce the equation for the functions $U_1(\phi)$, $Z_1(\phi)$ and $Y(\phi)$. In order to do that, we deduce the expression of the various vertices from the {\it ansatz} \eqref{ansatz}. We recall that we are working at linear order in $K$, and so, implicitly we are taking $K=0$ after differentiation.

First, the $\Gamma^{(1,1)}$ function is:
\begin{align}\label{Gamma11direct}
&\Gamma^{(1,1)}(y;x)=U_1'(\phi)\delta(x-y)-\partial_x^2\big(Y'(\phi)\delta(x-y)\big)\nonumber\\
&+\frac 1 2 Z_1'(\phi)(\partial^x_\mu\phi)^2 \delta(x-y)+Z_1(\phi) \partial^x_\mu(\phi)\partial^x_\mu \delta(x-y)
\end{align}

One can evaluate in a uniform field $\phi(x)=\phi$, at $y=0$ and Fourier-transforming to obtain:
\begin{equation}
\Gamma^{(1,1)}(p)=U_1'(\phi)+p^2 Y'(\phi)
\end{equation}

One can also differentiate (\ref{Gamma11direct}) with respect to $\varphi(x_2)$ to obtain in a uniform field (and for $x_2=0$) the Fourier transform:
\begin{equation}
\Gamma^{(2,1)}(p;l)=U_1''(\phi)+p\cdot(p+l)Z_1(\phi) +Y''(\phi)l^2
\end{equation}
In a similar manner one obtains as well the expression for any vertex $\Gamma^{(n,1)}$ at this order of the DE:
\begin{equation}
\begin{split}
\Gamma^{(n,1)}&(p_1,\dots,p_{n-1};l)=U_1^{(n)}(\phi)+Y^{(n)}(\phi)l^2\\
+\frac{1}{2}&\Big(\sum_{k=1}^{n-1} p_k^2+\big(l+\sum_{k=1}^{n-1} p_k\big)^2-l^2\Big) Z_1^{(n-2)}(\phi).
\end{split}
\end{equation}

It will be useful below to give the expression for $\hat{\Gamma}[\phi]$ defined in equation \eqref{eq:GammaHat}. At order $\mathcal{O}(\partial^2)$, one has:
\begin{equation}
\hat{\Gamma}[\phi]=\int_x \Big\{U_1(\phi)+\frac{1}{2}Z_1(\phi)(\partial_\mu\phi)^2\Big\}.
\end{equation}

\subsection{Equation for $U_1(\phi)$}

The equation for $U_1(\phi)$ (or more precisely for $U_1'(\phi)$) can be obtained from the dilatation equation for
$\Gamma^{(1,1)}(p)$ at $p=0$. One obtains:

\begin{align}
\label{V1eq}
&\big(D_\varphi-D_{\mathcal{O}}\Big)U_1'(\phi)+\phi D_\varphi U_1''(\phi)\nonumber\\
&=\int_q \dot{R}_k(q) G^2(q)
\Big\{\frac{1}{2}\big(U_1'''(\phi)+ q^2 Z_1'(\phi)\big)\nonumber\\
-&\big(U_1''(\phi)+q^2 Z_1(\phi)\big) G(q) \big(U_0'''(\phi)+q^2 Z_0'(\phi)\big)\Big\}
\end{align}

An alternative, and easier way to deduce this equation is to differenciate once the equation for $\hat{\Gamma}[\phi]$ with respect to $\phi$ and evaluate it at a uniform field.  It is important to stress that the equation for $U_1(\phi)$ is nothing but the equation
for the linear perturbation of the potential around the fixed point.

One must stress that special conformal transformations do not give new information for $U_1(\phi)$.

\subsection{Equation for $Z_1(\phi)$}

\begin{align}
\label{Z1eq}
&\big(2D_\varphi-D_{\mathcal{O}}+2\Big)Z_1(\phi)+\phi D_\varphi Z_1'(\phi)\nonumber\\
&=-\frac 1 2\int_q \dot{R}_k(q)G^2(q^2)\Big\{-\big(q^2 Z_1(\phi)+U_1''(\phi)\big) \nonumber\\
&\times \big(G'(q^2)\big)^2
\frac{8}{d} q^2 \big(q^2 Z_0'(\phi)+U_0^{(3)}(\phi)\big)^2 \nonumber\\
&-G'(q^2)\frac{4}{d} \Big(2 G(q^2) \big(q^2 Z_0'(\phi)+U_0^{(3)}(\phi)\big)\nonumber\\
&\times\Big(U_0^{(3)}(\phi) \big((d+2) q^2 Z_1(\phi)+d U_1''(\phi)\big)\nonumber\\
&+q^2
Z_0'(\phi) \big((d+6) q^2 Z_1(\phi)+(d+4) U_1''(\phi)\big)\Big)\nonumber\\
&-U_0^{(3)}(\phi) \big((d+2) q^2 Z_1'(\phi)+d U_1^{(3)}(\phi)\big)\nonumber\\
&-q^2
Z_0'(\phi) \big((d+4) q^2 Z_1'(\phi)+(d+2) U_1^{(3)}(\phi)\big)\Big)\nonumber\\
&-G''(q^2)\frac{8}{d} q^2 \big(q^2 Z_0'(\phi)+U_0^{(3)}(\phi)\big) \Big(-q^2 Z_1'(\phi)-U_1^{(3)}(\phi)\nonumber\\
&+2 G(q^2) \big(q^2 Z_0'(\phi)+U_0^{(3)}(\phi)\big) \big(q^2 Z_1(\phi)+U_1''(\phi)\big)\Big)\nonumber\\
&+\frac{2}{d} G^2(q^2) \Big(-2 U_0^{(3)}(\phi) Z_0'(\phi) \big(2 (2 d+1) q^2 Z_1(\phi)+3 d U_1''(\phi)\big)\nonumber\\
&+q^2 \big(Z_0'(\phi)\big)^2 \big(-7 (d+1) q^2
Z_1(\phi)-3 (2 d+1) U_1''(\phi)\big)\nonumber\\
&-d \big(U_0^{(3)}(\phi)\big)^2 Z_1(\phi)\Big)\nonumber\\
&+2 G(q^2) \Big(2 Z_0'(\phi) \big((2 d+1) q^2 Z_1'(\phi)+d
U_1^{(3)}(\phi)\big)\nonumber\\
&+d Z_0''(\phi) \big(q^2 Z_1(\phi)+U_1''(\phi)\big)+2 d U_0^{(3)}(\phi) Z_1'(\phi)\Big)-d Z_1''(\phi)
\Big\}
\end{align}

\subsection{Equation for $Y(\phi)$}

One can obtain the equation for $Y(\phi)$ by expanding the equation for $\Gamma^{(1,1)}(p)$ at order $p^2$. If one performs this calculation in the Ward identity for dilatation, one obtains:
\begin{align}
\label{Yeq}
\big(D_\varphi-&D_{\mathcal{O}}+2\big)Y'(\phi)+\phi D_\varphi Y''(\phi)
=\int_q \dot{R}_k(q) G^2(q^2) \nonumber\\
& \times \Big\{\frac{1}{2}Y'''(\phi)
-Y''(\phi) G(q^2) \big(U_0'''(\phi)+q^2 Z_0'(\phi)\big)\nonumber\\
-&\frac{q^2}{d} Z_1(\phi)\Big( G(q^2) Z_0'(\phi)+2 G'(q^2) \big(U_0'''(\phi)+q^2 Z_0'(\phi)\big)\Big)\nonumber\\
-&\big(U_1''(\phi)+q^2 Z_1(\phi)\big)\Big[Z_0'(\phi)\Big(G(q^2)+\frac{2}{d} q^2 G'(q^2)\Big)\nonumber\\
+&\Big(G'(q^2)+\frac{2 q^2}{d}G''(q^2)\Big)\big(U_0'''(\phi)+q^2 Z_0'(\phi)\big)+\Big]\Big\}
\end{align}
If one compares equation \eqref{Yeq} with the first two lines of equation for $U_1(\phi)$, Eq.~\eqref{V1eq}, one observes that they are identical except for
the fact that $U_1(\phi)$ is replaced by $Y(\phi)$, $D_{\mathcal{O}}$
is replaced by $D_{\mathcal{O}}-2$ and $Z_1$ is not present in the case of $Y$. This is not a surprise, because we have seen that the equation for $\hat{\Gamma}[\phi]$ is identical to that of $\hat{\Gamma}_{\mu\nu}[\phi]$ except that $D_{\mathcal{O}}$ is replaced by
$D_{\mathcal{O}}-2$. Now, given a solution of the eigenvalue equations, where $U_1(\phi)$ is the potential in $\hat{\Gamma}[\phi]$ one can construct a solution where $U_1(\phi)$ and $Z_1(\phi)$ are zero and where $Y(\phi)$ is the potential in translational invariant part of $\hat{\Gamma}_{\mu\mu}[\phi]$ (up to a normalization) so they must satisfy the same equation
(once we impose $U_1(\phi)=0$ and $Z_1(\phi)=0$ on it). The fact that the equivalent of $Z_1(\phi)$ is not present in the equation of $Y(\phi)$ should not be a surprise, given the fact that that would be a $\partial^4$ term in the effective action, so it has been correspondingly neglected.

The consequence of this analysis is that, if we are considering the leading even scalar perturbation, one may have $Y(\phi)\neq 0$, but we certainly must have $U_1(\phi)$ different from zero. Alternatively, one could consider the case $U_1(\phi)=Z_1(\phi)=0$ and $Y(\phi)$ different from zero. That would correspond simply to the local eigenoperator $\partial^2(\varphi_R^2(x))$ whose dimension is trivially related to the one of $\varphi_R^2(x)$.

It is important to stress that the operator $\partial^2(\varphi_R^2(x))$ is not a primary operator, so in order to impose conformal constraints to it, one must change the form of the special conformal transformations. Anyhow this is not necessary because its scaling properties can be obtained from $\varphi_R^2(x)$.

\subsection{Conformal equation at order $\mathcal{O}(\partial^2)$}

One can also obtain a non-trivial conformal equation at order $\mathcal{O}(\partial^2)$ by plugin the corresponding {\it ansatz} in the conformal identity for $\Gamma^{(1,1)}(p)$. In fact, imposing the conformal identity for $\Gamma^{(2,1)}(p;l)$ does not give further information. At this order of the DE, one has for this identity one term proportional to $p_{\mu}$ and one term proportional to $l_{\mu}$. The term proportional to $p_{\mu}$ can be evaluated at $l=0$ at this order and we proved before that the conformal identity at $l=0$ is nothing but the derivative of the dilatation identity. The term proportional to $l_{\mu}$ can be evaluated at this order at $p=0$ but it just give in that case the derivative with respect to $\phi$ of the equation for $\Gamma^{(1,1)}(p)$. Accordingly, a single non-trivial conformal identity (for each primary operator) is obtained at this order. The equation is presented in the main text in Eq.~\eqref{conf_eq_v2}.

\section{Compatibility of dilatation and conformal Ward identities for $\Gamma_{k}^{(2,0)}(p)$}
\label{gamma2confdil}

We now show that up to the vertex $\Gamma_{k}^{(2,0)}(p)$, which includes the functions  $U_0(\phi)$ and $Z_0(\phi)$, the Ward identities for dilatation and conformal invariance lead to the same constraints. To do this, we will prove that the special conformal Ward identity Eq.~\eqref{eq:G20Conf} is just a consequence of the dilatation Ward identity Eq.~\eqref{eq:G20Dil} (and rotation invariance). Let us start by noticing that
\begin{equation}\label{gamm2relation1}
\begin{split}
&\partial_{r^\mu}\Gamma_k^{(3,0)}(p,r)\big|_{r=0}=\partial_{r^\mu}\Gamma_k^{(3,0)}(-p-r,r)\big|_{r=0}\\
&=\partial_{r^\mu}\Gamma_k^{(3,0)}(p+r,-r)\big|_{r=0}\\
&=\partial_{p^\mu}\Gamma_k^{(3,0)}(p,0)-\partial_{r^\mu}\Gamma_k^{(3,0)}(p,r)\big|_{r=0}\\
&=\frac{1}{2}\partial_{p^\mu}\Gamma_k^{(3,0)}(p,0)=\frac{1}{2}\partial_{p^\mu}\partial_\phi\Gamma_k^{(2,0)}(p), 
\end{split}
\end{equation}
where in the last equality it was used equation \eqref{idGamma0momentum}. This transforms the term $2\phi D_\varphi\partial_{r^\mu}\Gamma^{(3,0)}(p,r)\big|_{r=0}$ into $\partial_{p^\mu}\phi D_\varphi \partial_\phi\Gamma^{(2,0)}(p)$.

At this point is useful to observe that the r.h.s. the dilatation identity for the vertex $\Gamma^{(n,m)}$ can be 
expressed in the form
\begin{equation}\label{eq:rhsdil}
   \frac 1 2 \int_q \partial_t R_k(q) G_k^2(q^2)H^{(n,m)}\big(p_1,\dots,p_{n-1},q,q';l_1,\dots,l_m\big)
\end{equation}
In a similar manner, r.h.s. of the conformal Ward identity can be written in the form 
\begin{equation}\label{eq:rhsconf}
\begin{split}
    -\frac 1 2&\int_q \partial_t R_k(q)G_k^2(q^2) \times\\
    &\big(\partial_{q_\mu}+\partial_{q'_\mu} \big) H^{(n,m)}\big(p_1,\dots,p_{n-1},q,q';l_1,\dots,l_m\big)\Big|_{q'=-q}.    
\end{split}
\end{equation}
For example, the explicit expression of $H^{(2,0)}\big(p,q,q'\big)$ can be extracted from Eq.~\eqref{eq:G20Conf}:
\begin{align}
&H^{(2,0)}\big(p,q,q'\big)=\Gamma_k^{(3,0)}(p,q) G (p+q)\Gamma_k^{(3,0)}(p+q,q')\nonumber\\
&+\Gamma_k^{(3,0)}(q,p+q') G (p+q')\Gamma_k^{(3,0)}(p,q')-\Gamma_k^{(4,0)}(p,q,q').
\end{align}
The functions $H^{(n,m)}$ appearing in equations \eqref{eq:rhsdil} and \eqref{eq:rhsconf} can be chosen with the following properties given the fact that it is inside the integral over $q$:
\begin{itemize}
	\item $H^{(n,m)}\big(p_1,\dots,p_{n-1},q,q';l_1,\dots,l_m\big)$ is completely symmetric on momenta in the set
 $\{p_1,\dots,p_{n-1},-p_1-\dots-p_{n-1}-l_1-\dots-l_{m}-q-q'\}$.
 \item $H^{(n,m)}\big(p_1,\dots,p_{n-1},q,q';l_1,\dots,l_m\big)$ is completely symmetric on momenta in the set
 $\{l_1,\dots,l_{n-1},l_{n}\}$.
	\item $H^{(n,m)}\big(p_1,\dots,p_{n-1},q,q';l_1,\dots,l_m\big)$ is symmetric under the exchange of $q$ and $q'$.
\end{itemize}
With these properties in mind, we can manipulate the right hand side of the special conformal Ward identity for the vertex function $\Gamma_k^{(2,0)}$ in the following manner:
\begin{equation}\label{gamm2relation2}
\begin{split}
&\big(\partial_{q_\mu}+\partial_{q'_\mu}\big)H^{(2,0)}(p,q,q')|_{q'=-q}\\
&=\big(\partial_{q_\mu}+\partial_{q'_\mu}\big)H^{(2,0)}(-p-q-q',q,q')|_{q'=-q}\\
&=\big(\partial_{q_\mu}+\partial_{q'_\mu}\big)H^{(2,0)}(p+q+q',-q,-q')|_{q'=-q}\\
&=2\partial_{p_\mu}H^{(2,0)}(p,q,-q)-\big(\partial_{q_\mu}+\partial_{q'_\mu}\big)H^{(2,0)}(p,q,q')|_{q'=-q}\\
&=\partial_{p_\mu}H^{(2,0)}(p,q,-q).
\end{split}
\end{equation}
The last piece of the puzzle is quite obvious now, we just apply a $p^\mu$ derivative on Eq.~\eqref{eq:G20Dil} for the $\Gamma_k^{(2,0)}$, this reads:
\begin{equation}
\begin{split}
&\partial_{p^\mu}\big(p_\nu\partial_{p^{\nu}}-d+2D_\varphi+\phi D_\varphi\partial_{\phi}\big)\Gamma^{(2,0)}(p)\\
&=\big(p_\nu\partial^2_{p^{\nu}p^\mu}+(2D_\varphi-d+1)\partial_{p^\mu}+\partial_{p^\mu}\phi D_\varphi\partial_{\phi}\big)\Gamma^{(2,0)}(p)\\
&=\big(2 p_\nu\partial^2_{p^{\nu}p^\mu}-p_\mu\partial^2_{p^\nu p^\nu}+2D_\varphi\partial_{p^\mu}+\partial_{p^\mu}\phi D_\varphi\partial_{\phi}\big)\Gamma^{(2,0)}(p)\\
&+\big(p_\mu\partial^2_{p^\nu p^\nu}- p_\nu\partial^2_{p^{\nu}p^\mu}+(-d+1)\partial_{p^\mu}\big)\Gamma^{(2,0)}(p)\\
&=\big(2 p_\nu\partial^2_{p^{\nu} p^\mu}-p_\mu\partial^2_{p^\nu p^\nu}+2D_\varphi\partial_{p^\mu}+\partial_{p^\mu}\phi D_\varphi\partial_{\phi}\big)\Gamma^{(2,0)}(p)\\
&=\partial_{p_\mu}\int_q \dot{R}(q) G^2(q) H^{(2,0)}(p,q,-q).
\end{split}
\end{equation}
The first to last equality is obtained after using the rotational Ward identity (as can be easily checked). Recognizing the terms already worked out in equations \eqref{gamm2relation1} and \eqref{gamm2relation2}, we check that we obtain Eq.~\eqref{eq:G20Dil}. In this way we verify that the special conformal Ward identity for $\Gamma_k^{(2,0)}$ is nothing but the derivative of the dilatation Ward identity for the same vertex.

\section{Numerical details}

After deducing the expressions that were analyzed in this study, the numerical procedure consists in two stages. The first stage consists in solving FRG flow equations in order to obtain the fixed point that governs the critical regime of the system. The second and final stage is that of performing a stability analysis of the fixed point or solving the equations for the eigenperturbations and to evaluate at this fixed point the conformal constraints for these eigenperturbations.

\subsection{Solving for the fixed point}
The FRG equations within the DE ansatz and with a generic regulator are partial integro-differential equations. To treat these type of equations we discretize the field dependence in an uniform $\rho=\phi^2/2$ grid with 201 points and consider discretized 7-points centered derivatives in this grid for the partial derivatives with respect to $\rho$. When getting close to the edges of the selected grid, the derivatives are taken as centered as possible.
For the momentum integrals, we employ a 21-point adaptive Gauss-Kronrod quadrature up to momentum $q_{\rm max}=10k$, where $k$ is the scale of the regulator. This ensures that integrals being considered are well converged due to the exponentially fast decay of the integrands in virtue of terms such as $n$-th powers of the regulator and $\partial_t R_k$. With this setup, we moved on to finding zeros of the beta functions. To do this one can proceed in two ways: to start from a microscopic initial condition and fine-tune on a temperature-like parameter or to start working in a dimension close to the upper critical one and to start a root-finding procedure using an approximate analytical expression and then to reduce the dimension parameter towards $d=3$. Either way, the fine-tuning procedure is not accurate enough on its own due to cumulative errors while flowing from a microscopic scale $k=\Lambda$ up to the Ginzburg scale $k_G$. Because of this issue, a root-finding procedure must be implemented after running the flow with a method of choice which serves as a good initial condition for the search of zeros of the beta function already at the dimension of interest $d=3$.
In any case, once this educated guess of fixed point is at our disposal, the root-finding procedure consists in a discretized Newton-Raphson or secant method.

\subsection{Linear analysis at the fixed point}

Once we are at the fixed point, we need to solve the eigenvalue problem to find the eigenperturbations. We do this by resorting again to a Newton-Raphson procedure. For this purpose, a simple linear stability analysis of the fixed point serves as a very good starting point for the search of 
eigenvectors and eigenvalues. Once a desired threshold of convergence is achieved for the method, we conclude the numerical study by evaluating the conformal constraint with the found eigenvalue ($D_{\mathcal{O}}$), eigenvector 
(functions $\big\{U_1(\rho),Z_1(\rho),Y(\rho)\big\}$) 
and the Wilson-Fisher fixed point solution (given by functions $\big\{U_0(\phi),Z_0(\phi)\big\}$).

\bibliographystyle{apsrev4-2}
\bibliography{articleConformalComposite}

\begin{thebibliography}{70}%
\makeatletter
\providecommand \@ifxundefined [1]{%
 \@ifx{#1\undefined}
}%
\providecommand \@ifnum [1]{%
 \ifnum #1\expandafter \@firstoftwo
 \else \expandafter \@secondoftwo
 \fi
}%
\providecommand \@ifx [1]{%
 \ifx #1\expandafter \@firstoftwo
 \else \expandafter \@secondoftwo
 \fi
}%
\providecommand \natexlab [1]{#1}%
\providecommand \enquote  [1]{``#1''}%
\providecommand \bibnamefont  [1]{#1}%
\providecommand \bibfnamefont [1]{#1}%
\providecommand \citenamefont [1]{#1}%
\providecommand \href@noop [0]{\@secondoftwo}%
\providecommand \href [0]{\begingroup \@sanitize@url \@href}%
\providecommand \@href[1]{\@@startlink{#1}\@@href}%
\providecommand \@@href[1]{\endgroup#1\@@endlink}%
\providecommand \@sanitize@url [0]{\catcode `\\12\catcode `\$12\catcode
  `\&12\catcode `\#12\catcode `\^12\catcode `\_12\catcode `\%12\relax}%
\providecommand \@@startlink[1]{}%
\providecommand \@@endlink[0]{}%
\providecommand \url  [0]{\begingroup\@sanitize@url \@url }%
\providecommand \@url [1]{\endgroup\@href {#1}{\urlprefix }}%
\providecommand \urlprefix  [0]{URL }%
\providecommand \Eprint [0]{\href }%
\providecommand \doibase [0]{https://doi.org/}%
\providecommand \selectlanguage [0]{\@gobble}%
\providecommand \bibinfo  [0]{\@secondoftwo}%
\providecommand \bibfield  [0]{\@secondoftwo}%
\providecommand \translation [1]{[#1]}%
\providecommand \BibitemOpen [0]{}%
\providecommand \bibitemStop [0]{}%
\providecommand \bibitemNoStop [0]{.\EOS\space}%
\providecommand \EOS [0]{\spacefactor3000\relax}%
\providecommand \BibitemShut  [1]{\csname bibitem#1\endcsname}%
\let\auto@bib@innerbib\@empty
\bibitem [{\citenamefont {Wilson}\ and\ \citenamefont
  {Fisher}(1972)}]{Wilson:1971dc}%
  \BibitemOpen
  \bibfield  {author} {\bibinfo {author} {\bibfnamefont {K.~G.}\ \bibnamefont
  {Wilson}}\ and\ \bibinfo {author} {\bibfnamefont {M.~E.}\ \bibnamefont
  {Fisher}},\ }\href {https://doi.org/10.1103/PhysRevLett.28.240} {\bibfield
  {journal} {\bibinfo  {journal} {Phys. Rev. Lett.}\ }\textbf {\bibinfo
  {volume} {28}},\ \bibinfo {pages} {240} (\bibinfo {year} {1972})}\BibitemShut
  {NoStop}%
\bibitem [{\citenamefont {Wilson}\ and\ \citenamefont
  {Kogut}(1974)}]{Wilson:1973jj}%
  \BibitemOpen
  \bibfield  {author} {\bibinfo {author} {\bibfnamefont {K.~G.}\ \bibnamefont
  {Wilson}}\ and\ \bibinfo {author} {\bibfnamefont {J.~B.}\ \bibnamefont
  {Kogut}},\ }\href {https://doi.org/10.1016/0370-1573(74)90023-4} {\bibfield
  {journal} {\bibinfo  {journal} {Phys. Rept.}\ }\textbf {\bibinfo {volume}
  {12}},\ \bibinfo {pages} {75} (\bibinfo {year} {1974})}\BibitemShut {NoStop}%
\bibitem [{\citenamefont {Polyakov}(1970)}]{Polyakov:1970xd}%
  \BibitemOpen
  \bibfield  {author} {\bibinfo {author} {\bibfnamefont {A.~M.}\ \bibnamefont
  {Polyakov}},\ }\href {https://ci.nii.ac.jp/naid/10006414995/en/} {\bibfield
  {journal} {\bibinfo  {journal} {JETP Lett.}\ }\textbf {\bibinfo {volume}
  {12}},\ \bibinfo {pages} {381} (\bibinfo {year} {1970})}\BibitemShut
  {NoStop}%
\bibitem [{\citenamefont {Migdal}(1971)}]{Migdal:1971xh}%
  \BibitemOpen
  \bibfield  {author} {\bibinfo {author} {\bibfnamefont {A.~A.}\ \bibnamefont
  {Migdal}},\ }\href {https://doi.org/10.1016/0370-2693(71)90583-1} {\bibfield
  {journal} {\bibinfo  {journal} {Phys. Lett.}\ }\textbf {\bibinfo {volume}
  {37B}},\ \bibinfo {pages} {98} (\bibinfo {year} {1971})}\BibitemShut
  {NoStop}%
\bibitem [{\citenamefont {Di~Francesco}\ \emph {et~al.}(1997)\citenamefont
  {Di~Francesco}, \citenamefont {Mathieu},\ and\ \citenamefont
  {Senechal}}]{DiFrancesco:1997nk}%
  \BibitemOpen
  \bibfield  {author} {\bibinfo {author} {\bibfnamefont {P.}~\bibnamefont
  {Di~Francesco}}, \bibinfo {author} {\bibfnamefont {P.}~\bibnamefont
  {Mathieu}},\ and\ \bibinfo {author} {\bibfnamefont {D.}~\bibnamefont
  {Senechal}},\ }\href {https://doi.org/10.1007/978-1-4612-2256-9} {\emph
  {\bibinfo {title} {{Conformal Field Theory}}}},\ Graduate Texts in
  Contemporary Physics\ (\bibinfo  {publisher} {Springer-Verlag},\ \bibinfo
  {address} {New York},\ \bibinfo {year} {1997})\BibitemShut {NoStop}%
\bibitem [{\citenamefont {Belavin}\ \emph {et~al.}(1984)\citenamefont
  {Belavin}, \citenamefont {Polyakov},\ and\ \citenamefont
  {Zamolodchikov}}]{Belavin:1984vu}%
  \BibitemOpen
  \bibfield  {author} {\bibinfo {author} {\bibfnamefont {A.~A.}\ \bibnamefont
  {Belavin}}, \bibinfo {author} {\bibfnamefont {A.~M.}\ \bibnamefont
  {Polyakov}},\ and\ \bibinfo {author} {\bibfnamefont {A.~B.}\ \bibnamefont
  {Zamolodchikov}},\ }\href {https://doi.org/10.1016/0550-3213(84)90052-X}
  {\bibfield  {journal} {\bibinfo  {journal} {Nucl. Phys.}\ }\textbf {\bibinfo
  {volume} {B241}},\ \bibinfo {pages} {333} (\bibinfo {year} {1984})},\
  \bibinfo {note} {[,605(1984)]}\BibitemShut {NoStop}%
\bibitem [{\citenamefont {Schnetz}(2018)}]{Schnetz:2016fhy}%
  \BibitemOpen
  \bibfield  {author} {\bibinfo {author} {\bibfnamefont {O.}~\bibnamefont
  {Schnetz}},\ }\href {https://doi.org/10.1103/PhysRevD.97.085018} {\bibfield
  {journal} {\bibinfo  {journal} {Phys. Rev. D}\ }\textbf {\bibinfo {volume}
  {97}},\ \bibinfo {pages} {085018} (\bibinfo {year} {2018})},\ \Eprint
  {https://arxiv.org/abs/1606.08598} {arXiv:1606.08598 [hep-th]} \BibitemShut
  {NoStop}%
\bibitem [{\citenamefont {Kompaniets}\ and\ \citenamefont
  {Panzer}(2017)}]{Kompaniets:2017yct}%
  \BibitemOpen
  \bibfield  {author} {\bibinfo {author} {\bibfnamefont {M.~V.}\ \bibnamefont
  {Kompaniets}}\ and\ \bibinfo {author} {\bibfnamefont {E.}~\bibnamefont
  {Panzer}},\ }\href {https://doi.org/10.1103/PhysRevD.96.036016} {\bibfield
  {journal} {\bibinfo  {journal} {Phys. Rev.}\ }\textbf {\bibinfo {volume}
  {D96}},\ \bibinfo {pages} {036016} (\bibinfo {year} {2017})},\ \Eprint
  {https://arxiv.org/abs/1705.06483} {arXiv:1705.06483 [hep-th]} \BibitemShut
  {NoStop}%
\bibitem [{\citenamefont {Shalaby}(2020)}]{Shalaby:2020faz}%
  \BibitemOpen
  \bibfield  {author} {\bibinfo {author} {\bibfnamefont {A.~M.}\ \bibnamefont
  {Shalaby}},\ }\href {https://doi.org/10.1103/PhysRevD.102.105017} {\bibfield
  {journal} {\bibinfo  {journal} {Phys. Rev. D}\ }\textbf {\bibinfo {volume}
  {102}},\ \bibinfo {pages} {105017} (\bibinfo {year} {2020})},\ \Eprint
  {https://arxiv.org/abs/2010.13097} {arXiv:2010.13097 [hep-th]} \BibitemShut
  {NoStop}%
\bibitem [{\citenamefont {{Abhignan}}\ and\ \citenamefont
  {{Sankaranarayanan}}(2021)}]{Abhignan06}%
  \BibitemOpen
  \bibfield  {author} {\bibinfo {author} {\bibfnamefont {V.}~\bibnamefont
  {{Abhignan}}}\ and\ \bibinfo {author} {\bibfnamefont {R.}~\bibnamefont
  {{Sankaranarayanan}}},\ }\href {https://doi.org/10.1007/s10955-021-02719-z}
  {\bibfield  {journal} {\bibinfo  {journal} {Journal of Statistical Physics}\
  }\textbf {\bibinfo {volume} {183}},\ \bibinfo {eid} {4} (\bibinfo {year}
  {2021})},\ \Eprint {https://arxiv.org/abs/2006.12064} {arXiv:2006.12064
  [cond-mat.stat-mech]} \BibitemShut {NoStop}%
\bibitem [{\citenamefont {Shalaby}(2021)}]{Shalaby:2020xvv}%
  \BibitemOpen
  \bibfield  {author} {\bibinfo {author} {\bibfnamefont {A.~M.}\ \bibnamefont
  {Shalaby}},\ }\href {https://doi.org/10.1140/epjc/s10052-021-08884-5}
  {\bibfield  {journal} {\bibinfo  {journal} {Eur. Phys. J. C}\ }\textbf
  {\bibinfo {volume} {81}},\ \bibinfo {pages} {87} (\bibinfo {year} {2021})},\
  \Eprint {https://arxiv.org/abs/2005.12714} {arXiv:2005.12714 [hep-th]}
  \BibitemShut {NoStop}%
\bibitem [{\citenamefont {Hasenbusch}(2019)}]{Hasenbusch:2019jkj}%
  \BibitemOpen
  \bibfield  {author} {\bibinfo {author} {\bibfnamefont {M.}~\bibnamefont
  {Hasenbusch}},\ }\href {https://doi.org/10.1103/PhysRevB.100.224517}
  {\bibfield  {journal} {\bibinfo  {journal} {Phys. Rev. B}\ }\textbf {\bibinfo
  {volume} {100}},\ \bibinfo {pages} {224517} (\bibinfo {year}
  {2019})}\BibitemShut {NoStop}%
\bibitem [{\citenamefont {{Hasenbusch}}(2020)}]{Hasenbusch2005}%
  \BibitemOpen
  \bibfield  {author} {\bibinfo {author} {\bibfnamefont {M.}~\bibnamefont
  {{Hasenbusch}}},\ }\href {https://doi.org/10.1103/PhysRevB.102.024406}
  {\bibfield  {journal} {\bibinfo  {journal} {\prb}\ }\textbf {\bibinfo
  {volume} {102}},\ \bibinfo {eid} {024406} (\bibinfo {year} {2020})},\ \Eprint
  {https://arxiv.org/abs/2005.04448} {arXiv:2005.04448 [cond-mat.stat-mech]}
  \BibitemShut {NoStop}%
\bibitem [{\citenamefont {Hasenbusch}(2021)}]{Hasenbusch:2021tei}%
  \BibitemOpen
  \bibfield  {author} {\bibinfo {author} {\bibfnamefont {M.}~\bibnamefont
  {Hasenbusch}},\ }\href {https://doi.org/10.1103/PhysRevB.104.014426}
  {\bibfield  {journal} {\bibinfo  {journal} {Phys. Rev. B}\ }\textbf {\bibinfo
  {volume} {104}},\ \bibinfo {pages} {014426} (\bibinfo {year} {2021})},\
  \Eprint {https://arxiv.org/abs/2105.09781} {arXiv:2105.09781
  [cond-mat.stat-mech]} \BibitemShut {NoStop}%
\bibitem [{\citenamefont {El-Showk}\ \emph {et~al.}(2012)\citenamefont
  {El-Showk}, \citenamefont {Paulos}, \citenamefont {Poland}, \citenamefont
  {Rychkov}, \citenamefont {Simmons-Duffin},\ and\ \citenamefont
  {Vichi}}]{ElShowk:2012ht}%
  \BibitemOpen
  \bibfield  {author} {\bibinfo {author} {\bibfnamefont {S.}~\bibnamefont
  {El-Showk}}, \bibinfo {author} {\bibfnamefont {M.~F.}\ \bibnamefont
  {Paulos}}, \bibinfo {author} {\bibfnamefont {D.}~\bibnamefont {Poland}},
  \bibinfo {author} {\bibfnamefont {S.}~\bibnamefont {Rychkov}}, \bibinfo
  {author} {\bibfnamefont {D.}~\bibnamefont {Simmons-Duffin}},\ and\ \bibinfo
  {author} {\bibfnamefont {A.}~\bibnamefont {Vichi}},\ }\href
  {https://doi.org/10.1103/PhysRevD.86.025022} {\bibfield  {journal} {\bibinfo
  {journal} {Phys. Rev.}\ }\textbf {\bibinfo {volume} {D86}},\ \bibinfo {pages}
  {025022} (\bibinfo {year} {2012})},\ \Eprint
  {https://arxiv.org/abs/1203.6064} {arXiv:1203.6064 [hep-th]} \BibitemShut
  {NoStop}%
\bibitem [{\citenamefont {El-Showk}\ \emph {et~al.}(2014)\citenamefont
  {El-Showk}, \citenamefont {Paulos}, \citenamefont {Poland}, \citenamefont
  {Rychkov}, \citenamefont {Simmons-Duffin},\ and\ \citenamefont
  {Vichi}}]{El-Showk:2014dwa}%
  \BibitemOpen
  \bibfield  {author} {\bibinfo {author} {\bibfnamefont {S.}~\bibnamefont
  {El-Showk}}, \bibinfo {author} {\bibfnamefont {M.~F.}\ \bibnamefont
  {Paulos}}, \bibinfo {author} {\bibfnamefont {D.}~\bibnamefont {Poland}},
  \bibinfo {author} {\bibfnamefont {S.}~\bibnamefont {Rychkov}}, \bibinfo
  {author} {\bibfnamefont {D.}~\bibnamefont {Simmons-Duffin}},\ and\ \bibinfo
  {author} {\bibfnamefont {A.}~\bibnamefont {Vichi}},\ }\href
  {https://doi.org/10.1007/s10955-014-1042-7} {\bibfield  {journal} {\bibinfo
  {journal} {J. Stat. Phys.}\ }\textbf {\bibinfo {volume} {157}},\ \bibinfo
  {pages} {869} (\bibinfo {year} {2014})},\ \Eprint
  {https://arxiv.org/abs/1403.4545} {arXiv:1403.4545 [hep-th]} \BibitemShut
  {NoStop}%
\bibitem [{\citenamefont {Kos}\ \emph {et~al.}(2014)\citenamefont {Kos},
  \citenamefont {Poland},\ and\ \citenamefont {Simmons-Duffin}}]{Kos:2014bka}%
  \BibitemOpen
  \bibfield  {author} {\bibinfo {author} {\bibfnamefont {F.}~\bibnamefont
  {Kos}}, \bibinfo {author} {\bibfnamefont {D.}~\bibnamefont {Poland}},\ and\
  \bibinfo {author} {\bibfnamefont {D.}~\bibnamefont {Simmons-Duffin}},\ }\href
  {https://doi.org/10.1007/JHEP11(2014)109} {\bibfield  {journal} {\bibinfo
  {journal} {JHEP}\ }\textbf {\bibinfo {volume} {11}},\ \bibinfo {pages}
  {109}},\ \Eprint {https://arxiv.org/abs/1406.4858} {arXiv:1406.4858 [hep-th]}
  \BibitemShut {NoStop}%
\bibitem [{\citenamefont {Chester}\ \emph {et~al.}(2019)\citenamefont
  {Chester}, \citenamefont {Landry}, \citenamefont {Liu}, \citenamefont
  {Poland}, \citenamefont {Simmons-Duffin}, \citenamefont {Su},\ and\
  \citenamefont {Vichi}}]{chester2019carving}%
  \BibitemOpen
  \bibfield  {author} {\bibinfo {author} {\bibfnamefont {S.~M.}\ \bibnamefont
  {Chester}}, \bibinfo {author} {\bibfnamefont {W.}~\bibnamefont {Landry}},
  \bibinfo {author} {\bibfnamefont {J.}~\bibnamefont {Liu}}, \bibinfo {author}
  {\bibfnamefont {D.}~\bibnamefont {Poland}}, \bibinfo {author} {\bibfnamefont
  {D.}~\bibnamefont {Simmons-Duffin}}, \bibinfo {author} {\bibfnamefont
  {N.}~\bibnamefont {Su}},\ and\ \bibinfo {author} {\bibfnamefont
  {A.}~\bibnamefont {Vichi}},\ }\href@noop {} {\bibinfo {title} {Carving out
  ope space and precise $o(2)$ model critical exponents}} (\bibinfo {year}
  {2019}),\ \Eprint {https://arxiv.org/abs/1912.03324} {arXiv:1912.03324
  [hep-th]} \BibitemShut {NoStop}%
\bibitem [{\citenamefont {Wetterich}(1993)}]{Wetterich:1992yh}%
  \BibitemOpen
  \bibfield  {author} {\bibinfo {author} {\bibfnamefont {C.}~\bibnamefont
  {Wetterich}},\ }\href
  {https://doi.org/https://doi.org/10.1016/0370-2693(93)90726-X} {\bibfield
  {journal} {\bibinfo  {journal} {Physics Letters B}\ }\textbf {\bibinfo
  {volume} {301}},\ \bibinfo {pages} {90 } (\bibinfo {year}
  {1993})}\BibitemShut {NoStop}%
\bibitem [{\citenamefont {Ellwanger}(1993)}]{Ellwanger:1993kk}%
  \BibitemOpen
  \bibfield  {author} {\bibinfo {author} {\bibfnamefont {U.}~\bibnamefont
  {Ellwanger}},\ }\href {https://doi.org/10.1007/BF01553022} {\bibfield
  {journal} {\bibinfo  {journal} {Z. Phys.}\ }\textbf {\bibinfo {volume}
  {C58}},\ \bibinfo {pages} {619} (\bibinfo {year} {1993})}\BibitemShut
  {NoStop}%
\bibitem [{\citenamefont {Morris}(1994{\natexlab{a}})}]{Morris:1993qb}%
  \BibitemOpen
  \bibfield  {author} {\bibinfo {author} {\bibfnamefont {T.~R.}\ \bibnamefont
  {Morris}},\ }\href {https://doi.org/10.1142/S0217751X94000972} {\bibfield
  {journal} {\bibinfo  {journal} {Int. J. Mod. Phys.}\ }\textbf {\bibinfo
  {volume} {A9}},\ \bibinfo {pages} {2411} (\bibinfo {year}
  {1994}{\natexlab{a}})},\ \Eprint {https://arxiv.org/abs/hep-ph/9308265}
  {arXiv:hep-ph/9308265 [hep-ph]} \BibitemShut {NoStop}%
\bibitem [{\citenamefont {Delamotte}(2012)}]{Delamotte:2007pf}%
  \BibitemOpen
  \bibfield  {author} {\bibinfo {author} {\bibfnamefont {B.}~\bibnamefont
  {Delamotte}},\ }\href {https://doi.org/10.1007/978-3-642-27320-9_2}
  {\bibfield  {journal} {\bibinfo  {journal} {Lect. Notes Phys.}\ }\textbf
  {\bibinfo {volume} {852}},\ \bibinfo {pages} {49} (\bibinfo {year} {2012})},\
  \Eprint {https://arxiv.org/abs/cond-mat/0702365} {arXiv:cond-mat/0702365
  [cond-mat.stat-mech]} \BibitemShut {NoStop}%
\bibitem [{\citenamefont {Dupuis}\ \emph {et~al.}(2021)\citenamefont {Dupuis},
  \citenamefont {Canet}, \citenamefont {Eichhorn}, \citenamefont {Metzner},
  \citenamefont {Pawlowski}, \citenamefont {Tissier},\ and\ \citenamefont
  {Wschebor}}]{Dupuis:2020fhh}%
  \BibitemOpen
  \bibfield  {author} {\bibinfo {author} {\bibfnamefont {N.}~\bibnamefont
  {Dupuis}}, \bibinfo {author} {\bibfnamefont {L.}~\bibnamefont {Canet}},
  \bibinfo {author} {\bibfnamefont {A.}~\bibnamefont {Eichhorn}}, \bibinfo
  {author} {\bibfnamefont {W.}~\bibnamefont {Metzner}}, \bibinfo {author}
  {\bibfnamefont {J.~M.}\ \bibnamefont {Pawlowski}}, \bibinfo {author}
  {\bibfnamefont {M.}~\bibnamefont {Tissier}},\ and\ \bibinfo {author}
  {\bibfnamefont {N.}~\bibnamefont {Wschebor}},\ }\href
  {https://doi.org/10.1016/j.physrep.2021.01.001} {\bibfield  {journal}
  {\bibinfo  {journal} {Phys. Rept.}\ }\textbf {\bibinfo {volume} {910}},\
  \bibinfo {pages} {1} (\bibinfo {year} {2021})},\ \Eprint
  {https://arxiv.org/abs/2006.04853} {arXiv:2006.04853 [cond-mat.stat-mech]}
  \BibitemShut {NoStop}%
\bibitem [{\citenamefont {Guida}\ and\ \citenamefont
  {Zinn-Justin}(1998)}]{Guida:1998bx}%
  \BibitemOpen
  \bibfield  {author} {\bibinfo {author} {\bibfnamefont {R.}~\bibnamefont
  {Guida}}\ and\ \bibinfo {author} {\bibfnamefont {J.}~\bibnamefont
  {Zinn-Justin}},\ }\href {https://doi.org/10.1088/0305-4470/31/40/006}
  {\bibfield  {journal} {\bibinfo  {journal} {J. Phys.}\ }\textbf {\bibinfo
  {volume} {A31}},\ \bibinfo {pages} {8103} (\bibinfo {year} {1998})},\ \Eprint
  {https://arxiv.org/abs/cond-mat/9803240} {arXiv:cond-mat/9803240 [cond-mat]}
  \BibitemShut {NoStop}%
\bibitem [{\citenamefont {Pelissetto}\ and\ \citenamefont
  {Vicari}(2002)}]{Pelissetto02}%
  \BibitemOpen
  \bibfield  {author} {\bibinfo {author} {\bibfnamefont {A.}~\bibnamefont
  {Pelissetto}}\ and\ \bibinfo {author} {\bibfnamefont {E.}~\bibnamefont
  {Vicari}},\ }\href
  {https://doi.org/http://dx.doi.org/10.1016/S0370-1573(02)00219-3} {\bibfield
  {journal} {\bibinfo  {journal} {Physics Reports}\ }\textbf {\bibinfo {volume}
  {368}},\ \bibinfo {pages} {549 } (\bibinfo {year} {2002})}\BibitemShut
  {NoStop}%
\bibitem [{\citenamefont {Zinn-Justin}(2002)}]{ZinnJustin:2002ru}%
  \BibitemOpen
  \bibfield  {author} {\bibinfo {author} {\bibfnamefont {J.}~\bibnamefont
  {Zinn-Justin}},\ }\href@noop {} {\emph {\bibinfo {title} {{Quantum field
  theory and critical phenomena}}}},\ Vol.\ \bibinfo {volume} {113}\ (\bibinfo
  {year} {2002})\ pp.\ \bibinfo {pages} {1--1054}\BibitemShut {NoStop}%
\bibitem [{\citenamefont {Tarjus}\ and\ \citenamefont
  {Tissier}(2004)}]{Tarjus:2004wyx}%
  \BibitemOpen
  \bibfield  {author} {\bibinfo {author} {\bibfnamefont {G.}~\bibnamefont
  {Tarjus}}\ and\ \bibinfo {author} {\bibfnamefont {M.}~\bibnamefont
  {Tissier}},\ }\href {https://doi.org/10.1103/PhysRevLett.93.267008}
  {\bibfield  {journal} {\bibinfo  {journal} {Phys. Rev. Lett.}\ }\textbf
  {\bibinfo {volume} {93}},\ \bibinfo {pages} {267008} (\bibinfo {year}
  {2004})},\ \Eprint {https://arxiv.org/abs/cond-mat/0410118}
  {arXiv:cond-mat/0410118 [cond-mat.dis-nn]} \BibitemShut {NoStop}%
\bibitem [{\citenamefont {Tissier}\ and\ \citenamefont
  {Tarjus}(2011)}]{Tissier:2011zz}%
  \BibitemOpen
  \bibfield  {author} {\bibinfo {author} {\bibfnamefont {M.}~\bibnamefont
  {Tissier}}\ and\ \bibinfo {author} {\bibfnamefont {G.}~\bibnamefont
  {Tarjus}},\ }\href {https://doi.org/10.1103/PhysRevLett.107.041601}
  {\bibfield  {journal} {\bibinfo  {journal} {Phys. Rev. Lett.}\ }\textbf
  {\bibinfo {volume} {107}},\ \bibinfo {pages} {041601} (\bibinfo {year}
  {2011})},\ \Eprint {https://arxiv.org/abs/1103.4812} {arXiv:1103.4812
  [cond-mat.stat-mech]} \BibitemShut {NoStop}%
\bibitem [{\citenamefont {Canet}\ \emph {et~al.}(2010)\citenamefont {Canet},
  \citenamefont {Chat\'e}, \citenamefont {Delamotte},\ and\ \citenamefont
  {Wschebor}}]{Canet10}%
  \BibitemOpen
  \bibfield  {author} {\bibinfo {author} {\bibfnamefont {L.}~\bibnamefont
  {Canet}}, \bibinfo {author} {\bibfnamefont {H.}~\bibnamefont {Chat\'e}},
  \bibinfo {author} {\bibfnamefont {B.}~\bibnamefont {Delamotte}},\ and\
  \bibinfo {author} {\bibfnamefont {N.}~\bibnamefont {Wschebor}},\ }\href
  {https://doi.org/10.1103/PhysRevLett.104.150601} {\bibfield  {journal}
  {\bibinfo  {journal} {Phys. Rev. Lett.}\ }\textbf {\bibinfo {volume} {104}},\
  \bibinfo {pages} {150601} (\bibinfo {year} {2010})}\BibitemShut {NoStop}%
\bibitem [{\citenamefont {Canet}\ \emph {et~al.}(2011)\citenamefont {Canet},
  \citenamefont {Chat\'e}, \citenamefont {Delamotte},\ and\ \citenamefont
  {Wschebor}}]{Canet11}%
  \BibitemOpen
  \bibfield  {author} {\bibinfo {author} {\bibfnamefont {L.}~\bibnamefont
  {Canet}}, \bibinfo {author} {\bibfnamefont {H.}~\bibnamefont {Chat\'e}},
  \bibinfo {author} {\bibfnamefont {B.}~\bibnamefont {Delamotte}},\ and\
  \bibinfo {author} {\bibfnamefont {N.}~\bibnamefont {Wschebor}},\ }\href
  {https://doi.org/10.1103/PhysRevE.84.061128} {\bibfield  {journal} {\bibinfo
  {journal} {Phys. Rev. E}\ }\textbf {\bibinfo {volume} {84}},\ \bibinfo
  {pages} {061128} (\bibinfo {year} {2011})}\BibitemShut {NoStop}%
\bibitem [{\citenamefont {Canet}\ \emph {et~al.}(2012)\citenamefont {Canet},
  \citenamefont {Chat\'e}, \citenamefont {Delamotte},\ and\ \citenamefont
  {Wschebor}}]{Canet11a}%
  \BibitemOpen
  \bibfield  {author} {\bibinfo {author} {\bibfnamefont {L.}~\bibnamefont
  {Canet}}, \bibinfo {author} {\bibfnamefont {H.}~\bibnamefont {Chat\'e}},
  \bibinfo {author} {\bibfnamefont {B.}~\bibnamefont {Delamotte}},\ and\
  \bibinfo {author} {\bibfnamefont {N.}~\bibnamefont {Wschebor}},\ }\href
  {https://doi.org/10.1103/PhysRevE.86.019904} {\bibfield  {journal} {\bibinfo
  {journal} {Phys. Rev. E}\ }\textbf {\bibinfo {volume} {86}},\ \bibinfo
  {pages} {019904} (\bibinfo {year} {2012})}\BibitemShut {NoStop}%
\bibitem [{\citenamefont {Coquand}\ \emph {et~al.}(2018)\citenamefont
  {Coquand}, \citenamefont {Essafi}, \citenamefont {Kownacki},\ and\
  \citenamefont {Mouhanna}}]{Coquand2017}%
  \BibitemOpen
  \bibfield  {author} {\bibinfo {author} {\bibfnamefont {O.}~\bibnamefont
  {Coquand}}, \bibinfo {author} {\bibfnamefont {K.}~\bibnamefont {Essafi}},
  \bibinfo {author} {\bibfnamefont {J.~P.}\ \bibnamefont {Kownacki}},\ and\
  \bibinfo {author} {\bibfnamefont {D.}~\bibnamefont {Mouhanna}},\ }\href
  {https://doi.org/10.1103/PhysRevE.97.030102} {\bibfield  {journal} {\bibinfo
  {journal} {Phys. Rev.}\ }\textbf {\bibinfo {volume} {E97}},\ \bibinfo {pages}
  {030102} (\bibinfo {year} {2018})},\ \Eprint
  {https://arxiv.org/abs/1708.08364} {arXiv:1708.08364 [cond-mat.dis-nn]}
  \BibitemShut {NoStop}%
\bibitem [{\citenamefont {L\'eonard}\ and\ \citenamefont
  {Delamotte}(2015)}]{Leonard15}%
  \BibitemOpen
  \bibfield  {author} {\bibinfo {author} {\bibfnamefont {F.}~\bibnamefont
  {L\'eonard}}\ and\ \bibinfo {author} {\bibfnamefont {B.}~\bibnamefont
  {Delamotte}},\ }\href {https://doi.org/10.1103/PhysRevLett.115.200601}
  {\bibfield  {journal} {\bibinfo  {journal} {Phys. Rev. Lett.}\ }\textbf
  {\bibinfo {volume} {115}},\ \bibinfo {pages} {200601} (\bibinfo {year}
  {2015})}\BibitemShut {NoStop}%
\bibitem [{\citenamefont {Canet}\ \emph
  {et~al.}(2004{\natexlab{a}})\citenamefont {Canet}, \citenamefont
  {Chat{\'{e}}},\ and\ \citenamefont {Delamotte}}]{Canet:2004je}%
  \BibitemOpen
  \bibfield  {author} {\bibinfo {author} {\bibfnamefont {L.}~\bibnamefont
  {Canet}}, \bibinfo {author} {\bibfnamefont {H.}~\bibnamefont {Chat{\'{e}}}},\
  and\ \bibinfo {author} {\bibfnamefont {B.}~\bibnamefont {Delamotte}},\ }\href
  {https://doi.org/10.1103/PhysRevLett.92.255703} {\bibfield  {journal}
  {\bibinfo  {journal} {Physical Review Letters}\ }\textbf {\bibinfo {volume}
  {92}},\ \bibinfo {pages} {255703} (\bibinfo {year} {2004}{\natexlab{a}})},\
  \Eprint {https://arxiv.org/abs/cond-mat/0403423} {arXiv:cond-mat/0403423
  [cond-mat]} \BibitemShut {NoStop}%
\bibitem [{\citenamefont {Canet}\ \emph
  {et~al.}(2004{\natexlab{b}})\citenamefont {Canet}, \citenamefont {Delamotte},
  \citenamefont {Deloubriere},\ and\ \citenamefont {Wschebor}}]{Canet:2003yu}%
  \BibitemOpen
  \bibfield  {author} {\bibinfo {author} {\bibfnamefont {L.}~\bibnamefont
  {Canet}}, \bibinfo {author} {\bibfnamefont {B.}~\bibnamefont {Delamotte}},
  \bibinfo {author} {\bibfnamefont {O.}~\bibnamefont {Deloubriere}},\ and\
  \bibinfo {author} {\bibfnamefont {N.}~\bibnamefont {Wschebor}},\ }\href
  {https://doi.org/10.1103/PhysRevLett.92.195703} {\bibfield  {journal}
  {\bibinfo  {journal} {Physical Review Letters}\ }\textbf {\bibinfo {volume}
  {92}},\ \bibinfo {pages} {195703} (\bibinfo {year} {2004}{\natexlab{b}})},\
  \Eprint {https://arxiv.org/abs/cond-mat/0309504} {arXiv:cond-mat/0309504
  [cond-mat.stat-mech]} \BibitemShut {NoStop}%
\bibitem [{\citenamefont {Machado}\ and\ \citenamefont
  {Dupuis}(2010)}]{Machado:2010wi}%
  \BibitemOpen
  \bibfield  {author} {\bibinfo {author} {\bibfnamefont {T.}~\bibnamefont
  {Machado}}\ and\ \bibinfo {author} {\bibfnamefont {N.}~\bibnamefont
  {Dupuis}},\ }\href {https://doi.org/10.1103/PhysRevE.82.041128} {\bibfield
  {journal} {\bibinfo  {journal} {Phys. Rev.}\ }\textbf {\bibinfo {volume}
  {E82}},\ \bibinfo {pages} {041128} (\bibinfo {year} {2010})},\ \Eprint
  {https://arxiv.org/abs/1004.3651} {arXiv:1004.3651 [cond-mat.stat-mech]}
  \BibitemShut {NoStop}%
\bibitem [{\citenamefont {Rose}\ \emph {et~al.}(2016)\citenamefont {Rose},
  \citenamefont {Benitez}, \citenamefont {L\'eonard},\ and\ \citenamefont
  {Delamotte}}]{Rose:2016wqz}%
  \BibitemOpen
  \bibfield  {author} {\bibinfo {author} {\bibfnamefont {F.}~\bibnamefont
  {Rose}}, \bibinfo {author} {\bibfnamefont {F.}~\bibnamefont {Benitez}},
  \bibinfo {author} {\bibfnamefont {F.}~\bibnamefont {L\'eonard}},\ and\
  \bibinfo {author} {\bibfnamefont {B.}~\bibnamefont {Delamotte}},\ }\href
  {https://doi.org/10.1103/PhysRevD.93.125018} {\bibfield  {journal} {\bibinfo
  {journal} {Phys. Rev.}\ }\textbf {\bibinfo {volume} {D93}},\ \bibinfo {pages}
  {125018} (\bibinfo {year} {2016})},\ \Eprint
  {https://arxiv.org/abs/1604.05285} {arXiv:1604.05285 [cond-mat.stat-mech]}
  \BibitemShut {NoStop}%
\bibitem [{\citenamefont {Meneses}\ \emph {et~al.}(2019)\citenamefont
  {Meneses}, \citenamefont {Penedones}, \citenamefont {Rychkov}, \citenamefont
  {Viana Parente~Lopes},\ and\ \citenamefont
  {Yvernay}}]{meneses2019structural}%
  \BibitemOpen
  \bibfield  {author} {\bibinfo {author} {\bibfnamefont {S.}~\bibnamefont
  {Meneses}}, \bibinfo {author} {\bibfnamefont {J.}~\bibnamefont {Penedones}},
  \bibinfo {author} {\bibfnamefont {S.}~\bibnamefont {Rychkov}}, \bibinfo
  {author} {\bibfnamefont {J.}~\bibnamefont {Viana Parente~Lopes}},\ and\
  \bibinfo {author} {\bibfnamefont {P.}~\bibnamefont {Yvernay}},\ }\href@noop
  {} {\bibfield  {journal} {\bibinfo  {journal} {Journal of High Energy
  Physics}\ }\textbf {\bibinfo {volume} {2019}},\ \bibinfo {pages} {1}
  (\bibinfo {year} {2019})}\BibitemShut {NoStop}%
\bibitem [{\citenamefont {Delamotte}\ \emph {et~al.}(2016)\citenamefont
  {Delamotte}, \citenamefont {Tissier},\ and\ \citenamefont
  {Wschebor}}]{delamotte2016scale}%
  \BibitemOpen
  \bibfield  {author} {\bibinfo {author} {\bibfnamefont {B.}~\bibnamefont
  {Delamotte}}, \bibinfo {author} {\bibfnamefont {M.}~\bibnamefont {Tissier}},\
  and\ \bibinfo {author} {\bibfnamefont {N.}~\bibnamefont {Wschebor}},\
  }\href@noop {} {\bibfield  {journal} {\bibinfo  {journal} {Physical Review
  E}\ }\textbf {\bibinfo {volume} {93}},\ \bibinfo {pages} {012144} (\bibinfo
  {year} {2016})}\BibitemShut {NoStop}%
\bibitem [{\citenamefont {Cardy}(1996)}]{cardy_1996}%
  \BibitemOpen
  \bibfield  {author} {\bibinfo {author} {\bibfnamefont {J.}~\bibnamefont
  {Cardy}},\ }\href {https://doi.org/10.1017/CBO9781316036440} {\emph {\bibinfo
  {title} {Scaling and Renormalization in Statistical Physics}}},\ Cambridge
  Lecture Notes in Physics\ (\bibinfo  {publisher} {Cambridge University
  Press},\ \bibinfo {year} {1996})\BibitemShut {NoStop}%
\bibitem [{\citenamefont {De~Polsi}\ \emph {et~al.}(2019)\citenamefont
  {De~Polsi}, \citenamefont {Tissier},\ and\ \citenamefont
  {Wschebor}}]{DePolsi2019}%
  \BibitemOpen
  \bibfield  {author} {\bibinfo {author} {\bibfnamefont {G.}~\bibnamefont
  {De~Polsi}}, \bibinfo {author} {\bibfnamefont {M.}~\bibnamefont {Tissier}},\
  and\ \bibinfo {author} {\bibfnamefont {N.}~\bibnamefont {Wschebor}},\ }\href
  {https://doi.org/10.1007/s10955-019-02411-3} {\bibfield  {journal} {\bibinfo
  {journal} {Journal of Statistical Physics}\ }\textbf {\bibinfo {volume}
  {177}},\ \bibinfo {pages} {1089} (\bibinfo {year} {2019})}\BibitemShut
  {NoStop}%
\bibitem [{\citenamefont {Balog}\ \emph {et~al.}(2020)\citenamefont {Balog},
  \citenamefont {De~Polsi}, \citenamefont {Tissier},\ and\ \citenamefont
  {Wschebor}}]{Balog2020}%
  \BibitemOpen
  \bibfield  {author} {\bibinfo {author} {\bibfnamefont {I.}~\bibnamefont
  {Balog}}, \bibinfo {author} {\bibfnamefont {G.}~\bibnamefont {De~Polsi}},
  \bibinfo {author} {\bibfnamefont {M.}~\bibnamefont {Tissier}},\ and\ \bibinfo
  {author} {\bibfnamefont {N.}~\bibnamefont {Wschebor}},\ }\href
  {https://doi.org/10.1103/PhysRevE.101.062146} {\bibfield  {journal} {\bibinfo
   {journal} {Phys. Rev. E}\ }\textbf {\bibinfo {volume} {101}},\ \bibinfo
  {pages} {062146} (\bibinfo {year} {2020})}\BibitemShut {NoStop}%
\bibitem [{\citenamefont {Balog}\ \emph
  {et~al.}(2019{\natexlab{a}})\citenamefont {Balog}, \citenamefont {Chat\'e},
  \citenamefont {Delamotte}, \citenamefont {Marohnic},\ and\ \citenamefont
  {Wschebor}}]{Balog:2019rrg}%
  \BibitemOpen
  \bibfield  {author} {\bibinfo {author} {\bibfnamefont {I.}~\bibnamefont
  {Balog}}, \bibinfo {author} {\bibfnamefont {H.}~\bibnamefont {Chat\'e}},
  \bibinfo {author} {\bibfnamefont {B.}~\bibnamefont {Delamotte}}, \bibinfo
  {author} {\bibfnamefont {M.}~\bibnamefont {Marohnic}},\ and\ \bibinfo
  {author} {\bibfnamefont {N.}~\bibnamefont {Wschebor}},\ }\href
  {https://doi.org/10.1103/PhysRevLett.123.240604} {\bibfield  {journal}
  {\bibinfo  {journal} {Phys. Rev. Lett.}\ }\textbf {\bibinfo {volume} {123}},\
  \bibinfo {pages} {240604} (\bibinfo {year} {2019}{\natexlab{a}})},\ \Eprint
  {https://arxiv.org/abs/1907.01829} {arXiv:1907.01829 [cond-mat.stat-mech]}
  \BibitemShut {NoStop}%
\bibitem [{\citenamefont {De~Polsi}\ \emph
  {et~al.}(2020{\natexlab{a}})\citenamefont {De~Polsi}, \citenamefont {Balog},
  \citenamefont {Tissier},\ and\ \citenamefont {Wschebor}}]{DePolsi2020}%
  \BibitemOpen
  \bibfield  {author} {\bibinfo {author} {\bibfnamefont {G.}~\bibnamefont
  {De~Polsi}}, \bibinfo {author} {\bibfnamefont {I.}~\bibnamefont {Balog}},
  \bibinfo {author} {\bibfnamefont {M.}~\bibnamefont {Tissier}},\ and\ \bibinfo
  {author} {\bibfnamefont {N.}~\bibnamefont {Wschebor}},\ }\href
  {https://doi.org/10.1103/PhysRevE.101.042113} {\bibfield  {journal} {\bibinfo
   {journal} {Phys. Rev. E}\ }\textbf {\bibinfo {volume} {101}},\ \bibinfo
  {pages} {042113} (\bibinfo {year} {2020}{\natexlab{a}})}\BibitemShut
  {NoStop}%
\bibitem [{\citenamefont {De~Polsi}\ \emph {et~al.}(2021)\citenamefont
  {De~Polsi}, \citenamefont {Hern\'andez-Chifflet},\ and\ \citenamefont
  {Wschebor}}]{DePolsi2021}%
  \BibitemOpen
  \bibfield  {author} {\bibinfo {author} {\bibfnamefont {G.}~\bibnamefont
  {De~Polsi}}, \bibinfo {author} {\bibfnamefont {G.}~\bibnamefont
  {Hern\'andez-Chifflet}},\ and\ \bibinfo {author} {\bibfnamefont
  {N.}~\bibnamefont {Wschebor}},\ }\href
  {https://doi.org/10.1103/PhysRevE.104.064101} {\bibfield  {journal} {\bibinfo
   {journal} {Phys. Rev. E}\ }\textbf {\bibinfo {volume} {104}},\ \bibinfo
  {pages} {064101} (\bibinfo {year} {2021})}\BibitemShut {NoStop}%
\bibitem [{\citenamefont {Berges}\ \emph
  {et~al.}(2002{\natexlab{a}})\citenamefont {Berges}, \citenamefont
  {Tetradis},\ and\ \citenamefont {Wetterich}}]{Berges:2000ew}%
  \BibitemOpen
  \bibfield  {author} {\bibinfo {author} {\bibfnamefont {J.}~\bibnamefont
  {Berges}}, \bibinfo {author} {\bibfnamefont {N.}~\bibnamefont {Tetradis}},\
  and\ \bibinfo {author} {\bibfnamefont {C.}~\bibnamefont {Wetterich}},\ }\href
  {https://doi.org/10.1016/S0370-1573(01)00098-9} {\bibfield  {journal}
  {\bibinfo  {journal} {Phys. Rept.}\ }\textbf {\bibinfo {volume} {363}},\
  \bibinfo {pages} {223} (\bibinfo {year} {2002}{\natexlab{a}})},\ \Eprint
  {https://arxiv.org/abs/hep-ph/0005122} {arXiv:hep-ph/0005122 [hep-ph]}
  \BibitemShut {NoStop}%
\bibitem [{\citenamefont {Rose}\ \emph {et~al.}(2015)\citenamefont {Rose},
  \citenamefont {L\'eonard},\ and\ \citenamefont {Dupuis}}]{Rose:2015bma}%
  \BibitemOpen
  \bibfield  {author} {\bibinfo {author} {\bibfnamefont {F.}~\bibnamefont
  {Rose}}, \bibinfo {author} {\bibfnamefont {F.}~\bibnamefont {L\'eonard}},\
  and\ \bibinfo {author} {\bibfnamefont {N.}~\bibnamefont {Dupuis}},\ }\href
  {https://doi.org/10.1103/PhysRevB.91.224501} {\bibfield  {journal} {\bibinfo
  {journal} {Phys. Rev. B}\ }\textbf {\bibinfo {volume} {91}},\ \bibinfo
  {pages} {224501} (\bibinfo {year} {2015})},\ \Eprint
  {https://arxiv.org/abs/1503.08688} {arXiv:1503.08688 [cond-mat.quant-gas]}
  \BibitemShut {NoStop}%
\bibitem [{\citenamefont {Rose}\ \emph {et~al.}(2022)\citenamefont {Rose},
  \citenamefont {Pagani},\ and\ \citenamefont {Dupuis}}]{Rose:2021zdk}%
  \BibitemOpen
  \bibfield  {author} {\bibinfo {author} {\bibfnamefont {F.}~\bibnamefont
  {Rose}}, \bibinfo {author} {\bibfnamefont {C.}~\bibnamefont {Pagani}},\ and\
  \bibinfo {author} {\bibfnamefont {N.}~\bibnamefont {Dupuis}},\ }\href
  {https://doi.org/10.1103/PhysRevD.105.065020} {\bibfield  {journal} {\bibinfo
   {journal} {Phys. Rev. D}\ }\textbf {\bibinfo {volume} {105}},\ \bibinfo
  {pages} {065020} (\bibinfo {year} {2022})},\ \Eprint
  {https://arxiv.org/abs/2110.13174} {arXiv:2110.13174 [hep-th]} \BibitemShut
  {NoStop}%
\bibitem [{\citenamefont {Polchinski}(1984)}]{Polchinski:1983gv}%
  \BibitemOpen
  \bibfield  {author} {\bibinfo {author} {\bibfnamefont {J.}~\bibnamefont
  {Polchinski}},\ }\href {https://doi.org/10.1016/0550-3213(84)90287-6}
  {\bibfield  {journal} {\bibinfo  {journal} {Nucl. Phys.}\ }\textbf {\bibinfo
  {volume} {B231}},\ \bibinfo {pages} {269} (\bibinfo {year}
  {1984})}\BibitemShut {NoStop}%
\bibitem [{\citenamefont {Pawlowski}(2007)}]{Pawlowski:2005xe}%
  \BibitemOpen
  \bibfield  {author} {\bibinfo {author} {\bibfnamefont {J.~M.}\ \bibnamefont
  {Pawlowski}},\ }\href {https://doi.org/10.1016/j.aop.2007.01.007} {\bibfield
  {journal} {\bibinfo  {journal} {Annals Phys.}\ }\textbf {\bibinfo {volume}
  {322}},\ \bibinfo {pages} {2831} (\bibinfo {year} {2007})},\ \Eprint
  {https://arxiv.org/abs/hep-th/0512261} {arXiv:hep-th/0512261} \BibitemShut
  {NoStop}%
\bibitem [{\citenamefont {Wilson}(1969)}]{Wilson1969}%
  \BibitemOpen
  \bibfield  {author} {\bibinfo {author} {\bibfnamefont {K.~G.}\ \bibnamefont
  {Wilson}},\ }\href {https://doi.org/10.1103/PhysRev.179.1499} {\bibfield
  {journal} {\bibinfo  {journal} {Phys. Rev.}\ }\textbf {\bibinfo {volume}
  {179}},\ \bibinfo {pages} {1499} (\bibinfo {year} {1969})}\BibitemShut
  {NoStop}%
\bibitem [{\citenamefont {Zimmermann}(1973)}]{Zimmerman1973}%
  \BibitemOpen
  \bibfield  {author} {\bibinfo {author} {\bibfnamefont {W.}~\bibnamefont
  {Zimmermann}},\ }\href
  {https://doi.org/https://doi.org/10.1016/0003-4916(73)90430-2} {\bibfield
  {journal} {\bibinfo  {journal} {Annals of Physics}\ }\textbf {\bibinfo
  {volume} {77}},\ \bibinfo {pages} {570} (\bibinfo {year} {1973})}\BibitemShut
  {NoStop}%
\bibitem [{\citenamefont {Wilson}(1964)}]{wilson1964products}%
  \BibitemOpen
  \bibfield  {author} {\bibinfo {author} {\bibfnamefont {K.}~\bibnamefont
  {Wilson}},\ }\href@noop {} {\bibfield  {journal} {\bibinfo  {journal}
  {Cornell report}\ }\textbf {\bibinfo {volume} {1364}} (\bibinfo {year}
  {1964})}\BibitemShut {NoStop}%
\bibitem [{\citenamefont {Morris}(1994{\natexlab{b}})}]{Morris:1994ie}%
  \BibitemOpen
  \bibfield  {author} {\bibinfo {author} {\bibfnamefont {T.~R.}\ \bibnamefont
  {Morris}},\ }\href {https://doi.org/10.1016/0370-2693(94)90767-6} {\bibfield
  {journal} {\bibinfo  {journal} {Phys. Lett.}\ }\textbf {\bibinfo {volume}
  {B329}},\ \bibinfo {pages} {241} (\bibinfo {year} {1994}{\natexlab{b}})},\
  \Eprint {https://arxiv.org/abs/hep-ph/9403340} {arXiv:hep-ph/9403340
  [hep-ph]} \BibitemShut {NoStop}%
\bibitem [{\citenamefont {Morris}\ and\ \citenamefont
  {Turner}(1998)}]{Morris:1997xj}%
  \BibitemOpen
  \bibfield  {author} {\bibinfo {author} {\bibfnamefont {T.~R.}\ \bibnamefont
  {Morris}}\ and\ \bibinfo {author} {\bibfnamefont {M.~D.}\ \bibnamefont
  {Turner}},\ }\href {https://doi.org/10.1016/S0550-3213(97)00640-8} {\bibfield
   {journal} {\bibinfo  {journal} {Nucl. Phys.}\ }\textbf {\bibinfo {volume}
  {B509}},\ \bibinfo {pages} {637} (\bibinfo {year} {1998})},\ \Eprint
  {https://arxiv.org/abs/hep-th/9704202} {arXiv:hep-th/9704202 [hep-th]}
  \BibitemShut {NoStop}%
\bibitem [{\citenamefont {Berges}\ \emph
  {et~al.}(2002{\natexlab{b}})\citenamefont {Berges}, \citenamefont
  {Tetradis},\ and\ \citenamefont {Wetterich}}]{Berges2002}%
  \BibitemOpen
  \bibfield  {author} {\bibinfo {author} {\bibfnamefont {J.~J.}\ \bibnamefont
  {Berges}}, \bibinfo {author} {\bibfnamefont {N.}~\bibnamefont {Tetradis}},\
  and\ \bibinfo {author} {\bibfnamefont {C.}~\bibnamefont {Wetterich}},\ }\href
  {https://doi.org/10.1016/S0370-1573(01)00098-9} {\bibfield  {journal}
  {\bibinfo  {journal} {Phys. Rept.}\ }\textbf {\bibinfo {volume} {363}},\
  \bibinfo {pages} {223} (\bibinfo {year} {2002}{\natexlab{b}})},\ \Eprint
  {https://arxiv.org/abs/hep-ph/0005122} {arXiv:hep-ph/0005122 [hep-ph]}
  \BibitemShut {NoStop}%
\bibitem [{\citenamefont {Canet}\ \emph
  {et~al.}(2003{\natexlab{a}})\citenamefont {Canet}, \citenamefont {Delamotte},
  \citenamefont {Mouhanna},\ and\ \citenamefont {Vidal}}]{Canet2003b}%
  \BibitemOpen
  \bibfield  {author} {\bibinfo {author} {\bibfnamefont {L.~L.}\ \bibnamefont
  {Canet}}, \bibinfo {author} {\bibfnamefont {B.}~\bibnamefont {Delamotte}},
  \bibinfo {author} {\bibfnamefont {D.}~\bibnamefont {Mouhanna}},\ and\
  \bibinfo {author} {\bibfnamefont {J.}~\bibnamefont {Vidal}},\ }\href
  {https://doi.org/10.1103/PhysRevB.68.064421} {\bibfield  {journal} {\bibinfo
  {journal} {Physical Review B - Condensed Matter and Materials Physics}\
  }\textbf {\bibinfo {volume} {68}},\ \bibinfo {pages} {64421} (\bibinfo {year}
  {2003}{\natexlab{a}})},\ \Eprint {https://arxiv.org/abs/hep-th/0302227}
  {arXiv:hep-th/0302227 [hep-th]} \BibitemShut {NoStop}%
\bibitem [{\citenamefont {Canet}\ \emph
  {et~al.}(2003{\natexlab{b}})\citenamefont {Canet}, \citenamefont {Delamotte},
  \citenamefont {Mouhanna},\ and\ \citenamefont {Vidal}}]{Canet2003}%
  \BibitemOpen
  \bibfield  {author} {\bibinfo {author} {\bibfnamefont {L.}~\bibnamefont
  {Canet}}, \bibinfo {author} {\bibfnamefont {B.}~\bibnamefont {Delamotte}},
  \bibinfo {author} {\bibfnamefont {D.}~\bibnamefont {Mouhanna}},\ and\
  \bibinfo {author} {\bibfnamefont {J.}~\bibnamefont {Vidal}},\ }\href
  {https://doi.org/10.1103/PhysRevD.67.065004} {\bibfield  {journal} {\bibinfo
  {journal} {Physical Review D - Particles, Fields, Gravitation and Cosmology}\
  }\textbf {\bibinfo {volume} {67}},\ \bibinfo {pages} {065004} (\bibinfo
  {year} {2003}{\natexlab{b}})}\BibitemShut {NoStop}%
\bibitem [{\citenamefont {Balog}\ \emph
  {et~al.}(2019{\natexlab{b}})\citenamefont {Balog}, \citenamefont
  {Chat{\'{e}}}, \citenamefont {Delamotte}, \citenamefont {Marohnic},
  \citenamefont {Wschebor}, \citenamefont {Marohni{\'{c}}},\ and\ \citenamefont
  {Wschebor}}]{Balog2019}%
  \BibitemOpen
  \bibfield  {author} {\bibinfo {author} {\bibfnamefont {I.}~\bibnamefont
  {Balog}}, \bibinfo {author} {\bibfnamefont {H.}~\bibnamefont {Chat{\'{e}}}},
  \bibinfo {author} {\bibfnamefont {B.}~\bibnamefont {Delamotte}}, \bibinfo
  {author} {\bibfnamefont {M.}~\bibnamefont {Marohnic}}, \bibinfo {author}
  {\bibfnamefont {N.}~\bibnamefont {Wschebor}}, \bibinfo {author}
  {\bibfnamefont {M.}~\bibnamefont {Marohni{\'{c}}}},\ and\ \bibinfo {author}
  {\bibfnamefont {N.}~\bibnamefont {Wschebor}},\ }\href
  {https://doi.org/10.1103/PhysRevLett.123.240604} {\bibfield  {journal}
  {\bibinfo  {journal} {Phys. Rev. Lett.}\ }\textbf {\bibinfo {volume} {123}},\
  \bibinfo {pages} {240604} (\bibinfo {year} {2019}{\natexlab{b}})},\ \Eprint
  {https://arxiv.org/abs/1907.01829} {arXiv:1907.01829 [cond-mat.stat-mech]}
  \BibitemShut {NoStop}%
\bibitem [{\citenamefont {De~Polsi}\ and\ \citenamefont
  {Wschebor}(2022)}]{DePolsi2022}%
  \BibitemOpen
  \bibfield  {author} {\bibinfo {author} {\bibfnamefont {G.}~\bibnamefont
  {De~Polsi}}\ and\ \bibinfo {author} {\bibfnamefont {N.}~\bibnamefont
  {Wschebor}},\ }\href {https://doi.org/10.1103/PhysRevE.106.024111} {\bibfield
   {journal} {\bibinfo  {journal} {Phys. Rev. E}\ }\textbf {\bibinfo {volume}
  {106}},\ \bibinfo {pages} {024111} (\bibinfo {year} {2022})}\BibitemShut
  {NoStop}%
\bibitem [{\citenamefont {Rosten}(2017)}]{Rosten:2014oja}%
  \BibitemOpen
  \bibfield  {author} {\bibinfo {author} {\bibfnamefont {O.~J.}\ \bibnamefont
  {Rosten}},\ }\href {https://doi.org/10.1140/epjc/s10052-017-5049-5}
  {\bibfield  {journal} {\bibinfo  {journal} {Eur. Phys. J.}\ }\textbf
  {\bibinfo {volume} {C77}},\ \bibinfo {pages} {477} (\bibinfo {year}
  {2017})},\ \Eprint {https://arxiv.org/abs/1411.2603} {arXiv:1411.2603
  [hep-th]} \BibitemShut {NoStop}%
\bibitem [{\citenamefont {Sonoda}(2015)}]{Sonoda:2015pva}%
  \BibitemOpen
  \bibfield  {author} {\bibinfo {author} {\bibfnamefont {H.}~\bibnamefont
  {Sonoda}},\ }\href {https://doi.org/10.1103/PhysRevD.92.065016} {\bibfield
  {journal} {\bibinfo  {journal} {Phys. Rev. D}\ }\textbf {\bibinfo {volume}
  {92}},\ \bibinfo {pages} {065016} (\bibinfo {year} {2015})},\ \Eprint
  {https://arxiv.org/abs/1504.02831} {arXiv:1504.02831 [hep-th]} \BibitemShut
  {NoStop}%
\bibitem [{\citenamefont {Rosten}(2019)}]{Rosten:2016zap}%
  \BibitemOpen
  \bibfield  {author} {\bibinfo {author} {\bibfnamefont {O.~J.}\ \bibnamefont
  {Rosten}},\ }\href {https://doi.org/10.1142/S0217751X19500271} {\bibfield
  {journal} {\bibinfo  {journal} {Int. J. Mod. Phys.}\ }\textbf {\bibinfo
  {volume} {A34}},\ \bibinfo {pages} {1950027} (\bibinfo {year} {2019})},\
  \Eprint {https://arxiv.org/abs/1605.01729} {arXiv:1605.01729 [hep-th]}
  \BibitemShut {NoStop}%
\bibitem [{\citenamefont {Simmons-Duffin}(2017)}]{Simmons-Duffin:2016wlq}%
  \BibitemOpen
  \bibfield  {author} {\bibinfo {author} {\bibfnamefont {D.}~\bibnamefont
  {Simmons-Duffin}},\ }\href {https://doi.org/10.1007/JHEP03(2017)086}
  {\bibfield  {journal} {\bibinfo  {journal} {JHEP}\ }\textbf {\bibinfo
  {volume} {03}},\ \bibinfo {pages} {086}},\ \Eprint
  {https://arxiv.org/abs/1612.08471} {arXiv:1612.08471 [hep-th]} \BibitemShut
  {NoStop}%
\bibitem [{\citenamefont {Ihssen}\ \emph {et~al.}(2023)\citenamefont {Ihssen},
  \citenamefont {Sattler},\ and\ \citenamefont {Wink}}]{wink23}%
  \BibitemOpen
  \bibfield  {author} {\bibinfo {author} {\bibfnamefont {F.}~\bibnamefont
  {Ihssen}}, \bibinfo {author} {\bibfnamefont {F.~R.}\ \bibnamefont
  {Sattler}},\ and\ \bibinfo {author} {\bibfnamefont {N.}~\bibnamefont
  {Wink}},\ }\href {https://doi.org/10.1103/PhysRevD.107.114009} {\bibfield
  {journal} {\bibinfo  {journal} {Phys. Rev. D}\ }\textbf {\bibinfo {volume}
  {107}},\ \bibinfo {pages} {114009} (\bibinfo {year} {2023})}\BibitemShut
  {NoStop}%
\bibitem [{\citenamefont {De~Polsi}\ \emph
  {et~al.}(2020{\natexlab{b}})\citenamefont {De~Polsi}, \citenamefont {Balog},
  \citenamefont {Tissier},\ and\ \citenamefont {Wschebor}}]{DePolsi:2020pjk}%
  \BibitemOpen
  \bibfield  {author} {\bibinfo {author} {\bibfnamefont {G.}~\bibnamefont
  {De~Polsi}}, \bibinfo {author} {\bibfnamefont {I.}~\bibnamefont {Balog}},
  \bibinfo {author} {\bibfnamefont {M.}~\bibnamefont {Tissier}},\ and\ \bibinfo
  {author} {\bibfnamefont {N.}~\bibnamefont {Wschebor}},\ }\href
  {https://doi.org/10.1103/PhysRevE.101.042113} {\bibfield  {journal} {\bibinfo
   {journal} {Phys. Rev. E}\ }\textbf {\bibinfo {volume} {101}},\ \bibinfo
  {pages} {042113} (\bibinfo {year} {2020}{\natexlab{b}})}\BibitemShut
  {NoStop}%
\bibitem [{\citenamefont {P\'eli}(2021)}]{Peli:2020yiz}%
  \BibitemOpen
  \bibfield  {author} {\bibinfo {author} {\bibfnamefont {Z.}~\bibnamefont
  {P\'eli}},\ }\href {https://doi.org/10.1103/PhysRevE.103.032135} {\bibfield
  {journal} {\bibinfo  {journal} {Phys. Rev. E}\ }\textbf {\bibinfo {volume}
  {103}},\ \bibinfo {pages} {032135} (\bibinfo {year} {2021})},\ \Eprint
  {https://arxiv.org/abs/2010.04020} {arXiv:2010.04020 [hep-th]} \BibitemShut
  {NoStop}%
\bibitem [{\citenamefont {Blaizot}\ \emph {et~al.}(2006)\citenamefont
  {Blaizot}, \citenamefont {M{\'{e}}ndez-Galain},\ and\ \citenamefont
  {Wschebor}}]{Blaizot2006}%
  \BibitemOpen
  \bibfield  {author} {\bibinfo {author} {\bibfnamefont {J.~P.}\ \bibnamefont
  {Blaizot}}, \bibinfo {author} {\bibfnamefont {R.}~\bibnamefont
  {M{\'{e}}ndez-Galain}},\ and\ \bibinfo {author} {\bibfnamefont
  {N.}~\bibnamefont {Wschebor}},\ }\href
  {https://doi.org/10.1016/j.physletb.2005.10.086} {\bibfield  {journal}
  {\bibinfo  {journal} {Physics Letters, Section B: Nuclear, Elementary
  Particle and High-Energy Physics}\ }\textbf {\bibinfo {volume} {632}},\
  \bibinfo {pages} {571} (\bibinfo {year} {2006})}\BibitemShut {NoStop}%
\bibitem [{\citenamefont {Benitez}\ \emph
  {et~al.}(2012{\natexlab{a}})\citenamefont {Benitez}, \citenamefont {Blaizot},
  \citenamefont {Chat\'e}, \citenamefont {Delamotte}, \citenamefont
  {M\'endez-Galain},\ and\ \citenamefont {Wschebor}}]{Benitez:2011xx}%
  \BibitemOpen
  \bibfield  {author} {\bibinfo {author} {\bibfnamefont {F.}~\bibnamefont
  {Benitez}}, \bibinfo {author} {\bibfnamefont {J.~P.}\ \bibnamefont
  {Blaizot}}, \bibinfo {author} {\bibfnamefont {H.}~\bibnamefont {Chat\'e}},
  \bibinfo {author} {\bibfnamefont {B.}~\bibnamefont {Delamotte}}, \bibinfo
  {author} {\bibfnamefont {R.}~\bibnamefont {M\'endez-Galain}},\ and\ \bibinfo
  {author} {\bibfnamefont {N.}~\bibnamefont {Wschebor}},\ }\href
  {https://doi.org/10.1103/PhysRevE.85.026707} {\bibfield  {journal} {\bibinfo
  {journal} {Phys. Rev.}\ }\textbf {\bibinfo {volume} {E85}},\ \bibinfo {pages}
  {026707} (\bibinfo {year} {2012}{\natexlab{a}})},\ \Eprint
  {https://arxiv.org/abs/1110.2665} {arXiv:1110.2665 [cond-mat.stat-mech]}
  \BibitemShut {NoStop}%
\bibitem [{\citenamefont {Benitez}\ \emph
  {et~al.}(2012{\natexlab{b}})\citenamefont {Benitez}, \citenamefont {Blaizot},
  \citenamefont {Chat{\'{e}}}, \citenamefont {Delamotte}, \citenamefont
  {M{\'{e}}ndez-Galain},\ and\ \citenamefont {Wschebor}}]{Benitez2012}%
  \BibitemOpen
  \bibfield  {author} {\bibinfo {author} {\bibfnamefont {F.}~\bibnamefont
  {Benitez}}, \bibinfo {author} {\bibfnamefont {J.~P.}\ \bibnamefont
  {Blaizot}}, \bibinfo {author} {\bibfnamefont {H.}~\bibnamefont
  {Chat{\'{e}}}}, \bibinfo {author} {\bibfnamefont {B.}~\bibnamefont
  {Delamotte}}, \bibinfo {author} {\bibfnamefont {R.}~\bibnamefont
  {M{\'{e}}ndez-Galain}},\ and\ \bibinfo {author} {\bibfnamefont
  {N.}~\bibnamefont {Wschebor}},\ }\bibfield  {journal} {\bibinfo  {journal}
  {Physical Review E - Statistical, Nonlinear, and Soft Matter Physics}\ }\href
  {https://doi.org/10.1103/PhysRevE.85.026707} {10.1103/PhysRevE.85.026707}
  (\bibinfo {year} {2012}{\natexlab{b}}),\ \Eprint
  {https://arxiv.org/abs/1110.2665} {arXiv:1110.2665} \BibitemShut {NoStop}%
\end{thebibliography}%

\end{document}